\documentclass{article}

\usepackage{arxiv}

\usepackage[utf8]{inputenc} 
\usepackage[T1]{fontenc}    
\usepackage{hyperref}       
\usepackage{url}            
\usepackage{booktabs}       
\usepackage{amsfonts}       
\usepackage{nicefrac}       
\usepackage{microtype}      
\usepackage{lipsum}
\usepackage{graphicx}
\graphicspath{ {./images/} }

\title{Cyberattacks and Countermeasures for In-Vehicle Networks}

\author{
 Emad Aliwa \\
  School of Computer Science and Informatics\\
  Cardiff University, UK \\
  \texttt{aliwaem@cardiff.ac.uk} \\
   \And
  Omer Rana \\
  School of Computer Science and Informatics\\
  Cardiff University, UK \\
  \texttt{RanaOF@cardiff.ac.uk} \\
  \And
  Charith Perera \\
  School of Computer Science and Informatics\\
  Cardiff University, UK \\
  \texttt{PereraC@cardiff.ac.uk} \\
  \And
   Peter Burnap \\
  School of Computer Science and Informatics\\
  Cardiff University, UK \\
  \texttt{BurnapP@cardiff.ac.uk} \\
}
\usepackage{subcaption}

\newenvironment{noindlist2}
{\begin{list}{\labelitemi}{\leftmargin=0.6em \itemindent=0em}}
	{\end{list}}
\begin{document}
\maketitle
\begin{abstract}
As connectivity between and within vehicles increases, so does concern about safety and security. Various automotive serial protocols are used inside vehicles such as Controller Area Network (CAN), Local Interconnect Network (LIN) and FlexRay. CAN bus is the most used in-vehicle network protocol to support exchange of vehicle parameters between Electronic Control Units (ECUs). This protocol lacks security mechanisms by design and is therefore vulnerable to various attacks. Furthermore, connectivity of vehicles has made the CAN bus not only vulnerable from within the vehicle but also from outside. With the rise of connected cars, more entry points and interfaces have been introduced on board vehicles, thereby also leading to a wider potential attack surface. Existing security mechanisms focus on the use of encryption, authentication and vehicle Intrusion Detection Systems (IDS), which operate under various constrains such as low bandwidth, small frame size (e.g. in the CAN protocol), limited availability of computational resources and real-time sensitivity. We survey In-Vehicle Network (IVN) attacks which have been grouped under: direct interfaces-initiated attacks, telematics and infotainment-initiated attacks, and sensor-initiated attacks. We survey and classify current cryptographic and IDS approaches and compare these approaches based on criteria such as real time constrains, types of hardware used, changes in CAN bus behaviour, types of attack mitigation and software/ hardware used to validate these approaches. We conclude with potential mitigation strategies and research challenges for the future. 
\end{abstract}


%


\maketitle

\section{Introduction}
In recent years, vehicles have become more connected (to other vehicles -- referred to as Vehicle-2-Vehicle (V2V) and external infrastructure -- referred to as Vehicle-2-Infrastructure (V2I)) and the cyberattack surface for these vehicles continues to increase. Cyberattacks have also become a real concern for vehicle manufacturers, especially where services need to be supported using networks outside a vehicle. These services can include Global Positioning Systems (GPS), On-Board Diagnostic (OBD-2) based cellular dongles and entertainment services. As a result, vehicles are now more vulnerable to different attacks not only from inside but also from outside the vehicle. For instance, recent report has indicated that two famous connected cars in Europe from Ford and Volkswagen are vulnerable to cyberattacks from infotainment unit \cite{IET2020}. As potential cyberattacks on vehicles have widened, more vulnerabilities and entry points have been discovered -- generally grouped under: direct interfaces-initiated attacks, infotainment-initiated attacks, telematics-initiated attacks and sensor-initiated attacks. This raises the need for better security mechanisms. Due to lack of suitable security support in the Controller Area Network (CAN) protocol itself,  mechanisms to secure communications between components within a vehicle is also limited. Attacks such as CAN bus Denial of Service (DoS) and bus injection attacks are common~\cite{Deng2017}. CAN bus security limitations have been investigated by various researchers over both laboratory-based environments and in real vehicles. The attacks demonstrate how attackers are able to successfully take control of various parts of a vehicle, such as brakes, lights, steering and gears \cite{Currie2015}. Such attacks and  malicious data on the CAN bus was generated from both on-board the vehicle and at remote locations. 

Serial protocols are used for in-vehicle networks to exchange parameters between Electronic Control Units (ECUs) and sensors. These protocols lack security mechanisms by design and are thus vulnerable to various attacks. Researchers have also shown how to attack vehicles from within a vehicle using direct interfaces and infotainment systems via the On-Board Diagnostics port (OBD-2), USB and CD player and from outside the vehicle using medium and long distance communication such as Wi-Fi, Bluetooth, mobile (phone) networks, and sensors signals such as keyless fob attacks and tyre pressure monitoring system sensors. These attacks have widened the potential attack entry points within a  connected vehicle -- suggesting the importance of protecting the CAN bus. This survey provides the following contributions:
\begin{itemize}
\item description of in-vehicle serial bus protocols (particularly the CAN bus);
\item evaluation of current cryptographic and IDS approaches used for protecting vehicular data;
\item comparison and assessment of current mitigation strategies to protect vehicles against cyberattacks;
\item challenges and potential future research directions for in-vehicular cybersecurity. 
\end{itemize}

The rest of this paper is structured as follows: an introduction to serial data exchange protocols within a vehicle is provided in~Section~\hyperref[Sec2:Automotive-Protocols]{2}. This material provides the context for the rest of this paper -- outlining key concepts and terminology. In Section~\hyperref[Sec3:CAN-Protocols]{3} the CAN bus protocol, bus architecture along with hardware and software used inside vehicles is described. In Section~\hyperref[Sec4:Connected-Cars]{4} we describe the connected car infrastructure, including various ECUs and sensors that can be used inside a vehicle. In Section~\hyperref[Sec5:Vulnerabilitiesof-IVN-CAN-bus]{5} attacks initiated using data interfaces, telematics, infotainment and sensor entry points and how such attacks can be generated are described. In Section~\hyperref[Sec6:CANbus-SecurityMechanism]{6}, we review security mechanisms reported in literature to secure the CAN bus, which includes encryption, message authentication and vehicular Intrusion Detection Systems (IDS). We evaluate existing approaches based on criteria such as real time data requirements, the types of network infrastructure required, computational resource constrains, modifications needed for vehicular hardware, change in the protocol behaviour, types of attacks mitigated and software/ hardware used to validate these approaches. Finally, in section~\hyperref[Sec7:MitigationStrategy-Evaluation]{7} we evaluate mitigation strategies mentioned in section~\hyperref[Sec6:CANbus-SecurityMechanism]{6} and discuss future research challenges.

%
%

\section{Automotive Serial Bus Protocols}
\label{Sec2:Automotive-Protocols}

\begin{figure}[ht]
  \centering
  \includegraphics[width=\linewidth]{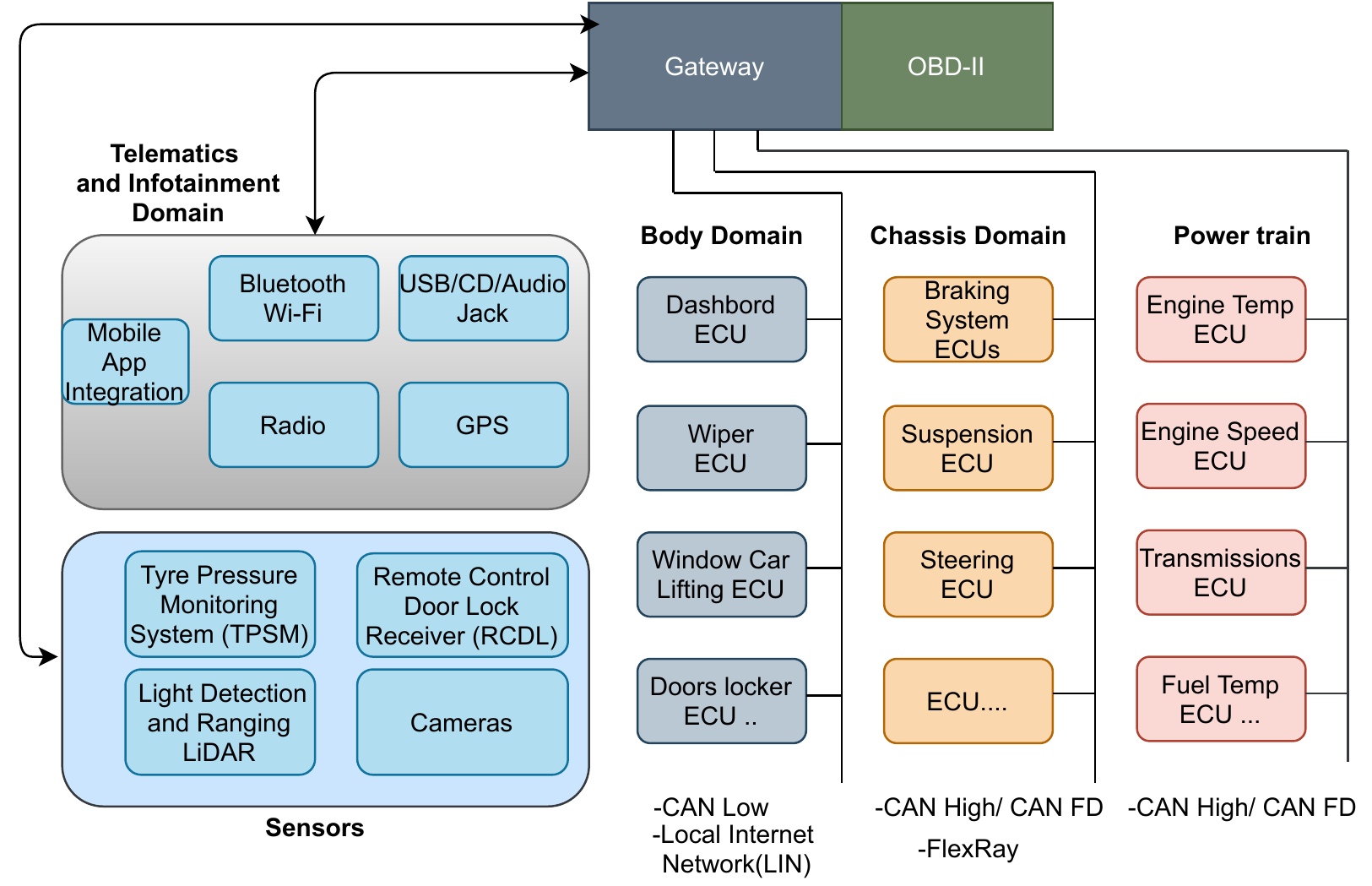}
  \caption{Serial bus protocols inside a vehicle -- from~\cite{Huang2017}. The figure focuses on the three most popular protocols: Controller Area Network (CAN), Local InterConnect Network(LIN) and FlexRay.}
  \label{Fig:Automotive-Serial-Protocols}
\end{figure}

Three key protocols are used inside vehicles for data commuications: CAN, FlexRay and LIN. Figure~\hyperref[Fig:Automotive-Serial-Protocols]{1}  shows the serial bus protocols used inside a vehicle. The CAN bus protocol is the most widely used to support critical functions such as Powertrain, engine management, anti-brake system and transmission. A vehicular system is divided into four domains, in terms of the function it performs and  whether it requires real time data~\cite{Navet2005}, as outlined below:

\begin{itemize}
\item  \textbf {Power train domain} such as engine and transmission functions. This domain is critical and needs real time response;  
\item  \textbf{Chassis domain} such as braking system, suspension and steering which also provides real time and safety critical functions inside the vehicle; 
\item  \textbf{Body domain} for functions such as dashboard, wipers, lights, windows and seats. These functions do not generally require real time response; 
\item  \textbf{Telematics and infotainment domain} which manages the various communications, information and entertainment  services inside a vehicle, e.g. in-car navigation, CD/ DVD players, rear seat entertainment systems, driving assistance and wireless interfaces.
\end{itemize}

These domains differ in terms of the functions they provide and the performance and quality of the network they require. Based on these differences, each domain has different performance and response time requirements. For example, a high speed CAN bus is used for real time and safety related domains inside the vehicle such as  transmission sensors. LIN and low speed CAN bus are however used for non-critical functions inside a vehicle, as they are less expensive compared with the high speed CAN and FlexRay buses. 
%

\subsection{Controller Area Network (CAN)}

CAN is serial communication protocol used for real-time, safety critical functions inside  road vehicles and other controlled applications \cite{TheInternationalOrganizationforStandardization2015}. It is a multi-master protocol and most widely used inside vehicles \cite{Kang2018}. Below are the main characteristics of the CAN protocol: 
 
\begin{itemize}
\item the maximum bitrate is 1Mbps in the {\it classical} CAN bus; 
\item high speed CAN bus bitrate can vary from 125kbps to 1Mbps, while low speed CAN bus bitrate from 5kbps to 125kbps; 
\item critical sensors can be connected to high speed CAN bus while less critical sensors can be connected to a low speed CAN bus;
\item in CAN Flexible Data rate (FD), the bitrate is up to 8 Mbps -- with a payload size of 8 bytes in Classical CAN and up to 64 bytes in CAN FD;
\item CAN bus protocol provides real time access for critical sensors via the Carrier Sense Multiple Access with Collision Avoidance and Arbitration Priority (CSMA/CA-AP) access/arbitration mechanism;
\item CAN bus is implemented physically using the twisted pair cable;  
\item CAN bus is used in real time automotive applications such as engine management, transmission, braking system and steering;
\item CAN bus provides error detection and correction mechanisms;
\item CAN XL is expected to be announced in 2020~\cite{CiA2020}, and is the third generation of CAN which provides up to 2048 bytes of data payload and bitrate of up to 10Mbps~\cite{RobertBoschGmbH2019}. This generation of the protocols is expected to be used with Internet Protocol (IP) based services.
\end{itemize}

As the CAN bus is the most widely used protocol inside a vehicle, a more detailed description of this protocol is provided in Section~\hyperref[Sec3:CAN-Protocols]{3}.

%
%

\subsection{Local Interconnect Network (LIN)}

The LIN protocol is used for low cost vehicular applications which have a low bit rate, asynchronous data requirement. It is used for non-critical applications such as door module and air condition systems \cite{ISO2016}. LIN is used where the reliability and robustness of the network is not critical \cite{Seung-Han2008}. It is standardised in International Organization for Standardization (ISO) 17987 series, which has been sub-divided into seven sub-standards -- describing the protocol functions aligned with the OSI layers such as physical layer, data link frame etc~\cite{ISO2016}. The LIN protocol has the following features: 
\begin{itemize}
\item a bit rate ranging from 1kbps to 20kbps;
\item single master with multiple slave nodes -- with support for up to 16 nodes;
\item a single physical wire to realise the bus;
\item standardised as ISO standard 17987.
\end{itemize}

\subsection{FlexRay}

It is a protocol used for high bitrate requirements, with error detection and correction, redundancy and safety \cite{Seung-Han2008}. It is used for high end applications inside vehicles such as power train and safety functions (adaptive cruise control and active suspension)~\cite {NationalInstruments2019}. The protocol is standardised under ISO 17458 series which describes the physical layer and data link layer characteristics~\cite {ISO10681-1:20102010}. The main features of the protocol are:

\begin{itemize}
\item A bit rate of up to 10 Mbps with half-duplex bus access;
\item standardised as ISO standard 17458;
\item support for fault tolerant mechanisms;
\item designed to work for high speed and safety critical applications (e.g. braking-by-wire and steering-by-wire)
\end{itemize}

Since the CAN bus protocol is the most widely used, this review paper will focus on this protocol in terms of attacks, vulnerabilities and potential countermeasures.
\begin{table}[ht!]
	\centering
	\small
	\caption{Automotive serial communication protocols. The table shows Controller Area Network (CAN and CAN FD), Local Internet Network (LIN) and FlexRay  }
	\label{tbl:Automotive-Serial-Communication-Protocols}
	\begin{tabular}{c|cccc} 
		\hline
		\textbf{Protocol }                                  & \textbf{Bit-rate }  & \textbf{Application }                                                                                    & \textbf{Domain}                                                          & \textbf{Standard }   \\ 
		\hline
		\begin{tabular}[c]{@{}c@{}}High \\ CAN \end{tabular} & 125Kbps to 1Mbps  & \begin{tabular}[c]{@{}c@{}}Real time critical applications\\ e.g engine and braking systems \end{tabular} & \begin{tabular}[c]{@{}c@{}}Powertrain \\ and Chassis train \end{tabular} & ISO 11898            \\ 
		\hline
		\begin{tabular}[c]{@{}c@{}}CAN\\ low \end{tabular}   & 5kbps to 125kbps   & \begin{tabular}[c]{@{}c@{}}Non-critical such as doors~ ~and \\ windows ~\end{tabular}               & Body Domain                                                              & ISO 11898            \\ 
		\hline
		\begin{tabular}[c]{@{}c@{}}CAN \\ FD \end{tabular}   & Up to 10Mbps        & \begin{tabular}[c]{@{}c@{}}Critical real time\\ applications \end{tabular}                               & \begin{tabular}[c]{@{}c@{}}Powertrain \\ and Chassis train \end{tabular} & ISO 11898            \\ 
		\hline
		LIN                                                  & 1kbps to 20 kbps    & \begin{tabular}[c]{@{}c@{}}Non-critical\\ applications \end{tabular}                              & Body Domain                                                              & ISO 17987            \\ 
		\hline
		FlexRay                                              & up to 10 Mbps       & \begin{tabular}[c]{@{}c@{}}Critical\\ applications \end{tabular}                                          & \begin{tabular}[c]{@{}c@{}}Powertrain \\ and Chassis \end{tabular}       & ISO 17458            \\
		\hline
	\end{tabular}
\end{table}
\section{Controller Area Network (CAN)}
\label{Sec3:CAN-Protocols}

Initial use of CAN bus within a vehicle did not consider security, as vehicles at that time were not connected to outside networks. The CAN protocol does not have security features and is vulnerable to attacks such as frame injection and denial of service. Nowadays vehicles are more connected, they have internal and external interfaces such as Wi-Fi, Bluetooth and mobile (phone) networks and thus cybersecurity has become a real concern. CAN is a protocol invented in the early 1980s by Bosch GmbH and used widely inside vehicles to send and receive data between ECUs and sensors \cite {Avatefipour2018}. It was standardised in ISO 11898 series and it works as a serial bus, indicating that any node on the network (e.g. an ECU) can use the network bus to send data via a multi-master mechanism using 2 wires~\cite{Dubitzky2013}. This reduces the cost of wiring compared to a point to point wiring mechanism and reduces the negative effects of external noise through its CAN-High and CAN-Low (signal differential) transmission~\cite{Avatefipour2018} \cite{Jafarnejad2015}. It works on the physical and data link layers, although it does not use Media Access Control addresses (MAC) and MAC tables to send and receive (route) frames. Instead, it uses message ID (does not have sender or receiver addresses compared to Ethernet) and a broadcast half duplex mechanism to transmit data over the bus. Also, it does not verify and use authentic messages as it sends data based on message id does not use a source address. CAN bus controller inside the vehicle connects critical parts of the vehicle such as engine and body control modules, such as gears, speed, brakes and so on. The CAN protocol itself does not provide message authentication and so it is vulnerable to cyberattacks such as CAN frame injections. The protocol consists of two versions: the classical CAN protocol and CAN FD protocol (Flexible Data rate) -- both protocols are defined and standardised  under ISO 11898 series \cite{CANinAutomation2013}. 

\subsection{Standard CAN bus Frame 2.0A}

The classical CAN bus was standardised in 1993 in ISO 11898. It consists of two versions based on the message identifier length. The standard CAN bus 2.0A has an 11-bit identifier while the extended CAN bus 2.0B has a 29-bit identifier. 

\subsection{Extended CAN bus Frame 2.0B}

\begin{figure}[ht]
	\centering
	\includegraphics[width=\linewidth]{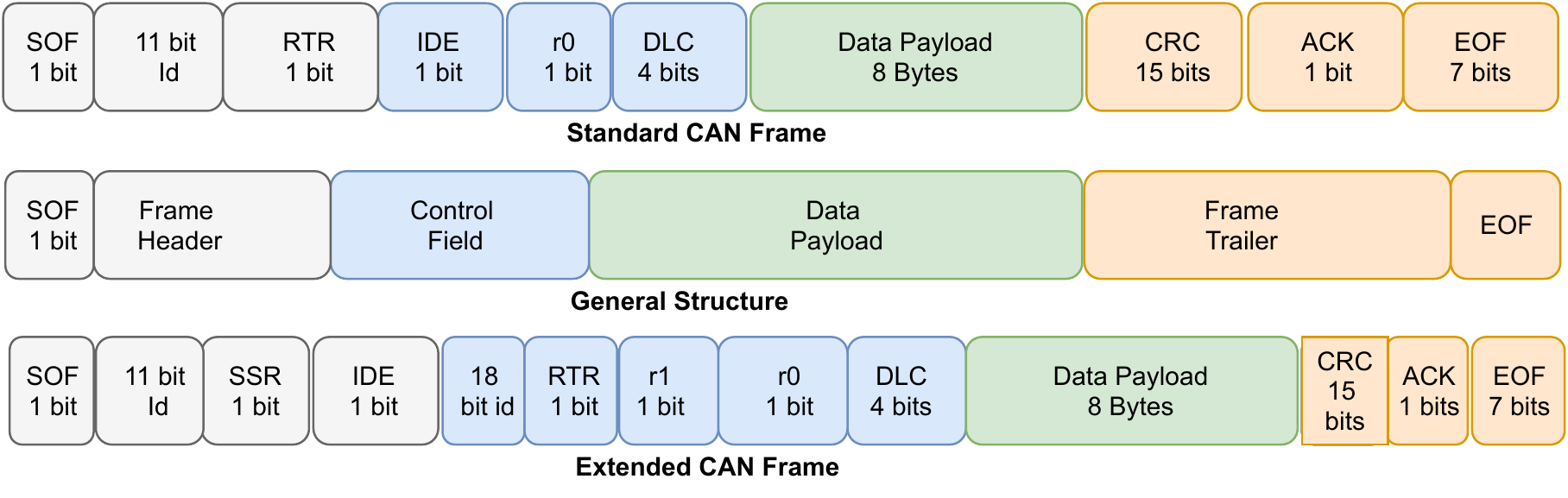}
	\caption{Standard and extended Controller Area Network CAN bus frame -- showing the Frame header fields, Control fields, Data payload fields and Frame trailer}
	\label{Fig:Classical-CAN-Frames}
\end{figure}

This extended CAN bus provides a 29-bit identifier which gives more message ids, and hence a greater number of potential nodes that can be supported. The data payload is up to 8bytes. CAN frame structure is illustrated in figure \hyperref[Fig:Classical-CAN-Frames]{2}

The CAN protocol frame consists of the fields shown in figure~\ref{Fig:Classical-CAN-Frames}.
A classical CAN frame~\cite{Gmbh1991}, \cite{Avatefipour2018} consists of:  the arbitration field, the message identifier field and the Remote Transmission Request (RTR) bit.
\begin{enumerate}
	\item \textbf{Frame header}: The frame header contains the message identifier (11-bits or 29-bits) which identifies the priority and the content of the message. It is worth noting that a CAN bus does not provide source and destination addresses like IP networks, instead it uses unique CAN identifier numbers in each message. 
	\begin{itemize}
		\item \textbf{Start of the Frame field (SOF)} has a dominant value of 1.
		\item \textbf{11-bit identifier} shows the message id which determines the priority of the message where the lowest value means the highest priority. Also, it is used to represent the content of the message.
		\item \textbf{29-bit identifier} -- as part of the extended frame has similar structure as the standard frame except additional fields in the frame header and control field such as the following:
		\textbf{The frame header} consists of a base identifier (11-bits) and the extended identifier 18-bits which both provide the priority of the frame and its content. 
		\item \textbf{Remote Transmission Request (RTR)}  is used to retrieve information from a node in the network.
	\end{itemize}
	\item \textbf{Control fields} 
	The control field consists of: Identifier Extension Flag (IDE), rO and Data length Control (DLC) flag:
		\begin{itemize}
		\item \textbf{IDE} Identifier Extension is used to distinguish between standard and extended frame
		\item \textbf{r0}field is a 1-bit value and reserved and it always has the recessive value of 0. 
		\item \textbf{DLC} field is a 4-bit value and indicates the length of the data in the data field
		\item \textbf{SSR} Substitute Remote Request 
		\item\textbf{IDE} Identifier Extension Flag is used to identify the frame as standard (dominant bit) and in the extended frame the value is a recessive bit  
		\item \textbf{R1 and r0} are reserved bits and they are always dominant. 
	\end{itemize}
	\item \textbf{Data field}:This field contain the payload of up to 8-bytes.
	\item \textbf{Frame trailer}: These fields are used to detect frame errors using checksum and correction mechanisms:
\begin{itemize}
	\item  \textbf{CRC} (Cyclic Redundancy Code) field consists of 15bits and is used for checksum error detection. \cite{KVASER}.
	\item \textbf{ACK} or acknowledgement field is used to indicate that messages have been sent without any errors. 
	\item \textbf{EOF} indicates the End of the Frame and the message sequence.
	\item 	\textbf{IFS} Inter Frame Space is 3bits in size and used to provide frame separation and initiate frame processing.
\end{itemize}
\end{enumerate}

\subsection{Controller Area Network with Flexible Data Rate (CAN FD)}

\begin{figure}[ht]
  \centering
  \includegraphics[width=\linewidth]{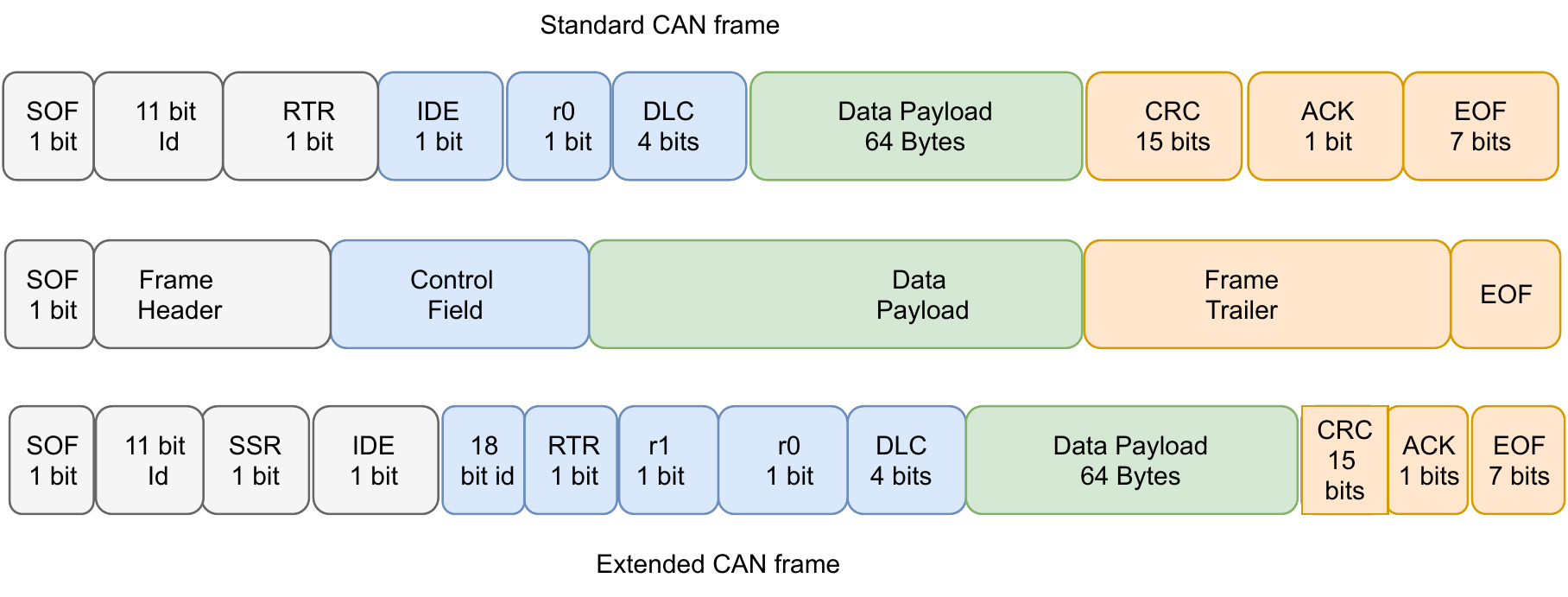}
  \caption{CAN FD 11- and 29-bit identifier frame structure}
  \label{Fig:CAN-FD-Frames}
\end{figure}

CAN FD is developed to meet the needs of higher speed and more data payload size. It can provide up to 64bytes of data payload along with up to 8Mbps of speed \cite{Cheon2013}. It comes with a standard and an extended version similar to the classical CAN protocol. It is standardised in ISO 11898 series as well. The frame structure of the CAN FD protocol is illustrated in figure~\ref{Fig:CAN-FD-Frames}.

\subsection{Frame Types of CAN bus Protocol}
The CAN bus is divided into four types of frames~\cite{Gmbh1991}:
\begin{enumerate} 

 \item \textbf{Data Frame}: contains the data payload.

 \item \textbf{Remote Frame}: used to ask for the transmission of data frame with the same identifier from another node on the bus. The difference between this frame and the data frame is that the RTR field inside the arbitration id is put in recessive and there is no data payload.
 
 \item \textbf{Error frame} indicates that there is an error in the bus and this frame can be used by any node.

 \item \textbf{Overload Frame} is used when a node on the bus is too busy to receive data from another node. When a CAN node transmits frames on the bus, it will be received by all other nodes connected to the bus due to its broadcast feature. The CAN controller board on a CAN node is responsible for handling the relevant frames. An error frame can be raised if an error occurs when receiving data and the remote frame is raised by the node to ask for a re-transmission of data.   
\end{enumerate}

\begin{figure}[ht]
  \centering
  \includegraphics[width=0.7\linewidth]{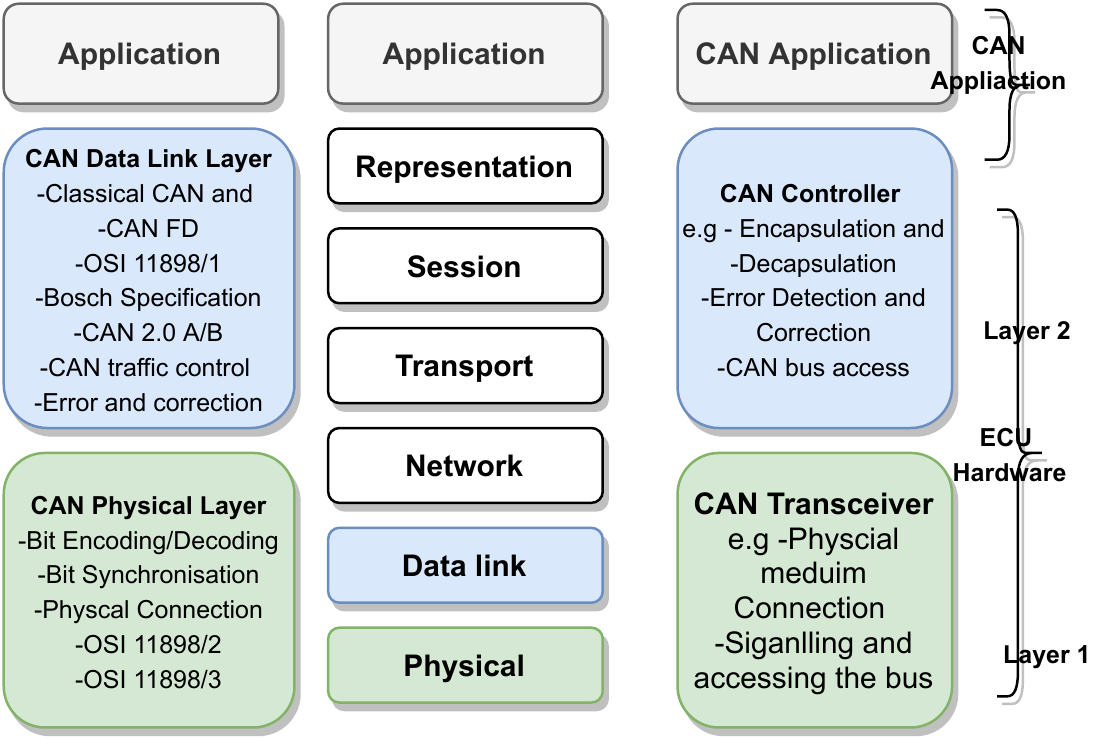}
  \caption{CAN bus protocol in OSI Model. CAN bus Hardware, Software and Standards in OSI layers}
  \label{Fig:CAN-bus-Protocol-OSI-Layers}
\end{figure}

\noindent The CAN standard was updated as ISO11898:2015 to include CAN FD~\cite{CANinAutomationa}. The 11898 series describes the CAN bus data link and physical layer functions such as described in figure~\ref{Fig:CAN-bus-Protocol-OSI-Layers}.

\subsection{Cybersecurity and Safety Standards for Vehicle Networks}

Various efforts already exist to provide security and safety for vehicles, from design to production and operation. Such efforts originate from standards organisations such as ISO, the Society of Automotive Engineering (SAE) and other organisations. Some of the main standards for automotive cybersecurity include: 
\begin{enumerate} 
	\item \textbf{ISO / SAE 21448:} This standard provides a guide from measurement to design, to enable the verification and validation of services inside vehicles \cite{InternationalOrganizationforStandardization2019}  
	
	\item \textbf{SAE J3061:} Provides a guide for cyber-physical systems within vehicles, and has been provided by the Society of Automotive Engineering \cite{StandardofAutomotiveEngineering2016}. 
	\item \textbf{SAE J3101} standard provides the required features to support hardware security protection for vehicular applications \cite{SAEInternational2012}
	
	\item \textbf{SAE J3138} (Diagnostic Link Connector Security):  A standard used for diagnostics and security purposes for direct interfaces such as the on-board diagnostics (OBD-2) port inside a vehicle \cite{J31382018}.  
	
	\item \textbf{ISO / SAE 21434:} This standard is used to support road vehicle cybersecurity engineering -- expected to be released in 2020~\cite{Barber2018}. It is a shared effort between ISO and SAE covering security for road vehicles, in-vehicle systems, components, software and communication with external networks and devices. This standard aims to provide guidelines for vehicle manufacturers and suppliers from design to production phases \cite{Barber2018}.     
\end{enumerate}

Other efforts to secure in-vehicle network infrastructure such as E-safety Vehicle Intrusion protected Applications (EVITA) and vehicle to infrastructure (V2I) such as Secure Vehicular Communication (SEVCOM):

\begin{enumerate} 

\item \textbf{E-safety Vehicle Intrusion Protected Applications (EVITA)(2008-2011)}: \cite{Seudie2009}: This  project focuses on securing in-vehicle network infrastructure from physical and remote attacks. 

\item \textbf{Secure Vehicular Communication (SEVCOM) (2006-2008)}:  \cite{SevecomSevecom2008} has focused on securing V2I networks, with an emphasis on wireless data communications that is used to transmit vehicle parameters to an outside network or device.

\item \textbf{Cooperative Vehicle-Infrastructure Systems (CVIS) (2006-2010)}: \cite{CVISCVIS2010} a framework to provide V2I security and privacy for drivers and connected vehicles.

\end{enumerate} 	

\subsection{CAN bus Network Infrastructure}

\begin{figure}[ht!]
  \centering
  \includegraphics[width=.7\linewidth]{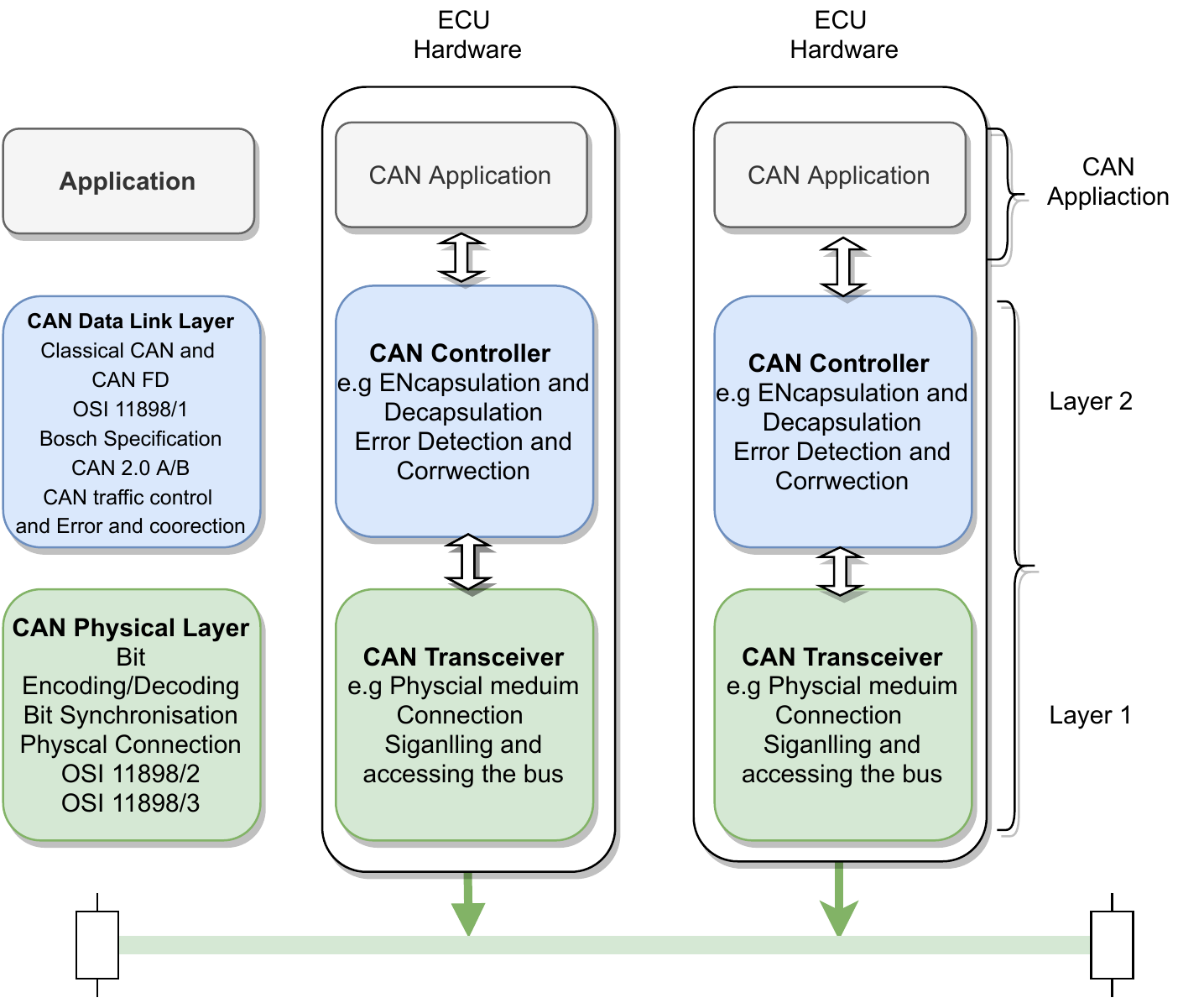}
  \caption{CAN Electronic Control Unit (ECU) hardware and software components, functions and standards}
  \label{Fig:CAN-ECU-Layer}
\end{figure}

The CAN protocol focuses on the physical and data link layers. Extended protocols from industry such as J1939 (for heavy vehicles) and OBD-2 (for vehicular diagnostics) are built on top of the CAN data link and physical layers.  The CAN layer functions can be identified \cite{Gmbh1991} as follows  and as shown in \hyperref[Fig:CAN-ECU-Layer]{5}: 

\textbf{CAN Physical Layer (CPL)} focuses on transmission of the signal across the CAN bus hardware. 

\textbf{CAN Data Link Layer protocol (CDLL)} defines the core protocol (realised using the CAN chipset) such as bit timing, message framing, synchronisation, arbitration logic and error detection (e.g. use of CRC and ACK). 

\textbf{CAN High Layer Protocols (CHLP)} (Application layer Protocols) use high speed CAN to provide real time information and diagnostic data exchange between ECUs. There are different high layer protocols in industry such as  J1939 by the Society of Automotive Engineers (SAE) and used in heavy duty vehicles \cite{SocietyofAutomotiveEngineers2007}. The J1939 protocol uses the 29bit version of the CAN bus protocol. Other protocols such as the protocol used with the On-Board Diagnostic (OBD-2) port are used for vehicle diagnostic and emissions analysis. 

\subsection{Automotive Application Layer Protocols}

\begin{figure}[t]
  \centering
  \includegraphics[width=.6\linewidth]{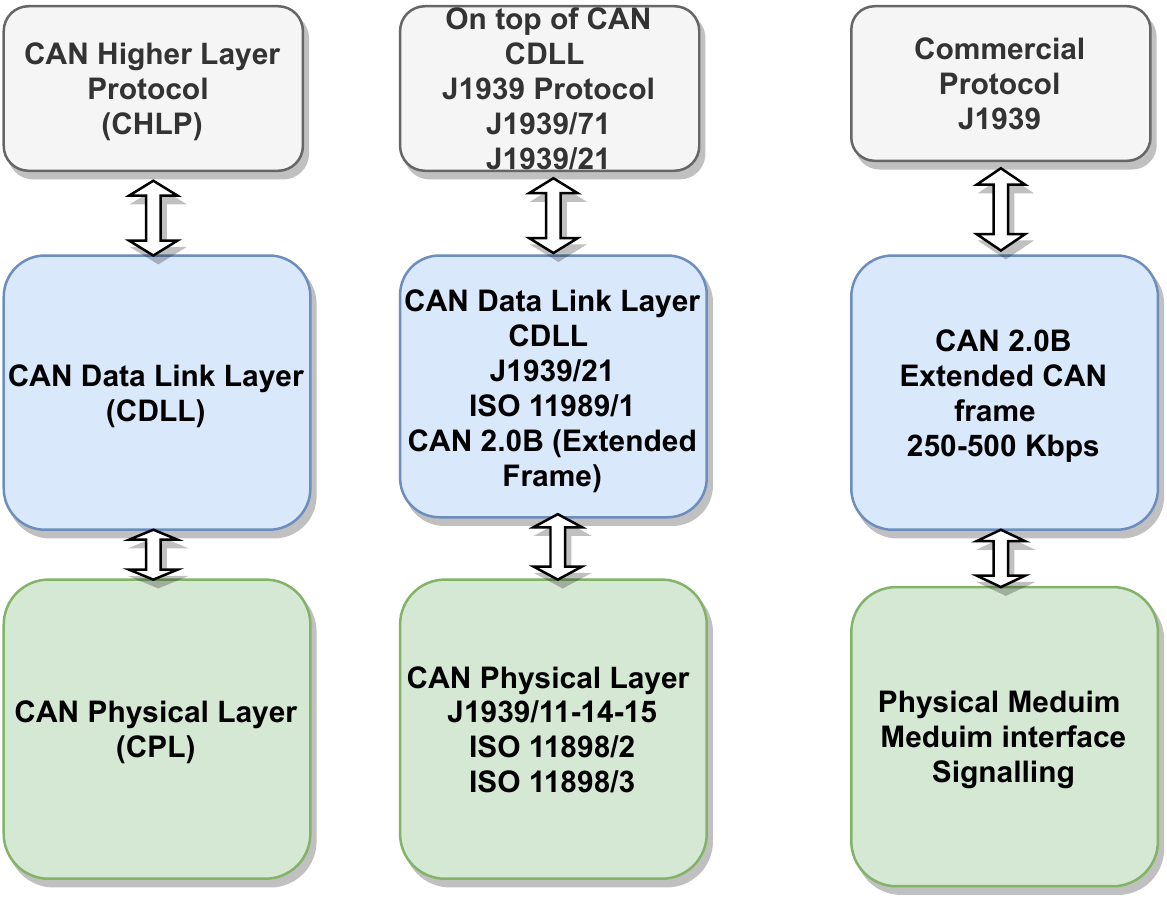}
  \caption{SAE J1939 CAN Application layer protocol}
  \label{Fig:J1939-layers}
\end{figure}

SAE J1939 is an application layer protocol widely used in commercial heavy vehicles such as coaches and agricultural vehicles. It works on top of the CAN 2.0B bus with a 29-bit extended frame providing a bitrate of 250Kbps to 500Kbps~\cite{VECTOR}. It also provides a standard message format and specification which allows using components from various manufacturers inside vehicles. Figure \hyperref[Fig:J1939-layers]{6} illustrates CAN bus application layer.

\subsection{Electronic Control Units (ECU)}

Modern cars have around 70 ECUs which control various functions of the car, such as breaks, gears and engine status~\cite{Seung-Han2008}. An ECU is primarily a microprocessor which contains a CAN controller used to support data link layer functions and a CAN transceiver used for physical layer functions such as frame delivery, error detection and correction and other data link layer tasks \cite {Microchip2003}. Figure \hyperref[Fig:CAN-ECU-Layer]{5} shows  CAN Electronic Control Unit (ECU) hardware and software components, functions and standards in OSI model. Also, table \hyperref[tbl:Some Electronic Control Units inside CAN bus Network]{2} illustrates some ECUs which are used inside vehicular network systems.



\subsection{CAN bus communication}

The CAN protocol uses a broadcast based mechanism for message exchange~\cite{Wampler2009} and each node can request use of the bus randomly. An arbitration mechanism is used to ensure priority on the bus \cite {Gmbh1991}, as ECUs with critical functions such as engine, transmission and braking systems usually have higher priority to access the bus and require least broadcast frequency \cite {Tomlinson2018a}. Priority is based on comparing the arbitration id of requesting nodes, and the node with higher priority is granted access to send data on the bus. 
Inside the vehicle, ECUs with critical functions (e.g. brakes, steering) can be connected to a high speed CAN bus while ECUs with low importance (e.g. windows) can be connected to low speed CAN bus \cite{Taylor2016a}\cite{Deng2017}. Both CAN buses then are connected through a gateway ECU \cite{Deng2017}.
\begin{figure}[ht]
  \centering
  \includegraphics[width=.7\linewidth]{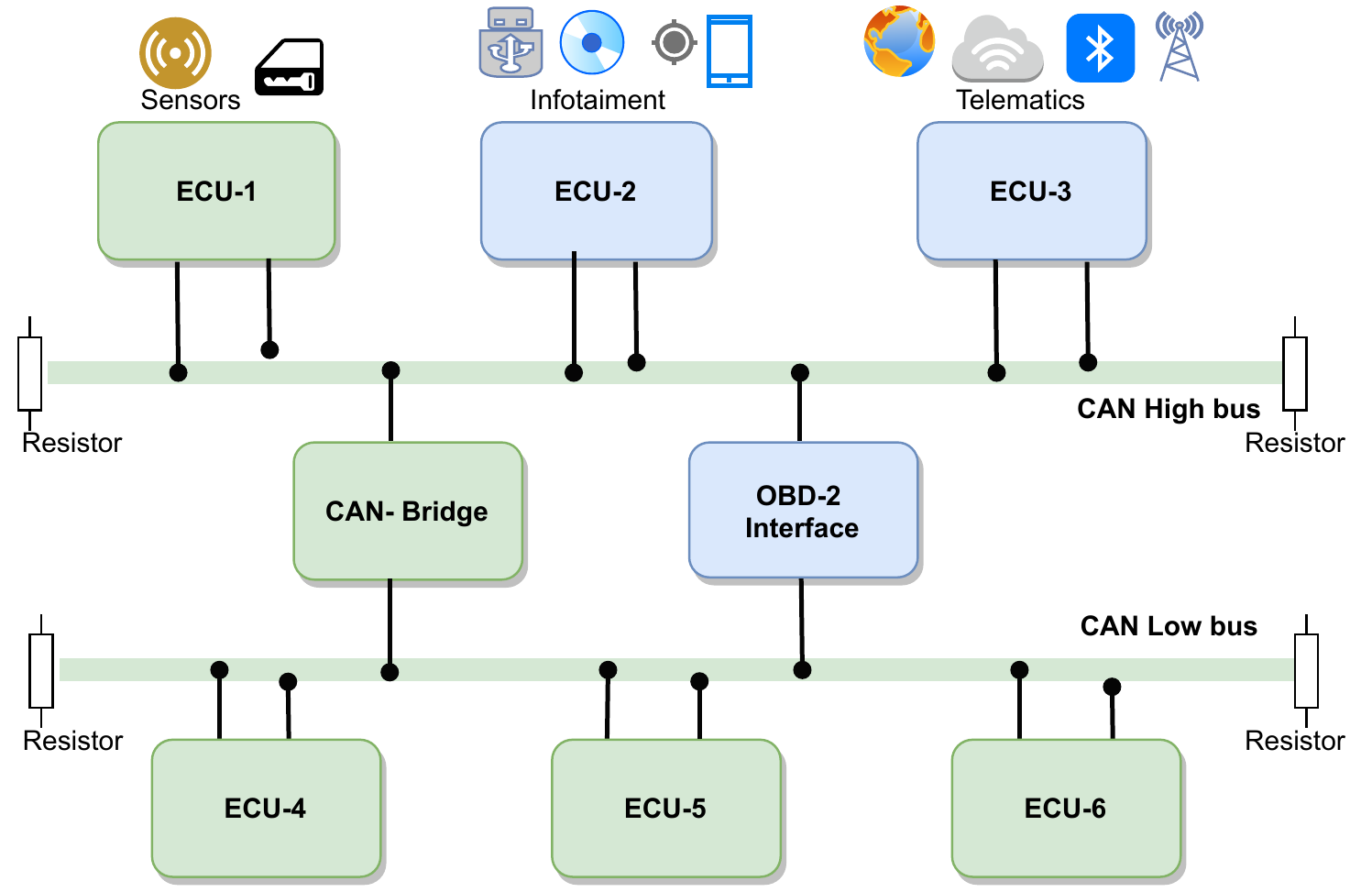}
  \caption{Two CAN buses connected to each other through a bridge -- based on~\cite{Wang2014}}
   \label{Fig:Two-CAN-bus-Networks-Via-Gateway}
\end{figure}

With this structure as outlined in figure \hyperref[Fig:Two-CAN-bus-Networks-Via-Gateway]{7}, any ECU with connection to the bus, using an OBD-2 port or Bluetooth, can sniff and inject data into both buses. These vulnerabilities have led to the development of Intrusion Detection Systems (IDSs) and firewalls to prevent unauthorised access, as well as using cryptography methods to provide confidentiality, integrity and authentication.  

\subsection{Protocol usage}

The CAN bus protocol has been used in many application areas due to its simplicity and offered bitrate. The ease of implementation, low cost and the small number of physical wires need to realise it makes it suitable for use in many embedded system \cite{Gmbh1991}. It is used inside vehicles, built environments (e.g. controlling elevators in buildings, building energy management systems), railway applications, medical devices and aircrafts \cite{Anon1999}. 
\begin{table}[ht]
	\centering
	\label{tbl:Some Electronic Control Units inside CAN bus Network}
	\caption{Some Electronic Control Units inside CAN bus Network -- based on~\cite{Koscher2010}}
	\begin{tabular}{l|l|l} 
		\hline
		\textbf{ Electronic Control Unit }                & \textbf{CAN bus Connection }  & \textbf{Critical }   \\ 
		\hline
		Engine Control Module (ECM)                        & High CAN bus                  & $\checkmark$         \\ 
		\hline
		Electronic Brake Control Module (EBCM)             & High CAN bus                  & $\checkmark$         \\ 
		\hline
		Transmission Control Module  (TCM)                 & High CAN bus                  & $\checkmark$         \\ 
		\hline
		Body Control Module   (BCM)                        & High and Low CAN bus          & $\times$             \\ 
		\hline
		Telematics Module (TM)                             & High and Low CAN bus          & $\times$             \\ 
		\hline
		Remote Control Door Lock Receiver (RCDLR)          & High CAN bus                  & $\checkmark$         \\ 
		\hline
		Heating, Ventilation, Air Conditioning             & High CAN bus                  & $\times$     \\ 
		\hline
		Sensing and Diagnostic Module (SDM)                & High CAN bus                  & $\checkmark$         \\ 
		\hline
		Instrument Panel Cluster/Driver Information Center & High CAN bus                  & $\times$             \\ 
		\hline
		Radio                                              & High CAN bus                  & $\times$             \\ 
		\hline
		Theft Deterrent Module  (TDM)                      & High CAN bus                  & $\checkmark$         \\
		\hline
	\end{tabular}
\end{table}
\section{Connected Car Environment }
\label{Sec4:Connected-Cars}
Connected cars can simply mean a vehicle connected to a network and providing services such as vehicle diagnostic parameters and GPS information to the vehicle owner. According to Juniper research, connected cars are expected to increase to 750 million by 2023 \cite{JuniperResearch}. This connectivity will be through telematics or by in-vehicle applications. Vehicles can be connected with either aftermarket tools such as OBD-2 cellular device, GPS device used in fleet management or hardware and software included from the vehicle Original Equipment Manufacturer (OEM). Components used in connected cars can be classified as: 

\textbf{Telematics Unit:} provides connectivity to the car using WiFi, Bluetooth, GPS and mobile data interfaces. 

\textbf{Infotainment Unit:} provides the information and entertainment to the driver through a head display unit such as CD, DVD player, USB and mobile applications integration with the head unit.

\textbf{Driver assistance Unit:} provides the driver and the vehicle with driving assistance hardware such as cameras and LiDAR sensors to provide safety on the road. Also, these sensors are used to support autonomous driving. Also, Adaptive Cruise Control and Park Assist are used for measuring parking space and auto park assistance.

\textbf{Vehicle 2 X}: connected cars can also provide communication to cars (V2V) and roadside infrastructure (V2I) using wireless communication  called Dedicated Short Range Communications (DSRC) which allow exchange  data such traffic conditions between cars and/or road side unit.
\begin{figure}[ht]
  \centering
  \includegraphics[width=.8\linewidth]{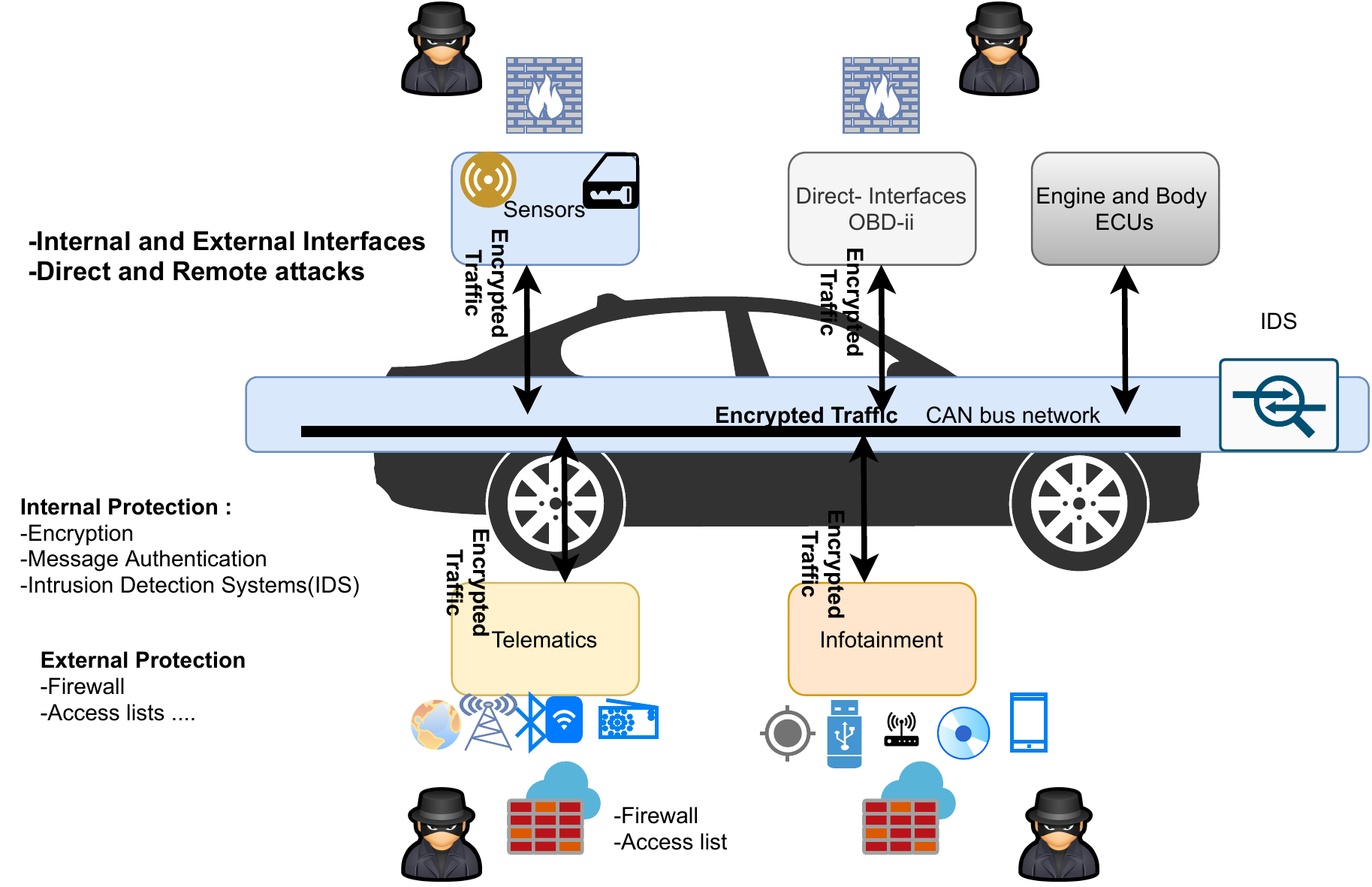}
  \caption{Connected Car environment with 
  four potential entry points for data injection: Telematics, Infotainment, Direct interfaces and Sensors. Also, security countermeasures to detect and prevent physical and remote attacks using Cryptography, Intrusion Detection System (IDS), Firewalls and Access Control Lists.}
\label{Fig:Connected-Car-Environement}
\end{figure}
\subsection{Connected Vehicle Interfaces and Sensors }

\textbf{Bluetooth} provides connectivity with mobile apps hosted on devices operated by passengers. Such a service involves pairing a mobiles phone(s) with head unit inside the vehicle \cite{Checkoway2011}. This can lead to vulnerabilities from the Bluetooth connection as legacy and vulnerable versions are still used -- as described in \cite{Cheah2017}\cite{Oka2014} \cite{Checkoway2011}.

\noindent \textbf{Wi-Fi:} Connected cars provide wireless connectivity for various services, such as providing internet through Wi-Fi on board. Wi-Fi has a number of vulnerabilities, e.g. via a Wi-Fi hotspot on a Jeep~\cite{Miller2015} and a Tesla S~\cite{Tesla2016}.

\noindent \textbf{Cellular/phone network}: modern cars can also provide mobile/phone/cellular connection which can be used to retrieve data such as weather conditions and traffic\cite{Miller2015}. Attacks on such networks have also been identified~\cite{Checkoway2011}. For example, \cite{Miller2015} have shown how a cellular network interface can be hacked inside a Jeep.

\noindent \textbf{OBD-2 (On-Board Diagnostic)} : It is a mandatory port which is used for capturing diagnostic and environmental (e.g. emissions) data. This interface is directly connected to the vehicle's CAN bus network and by using an aftermarket OBD dongle and attaching it to the OBD port, it is possible to initiate various attacks such as a DoS attach which can affect vehicle operation and driver safety \cite{Fowler2018}.  Various attacks have been demonstrated using direct connection to an OBD port and generating remote attacks using wireless OBD dongles~\cite{Checkoway2011}. In 2018, a remote attack on a vehicle was initiated using a custom hardware that was connected to a CAN bus over OBD-2 port. This customised board used a SIM card, and the attacker sent malicious SMS messages in order to inject this data into the CAN bus~\cite{Zorz2018}. 
 
\noindent \textbf{Global Position System (GPS):} is used to provide driving assistance and positioning for drivers. Further, it is used by fleet management to monitor vehicle location. This interface can provide an entry point for an attacker -- both for injecting and sniffing data~\cite{Oruganti2019}. In \cite{Miller2015} GPS information was retrieved from the head unit of a vehicle through unprotected 6667 port.

\noindent \textbf{Compact Disc (CD)} player is used in the head unit for entertainment purposes. It has been shown that this unit is directly connected to the internal data network of a vehicle, and also susceptible to cyberattacks, as described in~\cite{Checkoway2011}.

\noindent \textbf{Sensors:} sensors and actuators are used inside vehicles to support various functions such as sensing engine temperature. Physical availability attack can be initiated using signal jamming 
\cite{Loukas2017} to block data between the sensors and the CAN network. In correct sensor values can also be injected into the CAN bus to modify the behaviour of the ECUs that operate on this data. A particular type of sensors used for Tyre Pressure System Monitoring (TPSM) are connected to each tyre to monitor pressure and send real-time data to an ECU~\cite{Miller2015}. Attack on TPSM is described in~\cite{FiguerolaGarcia2012}, where the authors were able to perform eavesdropping attacks from 40 meters on a passing car.

\noindent \textbf{LiDAR and Camera:} Cameras and laser signals are used inside vehicles to provide safety and driving assistance. These components can be manipulated by various attacks such as signal jamming. In \cite{Petit2015a}, the authors performed signal jamming attacks on LiDAR and cameras. These components are also widely used inside autonomous vehicles.

\noindent \textbf{Keyless entry} : Miller and Valasek \cite{Miller2015} show how Remote Control  Door Lock Receiver (RCDL) within a vehicle is directly connected to the internal CAN bus. It receives the signal from the key fob to lock, unlock doors and trunk of the vehicle. Keyless entry attacks were initiated to steal vehicles in many occasions, and it has been shown as the most used attack between 2010--2019~\cite{Upstream2019}. This attack can be initiated by jamming the signal between the key fob and the vehicle to keep the doors open while the owner of the car thinks it is closed. Also, it can be initiated  by capturing the key fob signal and redirecting it to the vehicle. For example, in \cite{Eisenbarth2008} researchers were able to hack key fob block cipher and perform relay  signal attack, and were able to lock and unlock doors. The attacker needs to be in the range of the key fob to be able to intercept the signal for this type of attack.

\section{Vulnerability of In-Vehicle CAN bus}
\label{Sec5:Vulnerabilitiesof-IVN-CAN-bus}

The intention of using a CAN bus inside a vehicle was to reduce cost, simplify installation, and improve efficiency for real time communication. However as mentioned previously, a CAN bus has a number of security vulnerabilities~\cite{Currie2015}:
\begin{itemize}
\item The network is not segmented, as all nodes (ECUs) are connected to the same bus. The CAN bus protocol uses a broadcast mechanism to transfer data, which means all nodes on the network can send and receive the same messages.

\item There is no security mechanisms used for authentication and thus the CAN bus is vulnerable to message poisoning and denial of service (DoS) attacks.

\item The traffic on the CAN bus is not encrypted and can be easily read through a data sniffing attack. Every ECU connected to the bus can therefore sniff CAN frames due to the broadcast mechanism.

\item An ECU can make the CAN bus in domination status using the arbitration scheme (Message ID priority scheme) and thus prevent other ECUs from using the bus -- which can lead to DoS attack.

\item It is not possible to know whether an ECU has sent or received certain messages (non-repudiation).

\item Access to the CAN bus network via external interfaces and connections such as OBD-2, Wi-Fi and Bluetooth widens the potential attack surface (and entry points)~\cite{Wu2019a}. 
\end{itemize}

\begin{figure}[t]
	\centering
	\begin{subfigure}[b]{0.48\linewidth} 
		\centering
		\includegraphics[width=\textwidth]{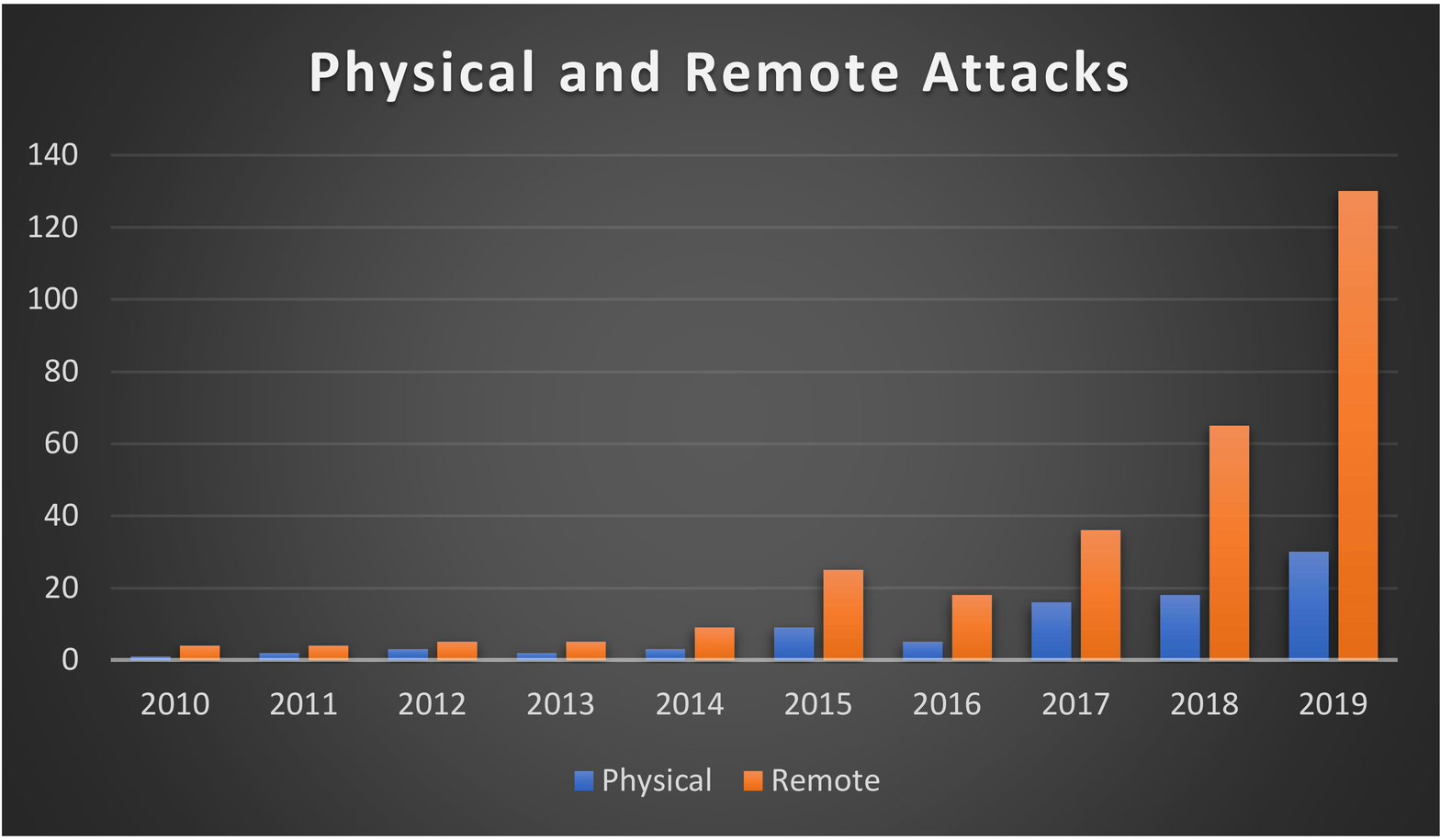}
		\caption{No. of physical \&  remote attacks: 2010--2019 \cite{Upstream2019}}
		\label{Fig:Physical and Remote attacks rate between 2010 and 2019}
	\end{subfigure}
	\quad
	\begin{subfigure}[b]{0.48\linewidth}
		\centering
		\includegraphics[width=\textwidth]{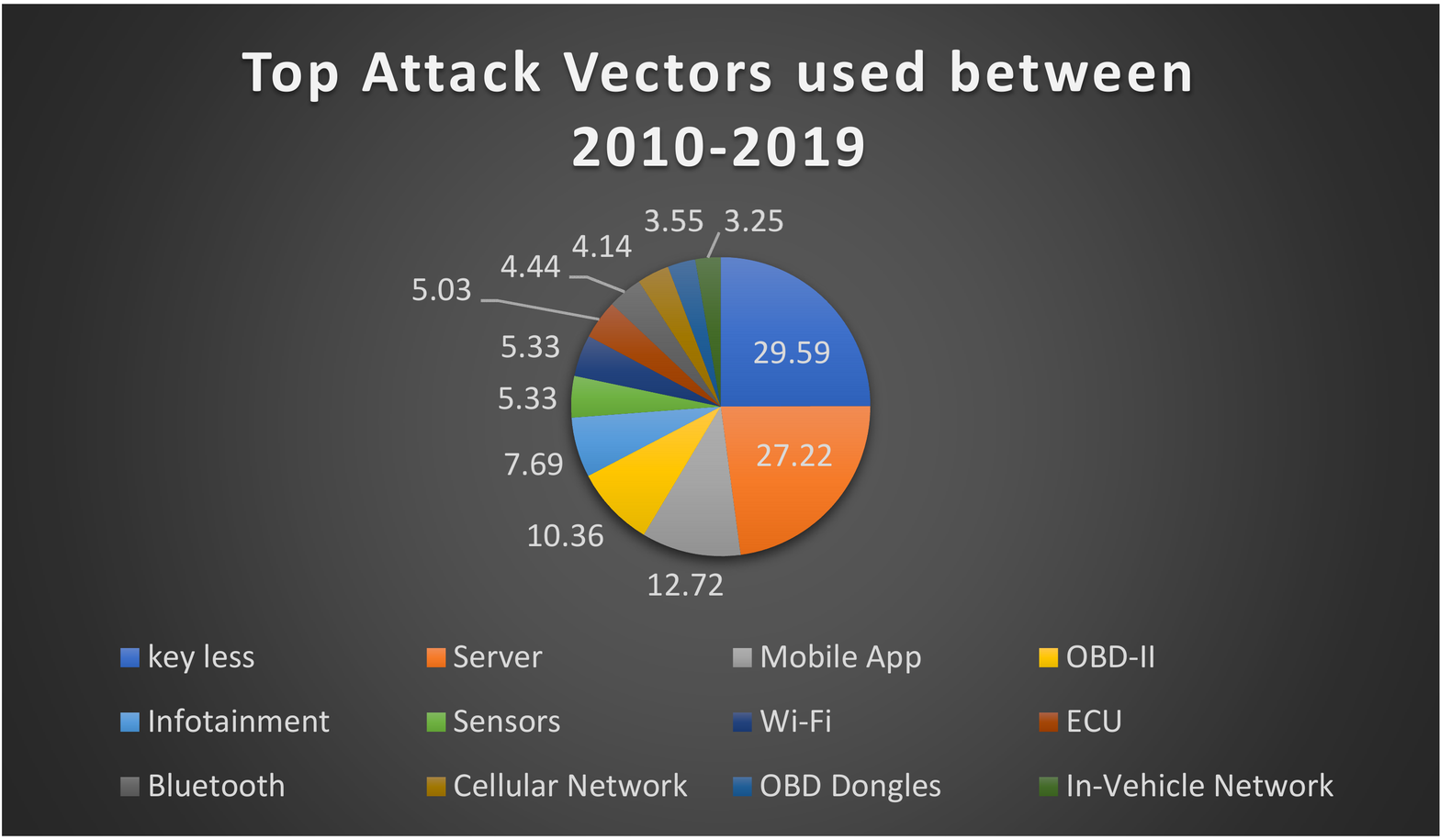}
		\caption{Top attack vectors: 2010--2019 \cite{Upstream2019}}
		\label{Fig:Top Attack Vectors between 2010 and 2019}
	\end{subfigure}
	\caption{Vehicle Attacks: 2010--2019}
	\label{FigVehicleAttacks20102019} 
\end{figure}

There has been an increase in the number of cyberattacks on vehicles, increasing 7 times in 2019 compared to 2010, and doubling in 2019 compared to 2018~\cite{Upstream2019} as shown in figure \hyperref[Fig:Top Attack Vectors between 2010 and 2019]{9}. The vulnerable points could be classified as direct, indirect, short-range and long-range attacks~\cite{Checkoway2011}. 

The CarShark software was used within a vehicle to sniff, analyse, observe and replay the data on the CAN bus using OBD-2 connector and then control the wheels, brakes and other ECUs and components of the vehicle~\cite{Koscher2010}. This work also reports on other entry points to the CAN bus inside the vehicle such as the audio jack, USB and Wi-Fi, and the use of these to perform various attacks. Similarly, other attacks were  performed from outside the car identifying potential vulnerabilities~\cite {Checkoway2011}. Figure~\ref{FigVehicleAttacks20102019} shows potential entry points for attacking a vehicle.

Other examples include attacks on Toyota Prius 2010~\cite{Miller2014} and Ford Escape 2012 vehicles by physically connecting to the OBD-2 port (and CAN bus) and controlling vehicle speed, brakes and steering. Other examples include remote attacks carried out on a Jeep Cherokee~\cite{Miller2015}. Another attack is the keyless fob attack used to forcibly unlock the doors of a vehicle~\cite{Eisenbarth2008}. A summary of in-vehicle network based attacks includes: 
\begin{table}[ht]
	\centering
	\tiny
	\caption{In-Vehicle EntryPoints }
	   \label{tbl:In-VehicleEntrypoints}
	\begin{tabular}{l|lllll} 
		\hline
		\begin{tabular}[c]{@{}l@{}}\textbf{ CAN bus }\\\textbf{initiated attack }\end{tabular}   & \begin{tabular}[c]{@{}l@{}}\textbf{Entry}\\\textbf{Points }\end{tabular}                                                                                               & \begin{tabular}[c]{@{}l@{}}\textbf{Physical}\\\textbf{remote }\end{tabular}                                                                                                      & \begin{tabular}[c]{@{}l@{}}\textbf{Attack}\\\textbf{Mechanism }\end{tabular}                                                                                                                                                                                                & \begin{tabular}[c]{@{}l@{}}\textbf{Position of the}\\\textbf{ attacker }\end{tabular}                                                                                            & \begin{tabular}[c]{@{}l@{}}\textbf{Result of }\\\textbf{the attacks}\end{tabular}                                                                                                                                                          \\ 
		\hline
		\begin{tabular}[c]{@{}l@{}}Interfaces \\Initiated \\attacks \end{tabular}                & \begin{tabular}[c]{@{}l@{}}\begin{tabular}{@{\labelitemi\hspace{\dimexpr\labelsep+0.5\tabcolsep}}l}OBD-2\\BT\\Wi-Fi \end{tabular}\end{tabular}                          & \begin{tabular}[c]{@{}l@{}}\begin{tabular}{@{\labelitemi\hspace{\dimexpr\labelsep+0.5\tabcolsep}}l}Physical\\Remote\\Remote \end{tabular}\end{tabular}                           & \begin{tabular}[c]{@{}l@{}}\begin{tabular}{@{\labelitemi\hspace{\dimexpr\labelsep+0.5\tabcolsep}}l}OBD-2 direct connection\\BT vulnerabilities\\WiFi on board access \end{tabular}\end{tabular}                                                                             & \begin{tabular}[c]{@{}l@{}}\begin{tabular}{@{\labelitemi\hspace{\dimexpr\labelsep+0.5\tabcolsep}}l}Inside/Outside\\Outside\\Outside \end{tabular}\end{tabular}                   & \begin{tabular}[c]{@{}l@{}}\begin{tabular}{@{\labelitemi\hspace{\dimexpr\labelsep+0.5\tabcolsep}}l}Full access \\Sniffing\\Injection\end{tabular}\end{tabular}                                                                             \\ 
		\hline
		\begin{tabular}[c]{@{}l@{}}Infotainment\\and telematics\\initiated attacks \end{tabular} & \begin{tabular}[c]{@{}l@{}}\begin{tabular}{@{\labelitemi\hspace{\dimexpr\labelsep+0.5\tabcolsep}}l}USB\\CD Player\\BT\\Wi-Fi\\Cellular\\GPS \end{tabular}\end{tabular} & \begin{tabular}[c]{@{}l@{}}\begin{tabular}{@{\labelitemi\hspace{\dimexpr\labelsep+0.5\tabcolsep}}l}Physical\\Physical\\Remote\\Remote\\Remote\\Remote \end{tabular}\end{tabular} & \begin{tabular}[c]{@{}l@{}}\begin{tabular}{@{\labelitemi\hspace{\dimexpr\labelsep+0.5\tabcolsep}}l}Direct Connection\\Direct connection\\Unauthorised access to BT\\Wi-Fi unauthorised access\\Access cellular interface\\Access GPS information \end{tabular}\end{tabular} & \begin{tabular}[c]{@{}l@{}}\begin{tabular}{@{\labelitemi\hspace{\dimexpr\labelsep+0.5\tabcolsep}}l}Inside\\Inside\\Outside\\Outside\\Outside\\Outside \end{tabular}\end{tabular} & \begin{tabular}[c]{@{}l@{}}\begin{tabular}{@{\labelitemi\hspace{\dimexpr\labelsep+0.5\tabcolsep}}l}Inject CAN\\Inject CAN \\Inject and sniffing\\Inject and sniffing\\inject and sniffing\\inject and sniffing \end{tabular}\end{tabular}  \\ 
		\hline
		\begin{tabular}[c]{@{}l@{}}Sensor \\initiated \\attacks \end{tabular}                    & \begin{tabular}[c]{@{}l@{}}\begin{tabular}{@{\labelitemi\hspace{\dimexpr\labelsep+0.5\tabcolsep}}l}TPSM\\Key fob\\LiDAR \end{tabular}\end{tabular}                     & \begin{tabular}[c]{@{}l@{}}\begin{tabular}{@{\labelitemi\hspace{\dimexpr\labelsep+0.5\tabcolsep}}l}Remote\\Remote\\Remote \end{tabular}\end{tabular}                             & \begin{tabular}[c]{@{}l@{}}\begin{tabular}{@{\labelitemi\hspace{\dimexpr\labelsep+0.5\tabcolsep}}l}Decode and replay\\Intercept and relay \\Jamming LiDAR signals \end{tabular}\end{tabular}                                                                                & \begin{tabular}[c]{@{}l@{}}\begin{tabular}{@{\labelitemi\hspace{\dimexpr\labelsep+0.5\tabcolsep}}l}Outside\\Outside\\Outside \end{tabular}\end{tabular}                          & \begin{tabular}[c]{@{}l@{}}\begin{tabular}{@{\labelitemi\hspace{\dimexpr\labelsep+0.5\tabcolsep}}l}Attack TPSM sensors\\Unlock doors \\Block driving assistance \end{tabular}\end{tabular}                                                 \\
		\hline
	\end{tabular}
	
	\textbf{OBD-2}: On-Board Diagnostics ; \textbf{BT}: Bluetooth; \textbf{USB}: Universal Serial Bus; \textbf{GPS}: Global Positioning System; \\ \textbf{TPSM}: Tyre Pressure Monitoring System; \textbf{LiDAR}: Light Detection and Ranging. 
\end{table}
Table~\ref{Figure:Attacks on In Vehicle Networks} identifies some of the attacks initiated on real vehicles and simulated environments. The table identifies entry points used, how attacks were initiated, the position of the attacker, the outcome of the attacks and the software/ hardware test environment used. 
\begin{table}[ht]
	\centering
	\tiny
	\caption{Attacks on In Vehicle Networks}
	\label{Figure:Attacks on In Vehicle Networks}
	\begin{tabular}{l|lllll} 
		\hline
		\textbf{Authors}                            & \begin{tabular}[c]{@{}l@{}}\textbf{Initiated }\\\textbf{Attacks} \end{tabular}                & \begin{tabular}[c]{@{}l@{}}\textbf{Entry }\\\textbf{Points} \end{tabular}                                                                                           & \begin{tabular}[c]{@{}l@{}}\textbf{Position of }\\\textbf{the Attackers} \end{tabular}                                                                                  & \begin{tabular}[c]{@{}l@{}}\textbf{Attack }\\\textbf{Result} \end{tabular}                                                                                                                & \begin{tabular}[c]{@{}l@{}}\textbf{Test }\\\textbf{Environment} \end{tabular}                                                                                                                                                               \\ 
		\hline
		\cite{Koscher2010}
		& \begin{tabular}[c]{@{}l@{}}Interfaces \\Infotainment\end{tabular}                             & \begin{tabular}[c]{@{}l@{}}\begin{tabular}{@{\labelitemi\hspace{\dimexpr\labelsep+0.5\tabcolsep}}l}OBD-2\\USB\\CD Player \end{tabular}\end{tabular}                 & \begin{tabular}[c]{@{}l@{}}\begin{tabular}{@{\labelitemi\hspace{\dimexpr\labelsep+0.5\tabcolsep}}l}Inside (Direct)\\Inside\\Inside \end{tabular}\end{tabular}           & \begin{tabular}[c]{@{}l@{}}\begin{tabular}{@{\labelitemi\hspace{\dimexpr\labelsep+0.5\tabcolsep}}l}CAN bus injection \\Full access \end{tabular}\end{tabular}                             & \begin{tabular}{@{\labelitemi\hspace{\dimexpr\labelsep+0.5\tabcolsep}}l}Real vehicles
		\end{tabular}
		\\ 
		\hline
		\cite{Miller2015}
		& Interfaces                                                                                    & \begin{tabular}{@{\labelitemi\hspace{\dimexpr\labelsep+0.5\tabcolsep}}l}OBD-2 \end{tabular}                                                                         & \begin{tabular}[c]{@{}l@{}}\begin{tabular}{@{\labelitemi\hspace{\dimexpr\labelsep+0.5\tabcolsep}}l}Inside\\Outside \end{tabular}\end{tabular}                           & \begin{tabular}[c]{@{}l@{}}\begin{tabular}{@{\labelitemi\hspace{\dimexpr\labelsep+0.5\tabcolsep}}l}Control brakes, \\Wheels and\\Get access to the CAN bus \end{tabular}\end{tabular}     & \begin{tabular}{@{\labelitemi\hspace{\dimexpr\labelsep+0.5\tabcolsep}}l}Real vehicles \end{tabular}                                                                                                                                         \\ 
		\hline
		\cite{Hoppe2011}
		& Interfaces                                                                                    & \begin{tabular}{@{\labelitemi\hspace{\dimexpr\labelsep+0.5\tabcolsep}}l}OBD-2 \end{tabular}                                                                         & \begin{tabular}{@{\labelitemi\hspace{\dimexpr\labelsep+0.5\tabcolsep}}l}Inside \end{tabular}                                                                            & \begin{tabular}[c]{@{}l@{}}\begin{tabular}{@{\labelitemi\hspace{\dimexpr\labelsep+0.5\tabcolsep}}l}Control Window car lifting\\Warning light \\and airbag \end{tabular}\end{tabular}      & \begin{tabular}[c]{@{}l@{}}\begin{tabular}{@{\labelitemi\hspace{\dimexpr\labelsep+0.5\tabcolsep}}l}Parts of a vehicle \\such as\\instrument cluster, \\window car lifting \\and head unit ECUs\\CANoe simulator \end{tabular}\end{tabular}  \\ 
		\hline
		\cite{Checkoway2011} 
		& \begin{tabular}[c]{@{}l@{}}Interfaces\\Infotainment\\Telematics\\~ ~ ~ ~ ~ ~ ~ ~\end{tabular} & \begin{tabular}[c]{@{}l@{}}\begin{tabular}{@{\labelitemi\hspace{\dimexpr\labelsep+0.5\tabcolsep}}l}OBD-2\\Cellular\\BT\\CD Player\\Radio \end{tabular}\end{tabular} & \begin{tabular}[c]{@{}l@{}}\begin{tabular}{@{\labelitemi\hspace{\dimexpr\labelsep+0.5\tabcolsep}}l}Inside\\Outside\\Outside\\Inside\\Outside \end{tabular}\end{tabular} & \begin{tabular}[c]{@{}l@{}}\begin{tabular}{@{\labelitemi\hspace{\dimexpr\labelsep+0.5\tabcolsep}}l}Get access to CAN bus\\Disable parts of the vehicle \end{tabular}\end{tabular}         & \begin{tabular}{@{\labelitemi\hspace{\dimexpr\labelsep+0.5\tabcolsep}}l}Real vehicles \end{tabular}                                                                                                                                         \\ 
		\hline
		\cite{Petit2015a}
		& Sensors                                                                                       & \begin{tabular}[c]{@{}l@{}}\begin{tabular}{@{\labelitemi\hspace{\dimexpr\labelsep+0.5\tabcolsep}}l}LiDAR\\Cameras \end{tabular}\end{tabular}                        & \begin{tabular}[c]{@{}l@{}}\begin{tabular}{@{\labelitemi\hspace{\dimexpr\labelsep+0.5\tabcolsep}}l}Outside\\Outside \end{tabular}\end{tabular}                          & \begin{tabular}{@{\labelitemi\hspace{\dimexpr\labelsep+0.5\tabcolsep}}l}Signal jamming \end{tabular}                                                                                      & \begin{tabular}[c]{@{}l@{}}\begin{tabular}{@{\labelitemi\hspace{\dimexpr\labelsep+0.5\tabcolsep}}l}LiDAR Hardware \\CAN software \end{tabular}\end{tabular}                                                                                 \\ 
		\hline
		\cite{Rouf2012}  
		\begin{tabular}[c]{@{}l@{}} \\ \end{tabular} & Sensors                                                                                       & \begin{tabular}{@{\labelitemi\hspace{\dimexpr\labelsep+0.5\tabcolsep}}l}TPSM \end{tabular}                                                                          & \begin{tabular}{@{\labelitemi\hspace{\dimexpr\labelsep+0.5\tabcolsep}}l}Outside \end{tabular}                                                                           & \begin{tabular}[c]{@{}l@{}}\begin{tabular}{@{\labelitemi\hspace{\dimexpr\labelsep+0.5\tabcolsep}}l}Inject with\\False TPSM values\\and signal jamming \end{tabular}\end{tabular}          & \begin{tabular}{@{\labelitemi\hspace{\dimexpr\labelsep+0.5\tabcolsep}}l}Real vehicles \end{tabular}                                                                                                                                         \\ 
		\hline
		\cite{Eisenbarth2008}
		& Sensors                                                                                       & \begin{tabular}{@{\labelitemi\hspace{\dimexpr\labelsep+0.5\tabcolsep}}l}Keyfob Keyless entry system \end{tabular}                                                   & \begin{tabular}{@{\labelitemi\hspace{\dimexpr\labelsep+0.5\tabcolsep}}l}Outside \end{tabular}                                                                           & \begin{tabular}[c]{@{}l@{}}\begin{tabular}{@{\labelitemi\hspace{\dimexpr\labelsep+0.5\tabcolsep}}l}Lock,unlock door \\and start the engine \end{tabular}\end{tabular}                     & \begin{tabular}{@{\labelitemi\hspace{\dimexpr\labelsep+0.5\tabcolsep}}l}Real vehicles \end{tabular}                                                                                                                                         \\ 
		\hline
		\cite{Miller2015}
		& Telematics                                                                                    & \begin{tabular}{@{\labelitemi\hspace{\dimexpr\labelsep+0.5\tabcolsep}}l}WiFi \end{tabular}                                                                          & \begin{tabular}{@{\labelitemi\hspace{\dimexpr\labelsep+0.5\tabcolsep}}l}Outside \end{tabular}                                                                           & \begin{tabular}[c]{@{}l@{}}\begin{tabular}{@{\labelitemi\hspace{\dimexpr\labelsep+0.5\tabcolsep}}l}Unauthorised access\\Inject CAN message\\to stop the engine \end{tabular}\end{tabular} & \begin{tabular}{@{\labelitemi\hspace{\dimexpr\labelsep+0.5\tabcolsep}}l}Jeep Cherokee \end{tabular}                                                                                                                                         \\ 
		\hline
		\cite{Tesla2016}
		& Telematics                                                                                    & \begin{tabular}{@{\labelitemi\hspace{\dimexpr\labelsep+0.5\tabcolsep}}l}WiFi\end{tabular}                                                                           & \begin{tabular}{@{\labelitemi\hspace{\dimexpr\labelsep+0.5\tabcolsep}}l}Outside \end{tabular}                                                                           & \begin{tabular}{@{\labelitemi\hspace{\dimexpr\labelsep+0.5\tabcolsep}}l}Full access to CAN bus \end{tabular}                                                                              & \begin{tabular}{@{\labelitemi\hspace{\dimexpr\labelsep+0.5\tabcolsep}}l}Tesla model S \end{tabular}                                                                                                                                         \\ 
		\hline
		\cite{Zorz2018}
		& Interfaces                                                                                    & \begin{tabular}{@{\labelitemi\hspace{\dimexpr\labelsep+0.5\tabcolsep}}l}OBD-2 Cellular Dongle \end{tabular}                                                         & \begin{tabular}{@{\labelitemi\hspace{\dimexpr\labelsep+0.5\tabcolsep}}l}Outside \end{tabular}                                                                           & \begin{tabular}{@{\labelitemi\hspace{\dimexpr\labelsep+0.5\tabcolsep}}l}CAN bus injection \end{tabular}                                                                                   & \begin{tabular}{@{\labelitemi\hspace{\dimexpr\labelsep+0.5\tabcolsep}}l}Real vehicle \end{tabular}                                                                                                                                          \\
		\hline
	\end{tabular}
\end{table}
\subsection{Attacks against the CAN Bus}

The classical CAN and CAN FD buses are vulnerable to various attacks. Once attackers have access from either inside or outside the vehicle, they can generate various attacks on the CAN bus network such as CAN sniffing, CAN fuzzing, CAN replay and DoS attacks. Some of the mechanisms for initiating these attacks include:
\begin{figure}[ht]
  \centering
  \includegraphics[width=.9\linewidth]{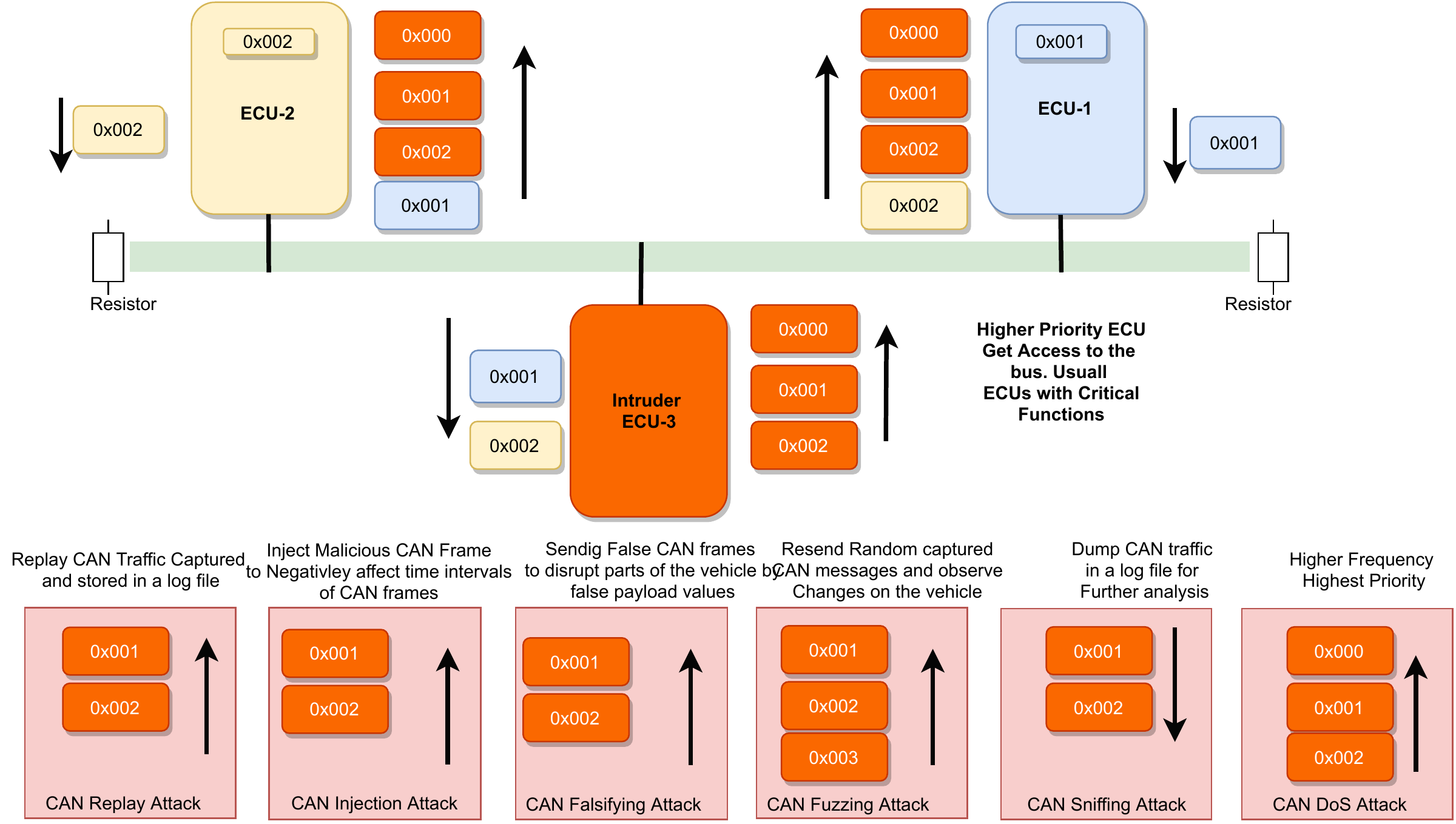}
  \caption{Overview of some In Vehicle CAN bus Network (IVN) attacks}
  \label{Figure:Overview of some IVN CAN bus attacks}
\end{figure}
\noindent \textbf{CAN bus sniffing}:
With no authentication mechanisms, encryption and broadcast transmission, it is possible to sniff the data on the CAN bus~\cite{Currie2015}. Using  off the shelf OBD2 sniffer such as CANdo board, it is possible to read and analyse the data on the bus to manipulate and generate similar messages \cite{Miller2015}. This attack can be avoided by implementing encryption to prevent exposing CAN frames. This attack is difficult to detect due to the passive nature of sniffing traffic. The next step is to reverse engineer the raw CAN messages so that they can be used to target specific parts of the vehicle. This an important step since manufacturers tend not to publish their CAN message specification~\cite{Kang2018}.

\noindent \textbf{CAN bus fuzzing attack}
CAN bus protocol lacks authentication and data integrity checking and as a result ECUs accept CAN messages and respond to them. This attack is used to send random CAN data frames, checking the bus and observing changes on the instrument panel of the vehicle. This attack looks at the impact of CAN frames on the ECUs such as observing the change in vehicle speed while injecting CAN frames~\cite{Khan2017}. It usually happens after sniffing and analysing captured CAN messages. Also, it can be generated using a black-box, where CAN id and payload values are generated randomly without prior knowledge of the actual CAN id used. It involves sending randomly captured CAN frames and recording the outcome. Encryption is needed to prevent analysis of the captured data, along with authentication to only accept CAN frames from legitimate ECUs.  

\noindent \textbf{CAN bus frame falsifying attack}
This attack is used to modify CAN message payload by inserting incorrect values. For example, the attacker can inject a vehicle with incorrect parameter values. This type of modification attack is used when the CAN id is known, and the intention is to provide incorrect data payload to disrupt vehicle services. This happens due to the lack of data integrity and authentication support in the CAN bus protocol. In order to prevent this attack, CAN bus should provide authentication to verify the source of the data before acting upon it. Usually this attack involves a small amount of data, making it difficult to detect and monitor. To detect this attack, a system should consider checking CAN id and data payload consistency in a time window.  

\noindent \textbf{CAN bus injection attack}
Injecting data into a CAN bus can be used to send messages at an abnormal rate \cite{Marchetti2017a}. The purpose of this attack is to change frequency and amount of CAN frames on the bus, and change the sequence of legitimate CAN frames and data payload. Since CAN bus does not provide authentication to check if the sender is legitimate, this attack will inject the bus with abnormal CAN traffic targeting the vehicle speed. Lack of encryption also enables arbitrary nodes to connect to the bus. The data on the bus can then be monitored to obtain the arbitration and data field, and and generate messages to simulate events~\cite{Culling2017}. This could lead to generation of fake events that cause parts of the vehicle to behave as required by the attacker. This attack can be prevented using authentication and integrity mechanisms. The result of the attack can increase the broadcast frequency of certain CAN id which can be detected through abnormal broadcast behaviour.

\noindent \textbf{CAN bus DoS attack:}
Classical CAN and CAN FD use the same mechanism to access the medium with multi access using the CAN id  priority \cite{Huang2017}. The nodes on the CAN bus use the arbitration field to determine the priority of the message and which node can occupy the bus and send data. In this case, a DoS attacked can be lunched using highest attribution id such as 0x000 to occupy the bus and make it busy by using CAN frame priority arbitration scheme and send too many highest priority frames so that other nodes cannot use the bus \cite{Miller2015}. Also, it can use the same CAN message id of an existed ECU and by knowing its transmission rate, a DoS can be performed by incrementing the frequency time. For example, if an ECU sends a message every 200 ms, the attacker can increase the frequency by injecting the same message with higher frequency which can lead to disruption of the sensor part. 

\noindent \textbf{ECU impersonation:} 
Once an attacker has access to the CAN bus network, the attacker can receive all the traffic broadcast on the bus. With a focused analysis of the traffic, attackers can learn the behaviour of each ECU such as it's CAN ID, payload range and transmission rate. In this way, they can simulate ECU behaviour by sending the same data with the same frequency. An increase in the CAN messages rate will occur which generates an attack. However, if the attack was more focused, they could initiate an attack to disable particular ECUs. For example, Iehira et al.~\cite{Iehira2018} introduced a sophisticated spoofing ECU attack by first performing an attack on an ECU by taking advantage of the error handling mechanism of the CAN bus protocol. This attack works by mimicking the target ECU behaviour, CAN ID and frequency. Then, the attacker ECU contradicts the target ECU by sending a dominant bit while the original ECU sends a recessive bit. This would raise an error in the ECU controller which leads, at a certain point, to disconnecting the ECU from the bus and dropping all the CAN bus communication. This enables an attacker to perform various attacks, such as an ECU impersonation attack, which is difficult to detect. 

\section{CAN bus Security Mechanisms}
\label{Sec6:CANbus-SecurityMechanism}

Implementing and testing security of CAN bus traffic has been conducted by many researchers. 
In this section we identify current countermeasures used and divide them based on the mechanisms they use, and whether these are used from within or outside the vehicle. We also consider factors such as the test environment, security metric being considered, countermeasure used, the type of mitigated attacks, overhead of supporting the countermeasure and utilization. 

\subsection{In Vehicle Network Cybersecurity}

Given the limited capacity of the CAN bus, any countermeasures used to address its vulnerabilities should consider this limitation and not overload the bus. Security solutions for a CAN bus can be divided into encryption, authentication and redesign of the protocol stack by replacing fields in the frame, splitting the message to multiple frames, or by adding nodes and components to the bus to realise additional capability. These approaches can be costly to deploy. Cryptography-based methods have focused on securing the CAN bus from malicious messages, while Intrusion Detection Systems (IDS) focus on the detection of malicious messages. Firewall and Intrusion Prevention Systems (IPS) can be used in external interfaces to block access to the bus. Implementing a dedicated node to realise the IDS and firewall may be required.


\subsection{Using Cryptography }

Implementing cryptography in the CAN bus requires additional computational resources in the ECUs and the CAN bus controller. Cryptography can be used to provide authentication and data integrity through Message Authentication Code (MAC)  and confidentiality through symmetric and asymmetric cryptosystems. For in-vehicle networks, a key challenge is to create a secure method that does not alter the payload size (e.g. splitting the message can lead to more load on the bus) and response time latency which would affect vehicle safety.  CAN bus also provides a checksum calculation using Cyclic Redundancy Code (CRC) to check if there is a change in the frame during transmission, but it only provides error detection not the integrity and authentication of the frame. An ACK field is used for error detection and correction purposes. Implementing cryptography in the CAN bus should consider the following \cite{Herrewege2011}, \cite{Nurnberger2016}, \cite{Radu2016}:

\begin{itemize}
\item Limited frame size and capacity of the bus;
\item limited speed of response and high latency;
\item broadcast nature of the bus -- and lack of support for confidentiality, integrity and authentication by design;
\item no backward compatibility;
\item limited computational capacity within ECUs.
\end{itemize}

Lightweight encryption is needed in such embedded systems, due to limited computational capacity within ECUs inside the vehicle. The approach used in the classical CAN bus protocol involves creating a small MAC tag size, less than 8bytes, and inserting it along with the actual data payload. This tag provides integrity and authentication as it is encrypted by a shared secret key. Session keys are used for authentication and to prevent subsequent re-play attacks. Key distribution is a concern in CAN bus broadcast environments and therefore a pre-loaded key in each ECU can be used to establish key exchange and freshness to tackle data broadcast and to avoid bus loading due to key exchange. Also, to tackle the issue of low computing resources, Hardware Security Module (HSM) can be used in resource-constrained ECUs to provide better encryption/ decryption time. However, these approaches can still be costly to realise within existing vehicles. In summary, the adopted approaches involve: 

\begin{figure}[ht]
  \centering
  \includegraphics[width=\linewidth]{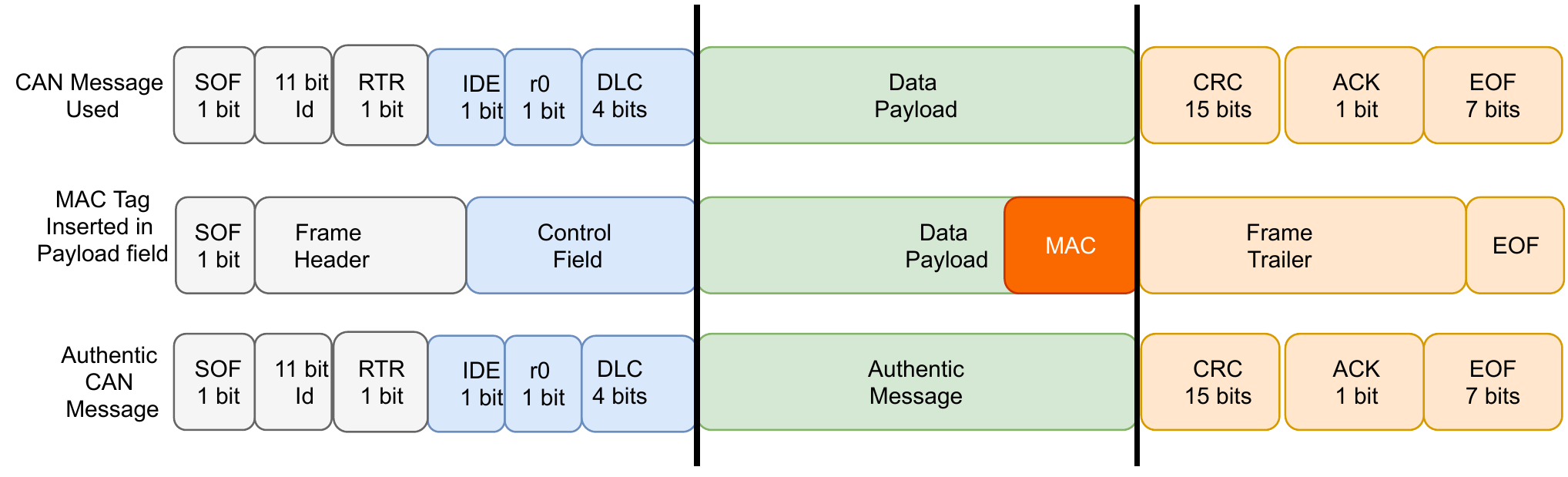}
  \caption{MAC signature inside CAN frame
  \cite {Ueda2015}}
  \label{Fig:MAC generated Tag in CAN frame}
\end{figure}



\begin{itemize}
\item Using lightweight Message Authentication Code (MAC) to overcome resource constraints in ECUs, and making use of a small key size and MAC signature.
\item Implementing changes in the CAN protocol standard by replacing fields such as CRC with MAC signature, or by extending the protocol  data field (called CAN+) and extending the data payload to 16 bytes to give more space for the MAC signature. However, this approach leads to compatibility issues.
\item A Hardware Security Accelerator can be used (as an additional hardware) to overcome computational resource limitations -- however, it may not be a cost-effective approach. 
\end{itemize}

According to the National Institute of Standard and Technology (NIST), HMAC \cite{FIPS-198-12008} provides authentication and integrity of the data through a hash function and a shared secret key between sender and receiver. Hash algorithms such as SHA-1, SHA-224 and SHA-256 produce message digest or MacTag of 160, 224 and 256 bits according to \cite{Dang2012}. The size of this tag is exceeds the maximum data payload of CAN frame (of 64 bits) and thus a truncated tag is used by deriving a smaller size from the MAC computation. 

\begin{figure}[t]
  \centering
  \includegraphics[width=.9\linewidth]{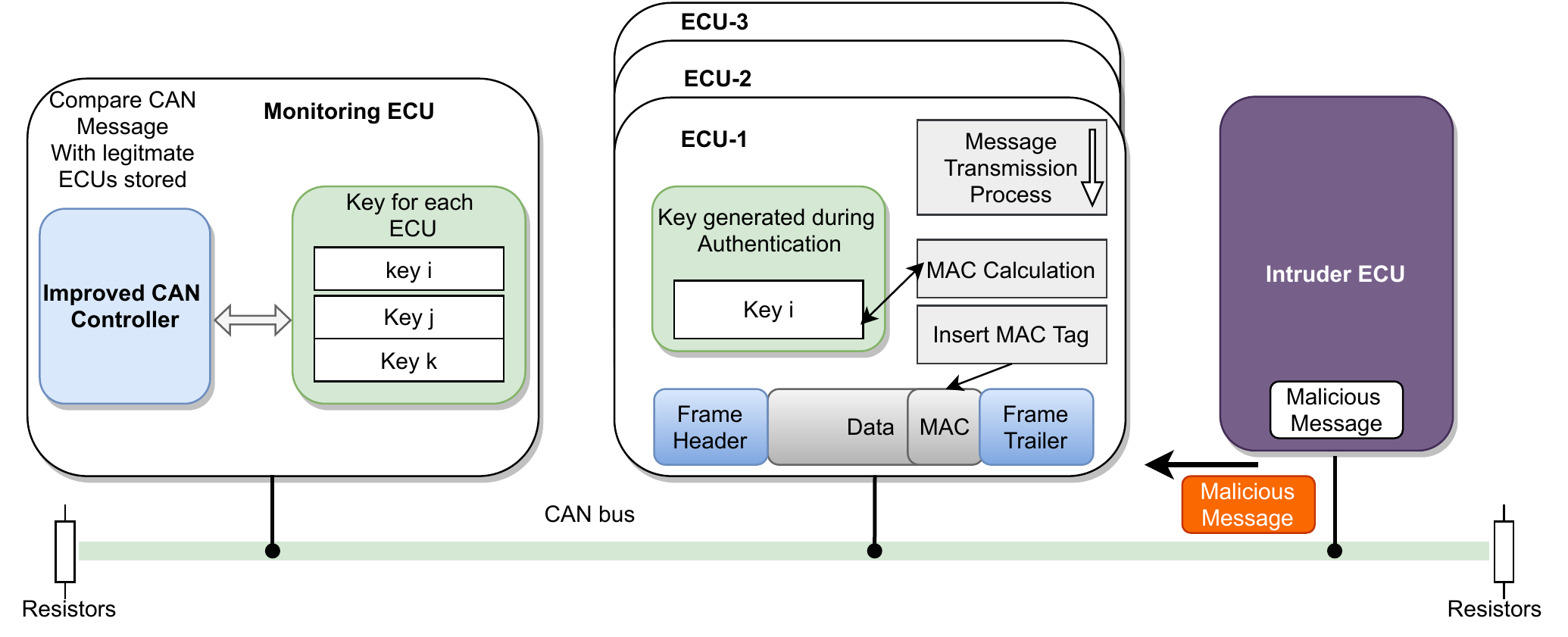}
  \caption{ CAN message authentication and dedicated monitoring ECU. Derived from \cite {Ueda2015}}
   \label{Fig:CAN message authentication and dedicated monitoring ECU}
\end{figure}

\subsection{CAN Frame Authentication}

Authentication mechanisms can be used to provides authentication and integrity of the data for an in-vehicle network. This mechanism is called Message Authentication Code (MAC). However, this approach does not provide confidentiality which means CAN traffic is still exposed to sniffing and reverse engineering attacks. Thus, a combination of MAC and encryption is needed to provide better security. The following approaches can provide authentication and integrity of CAN bus data, but either change the behaviour of the protocol by splitting CAN frames, replace fields, increase CAN frame size or increase bus payload and response time. Some approaches also require additional hardware which can increase the cost of implementation and lead to incompatibility with current vehicles. 

-Nilsson et al \cite{Nilsson2008} introduced an approach based on a shared 128-bit key between ECUs and a Cipher-Block Chaining Message Authentication Code (CBC-MAC) using KASUMI encryption algorithm. Their approach provides integrity and authentication through a 64-bit MAC tag. However, it splits a single CAN ID message into multiple messages in order to insert 16-bits inside the CRC field. This means that 4 messages are needed in order to send the 64-bit tag. As a result, their approach increases the bus load by increasing the number of CAN bus messages and it changes the CAN protocol behaviour by replacing the CRC field with MAC tags. Furthermore, it causes more latency through CAN message splitting.

-Wang and Sawhney \cite{Wang2014} proposed a trusted group-based technique to enforce access control while minimizing the distribution of keys through a pre-calculated cryptographic function. Their approach was able to successfully prevent message injection attacks while message processing delay was approx. 50$\mu$s. They used Freescale’s automotive boards to test their solutions. 
Trusted group ECUs share a secret symmetric key (K h) -- and
ECUs in the trusted group can hold keys of other groups if needed. This approach is effective since it separates telematics and OBD-2 ports which are the main entry points for attackers. However, their authentication approach is achieved by sending data and authentication messages for each CAN ID which doubles the bus load. 

-Another approach CANAuth used light weight encryption to mitigate sniffing and poisoning attacks~\cite{Herrewege2011}. The authors considered the limitations of the CAN bus protocol and used lightweight encryption mechanisms to mitigate attacks, but DoS attacks were not investigated and a CAN+ protocol (16bits) was used which is incompatible with standard CAN bus specifications.  The proposed approach uses HMAC function with pre-shared symmetric and group keys for key distribution. It uses 15 bytes for HMAC flag and key transmission while 1 byte is used for the actual data. 

-LCAP in \cite{Hazem2012} used a one way function to provide a 2byte magic number. The authors proposed using their approach in the data field (2 bytes out of 8 bytes) in the standard CAN frames 2.0A. In the extended CAN, they used magic number of 16 bit in the header field of the extended header 29 bit CAN frame 2.0B. Their approach provides authentication and integrity through symmetric keys and HMAC magic number. LCAP provides protection against re-play attacks due to the changeable magic number and session keys. A drawback is that it is based on the CAN+ (16 bytes of data) which raises compatibility issues as the CAN transceiver hardware needs modification to handle the 16 bytes of data payload. 

-LibrA-CAN is a broadcast authentication protocol using the MD5 Message Digest, compatible with CAN+ specification introduced in \cite{Groza2012}. Multiple receivers can hold keys and provide authentication roles based on monitoring their own message ID usage inside the CAN bus. This approach splits CAN messages into normal CAN messages and authentication tag messages -- which increases bus traffic~\cite{Bella2019}. For improvement, they suggest using CAN+, but the CAN transceiver hardware should be changed to be able to handle the new CAN+ frame.  

-In\cite{Bittl2014}, the authors used a combination of SHA3 and HMAC function along with session keys to avoid re-play attacks. This method used the length in the CRC fields to insert a cryptographic checksum. The processing time of sending and receiving a message was not provided in this approach. Also, there is a change in compatibility of the CAN frame specification by replacing CRC field with MAC tag field.

-CaCAN in \cite{Kurachi2014} is used to carry out authentication and validate integrity of CAN messages. This is achieved by using a main ``Monitor" ECU that shares keys with other ECUs. Using the broadcast behaviour of the CAN us, it receives all messages and can detect and overwrite unauthorised messages. This approach does not provide confidentiality and additional hardware is needed as a monitoring ECU node.

-LeiA was introduced in \cite{Radu2016} which used 128-bit key, MAC and counter based algorithms to authenticate data and generate counters to mitigate re-play attacks. This algorithm does not require any changes to the hardware and topology of the CAN Bus. However, it is compatible only with CAN 2.0B 29-bit extended frame and changes the CAN header by replacing the 18-bit identifier with a counter -- potentially leading to incompatibility issues with current vehicle networks.

-In \cite{Kang2017}, the authors introduced a one-way hash chain using HMAC-MD5 and AES-128 to provide authentication. they tested their approach on simulated ECUs using CANoe Vector tool and Freescale S12XF as CAN hardware. Re-play and spoofing attacks were considered in this approach. They used the symmetric key and Authentication Key Exchange Protocol2 (AKEP2), and assume that the symmetric key and the ID of the sender are stored during manufacture. They have demonstrated only a limited overhead (bus load and latency) when using their approach.

-In \cite{Ueda2015}, the authors have used HMAC-SHA256 to provide ECU authentication and data integrity. Their MAC tag size is 1byte and it is inserted along with the actual data payload and a counter size of 4-bit to prevent re-play attacks. They include a `Monitor' ECU that receives all the messages on the bus and checks if they are legitimate, by holding all the keys of the ECUs. In case of an illegitimate message, they send a remote frame to overwrite the malicious message.
\begin{table}[ht!]
	\centering
	\tiny

	\caption{ Frames Authentication for Controller Area Network}
	\label{tbl:Message Authentication for CAN bus frmaes}
	\begin{tabular}{l|l|llllll} 
		\hline
		\textbf{ Authors }  & \textbf{Method }                                                                                 & \begin{tabular}[c]{@{}l@{}}\textbf{Attacks}\\\textbf{mitigated } \end{tabular}                                                                                         & \begin{tabular}[c]{@{}l@{}}\textbf{Hardware}\\\textbf{Security }\\\textbf{Module } \end{tabular} & \begin{tabular}[c]{@{}l@{}}\textbf{Real }\\\textbf{time } \end{tabular}                     & \begin{tabular}[c]{@{}l@{}}\textbf{Bus}\\\textbf{load } \end{tabular}        & \begin{tabular}[c]{@{}l@{}}\textbf{Change }\\\textbf{CAN bus } \end{tabular}                   & \begin{tabular}[c]{@{}l@{}}\textbf{Test }\\\textbf{bed }\\\textbf{environment } \end{tabular}          \\ 
		\hline
		\cite{Nilsson2008}
		&
		 CBC-MAC                                                                                          & \begin{tabular}[c]{@{}l@{}}\begin{tabular}{@{\labelitemi\hspace{\dimexpr\labelsep+0.5\tabcolsep}}l}Injection \\ spoofing \end{tabular}\end{tabular}                    & No                                                                                               & \begin{tabular}[c]{@{}l@{}}Delayed\\ authentication \end{tabular}                           & \begin{tabular}[c]{@{}l@{}}Multiple\\CAN\\ frames \end{tabular}              & \begin{tabular}[c]{@{}l@{}}Splitting~\\CAN frame\\for authentication \\purposes \end{tabular}  & Theoretical                                                                                            \\ 
		\hline
		\cite{Wang2014}
		& \begin{tabular}[c]{@{}l@{}}Trusted Group\\HMAC\\Symmetric key\end{tabular}                       & \begin{tabular}[c]{@{}l@{}}\begin{tabular}{@{\labelitemi\hspace{\dimexpr\labelsep+0.5\tabcolsep}}l}Sniffing,\\Spoofing \\Injection \end{tabular}\end{tabular}          & \begin{tabular}[c]{@{}l@{}}Pre-load keys\\During \\manufacturing \end{tabular}                   & Yes                                                                                         & \begin{tabular}[c]{@{}l@{}}Message\\Splitting \end{tabular}                  & \begin{tabular}[c]{@{}l@{}}Split CAN frames\\ for authentication\\ purposes \end{tabular}      & \begin{tabular}[c]{@{}l@{}}Freescale’s \\automotive\\ boards \end{tabular}                             \\ 
		\hline
		\cite{Herrewege2011}
		& \begin{tabular}[c]{@{}l@{}}HMAC\\Symmetric~key\\Counters\end{tabular}                            & \begin{tabular}[c]{@{}l@{}}\begin{tabular}{@{\labelitemi\hspace{\dimexpr\labelsep+0.5\tabcolsep}}l}Sniffing, \\Spoofing\\Injection \\Replay \end{tabular}\end{tabular} & No                                                                                               & Yes                                                                                         & \begin{tabular}[c]{@{}l@{}}16 bytes of \\data payload \end{tabular}          & \begin{tabular}[c]{@{}l@{}}CAN+ 16 Bytes\\All nodes must~\\Know pre-shared key\\ \end{tabular} & Theoretical                                                                                            \\ 
		\hline
		\cite{Hazem2012} 
		& \begin{tabular}[c]{@{}l@{}}One way function\\Magic number\\of 2 bytes\\Session keys\end{tabular} & \begin{tabular}[c]{@{}l@{}}\begin{tabular}{@{\labelitemi\hspace{\dimexpr\labelsep+0.5\tabcolsep}}l}Replay \\ Injection \end{tabular}\end{tabular}                      & No                                                                                               & \begin{tabular}[c]{@{}l@{}}-Yes, but~\\Consume time\\during key\\~distribution\end{tabular} & \begin{tabular}[c]{@{}l@{}}Add extra~\\2 Bytes in~\\the payload\end{tabular} & \begin{tabular}[c]{@{}l@{}}HMAC tag\\ in the 2.0B CAN\\frame~header\\\end{tabular}             & \begin{tabular}[c]{@{}l@{}}Starter-TRAK \\TRK-MPC5604B\\~board\end{tabular}                            \\ 
		\hline
		\cite{Kurachi2014}
		& \begin{tabular}[c]{@{}l@{}}HMAC\\Symmetric keys~\\Counter\end{tabular}                           & \begin{tabular}[c]{@{}l@{}}\begin{tabular}{@{\labelitemi\hspace{\dimexpr\labelsep+0.5\tabcolsep}}l}Replay\\Spoofing \\Injection \end{tabular}\end{tabular}             & \begin{tabular}[c]{@{}l@{}}ECU \\server \end{tabular}                                            & Yes                                                                                         & No                                                                           & \begin{tabular}[c]{@{}l@{}}Special ECU\\server\\hardware\end{tabular}                          & \begin{tabular}[c]{@{}l@{}}Altera FPGA \\ board\\CAN \\transceiver board \end{tabular}                 \\ 
		\hline
		\cite{Bittl2014}
		& \begin{tabular}[c]{@{}l@{}}SAH3\\HMAC\end{tabular}                                               & \begin{tabular}[c]{@{}l@{}}\begin{tabular}{@{\labelitemi\hspace{\dimexpr\labelsep+0.5\tabcolsep}}l}Replay \\Injection \end{tabular}\end{tabular}                       & No                                                                                               & Yes                                                                                         & No                                                                           & \begin{tabular}[c]{@{}l@{}}Replace CRC\\field \end{tabular}                                    & Theoretical                                                                                            \\ 
		\hline
		\cite{Groza2012}
		& \begin{tabular}[c]{@{}l@{}}MD5\\LMAC\end{tabular}                                                & \begin{tabular}[c]{@{}l@{}}\begin{tabular}{@{\labelitemi\hspace{\dimexpr\labelsep+0.5\tabcolsep}}l}Replay\\Injection \end{tabular}\end{tabular}                        & No                                                                                               & No                                                                                          & Yes                                                                          & \begin{tabular}[c]{@{}l@{}}Split CAN~messages\\Using CAN+ \end{tabular}                        & Theoretical                                                                                            \\ 
		\hline
		\cite{Radu2016} 
		& \begin{tabular}[c]{@{}l@{}}128-bit key\\MAC\\Counter\end{tabular}                                & \begin{tabular}[c]{@{}l@{}}\begin{tabular}{@{\labelitemi\hspace{\dimexpr\labelsep+0.5\tabcolsep}}l}Replay \\Injection\\Spoofing \end{tabular}\end{tabular}             & No                                                                                               & \begin{tabular}[c]{@{}l@{}}No\\ \end{tabular}                                               & \begin{tabular}[c]{@{}l@{}}No\\ \end{tabular}                                & \begin{tabular}[c]{@{}l@{}}16-bit counter\\ in the CAN 2.0b \\frame header \end{tabular}       & \begin{tabular}[c]{@{}l@{}}FreescaleS12X\\and Infineon TriCore \end{tabular}                           \\ 
		\hline
		\cite{Kang2017}
		& \begin{tabular}[c]{@{}l@{}}HMAC\\MD5\\AES-128\end{tabular}                                       & \begin{tabular}[c]{@{}l@{}}\begin{tabular}{@{\labelitemi\hspace{\dimexpr\labelsep+0.5\tabcolsep}}l}Spoofing \\Replay \end{tabular}\end{tabular}                        & No                                                                                               & Yes                                                                                         & No                                                                           & \begin{tabular}[c]{@{}l@{}}Insert MAC tag \\in 18-bit filed \\in CAN 2.0B header \end{tabular} & \begin{tabular}[c]{@{}l@{}}CANoeVector tool\\Freescale \\S12XF board \end{tabular}                     \\ 
		\hline
		\cite{Ueda2015}
		& \begin{tabular}[c]{@{}l@{}}HMAC\\SHA 256\end{tabular}                                            & \begin{tabular}[c]{@{}l@{}}\begin{tabular}{@{\labelitemi\hspace{\dimexpr\labelsep+0.5\tabcolsep}}l}Replay \\Spoofing \end{tabular}\end{tabular}                        & \begin{tabular}[c]{@{}l@{}}Dedicated\\ECU-FPGA \\board\\with built-in \\HMAC \end{tabular}       & Yes                                                                                         & No                                                                           & \begin{tabular}[c]{@{}l@{}}Insert 8-bit \\MAC tag\\ 4-bit counter \end{tabular}                & \begin{tabular}[c]{@{}l@{}}Altera FPGA \\development\\~board and CAN\\~transceiver board\end{tabular}  \\
		\hline
	\end{tabular}
	
	\textbf{HMAC}, Hash Message Authentication Code; \textbf{CBC-MAC}, Cipher Block Chaining Message Authentication Code; \\ \textbf{LM-MAC}, Linearly Mixed MAC; \textbf{MD5}, Message Digest; \textbf{SHA3}, Secure Hash Algorithm3.
\end{table}

While the above approaches provide authentication and integrity for the CAN bus protocol, they suffer from other limitations such as backward incompatibility, real time constrains (delayed authentication) or cost of implementations by using dedicated hardware. Therefore, a software-based approach should focus on providing authentication and integrity without failing in these shortcomings.

-The approach proposed by Fassak et al.~\cite{Fassak2017} made use of an asymmetric key.  HMAC is then used with changeable session keys. The authors assume that both public and private keys are pre-installed in the ECUs during manufacture. Also, the performance of their approach was validated analytically using a commercial bus load calculator by OptimumG. The security of the algorithm was validated using the AVISPA software. However, their approach was not tested in a realistic test environment, and it is not compatible with current vehicles -- as their assumption is to embed the key during manufacture.
 
- Groza and Murvay~\cite{Groza2013} provide a secure broadcast protocol for the CAN bus. It uses a central ECU to manage and distribute the keys between the sender and the receiver. They validated their approach using Freescale and S12X (16-bit) and TriCore (32bit) microcontrollers. However, their approach is based on a delayed authentication approach which is difficult to support in real time for a CAN bus~\cite{Woo2015}.

- In \cite{Bella2019}, the authors introduced ``TOUCAN'' which provides authentication, integrity and encryption for a CAN bus. They use AES 128bit and Chaskey hashing for MAC authentication without the need for ECU upgrade or additional hardware on the bus. They tested their approach on STM32F407 CAN boards. The actual data in the payload is 40bits while the remaining 24bits are used for the hashing value. In their approach,  (using AES 128 and Chaskey hashing) the overall execution times are approx. 12ms.  

-In \cite{Wu2016}, the authors used AES 128 encryption and HMAC for authentication. They also use a compression algorithm to improve the efficiency of their approach by reducing the delay time and bus load. They have used Vector CANoe software to validate their approach, showing that the average message delay is 0.13ms.

-In \cite{Lin2012}, the authors focus on preventing re-play and spoofing attacks by using various approaches such as message counters, CAN ID tables to look up the ID of each ECU and which MAC to use for each ECU ID. Pair-wise symmetric key is used as each ECU stores the shared key with other ECUs. Also, the MAC tag is previously generated and stored in the look up ID table where the ECU uses this ID table to link the receiving ECU with the correspondent MAC tag. Their approach does not need any hardware modification and has a low message latency and bus load. However, it does not provide confidentiality.
\begin{table}[ht]
	\centering
	\tiny
	\caption{Message Authentication for CAN frames}
	\label{tbl:Message Authentication for CAN bus frmae}
	\begin{tabular}{l|l|llllll} 
		\hline
		\textbf{ Authors }  & \textbf{Method }                                                                                                                                                      & \begin{tabular}[c]{@{}l@{}}\textbf{Attacks}\\\textbf{mitigated } \end{tabular}                                                                    & \begin{tabular}[c]{@{}l@{}}\textbf{Hardware}\\\textbf{Security }\\\textbf{Module } \end{tabular} & \begin{tabular}[c]{@{}l@{}}\textbf{Real }\\\textbf{time } \end{tabular} & \begin{tabular}[c]{@{}l@{}}\textbf{Bus}\\\textbf{load } \end{tabular}         & \begin{tabular}[c]{@{}l@{}}\textbf{Change }\\\textbf{CAN bus } \end{tabular}                      & \begin{tabular}[c]{@{}l@{}}\textbf{Test }\\\textbf{bed }\\\textbf{environment } \end{tabular}               \\ 
		\hline
		\cite{Fassak2017}
		& \begin{tabular}[c]{@{}l@{}}\begin{tabular}{@{\labelitemi\hspace{\dimexpr\labelsep+0.5\tabcolsep}}l}Asymmetric key\\HMAC\\Changeable keys \end{tabular}\end{tabular}   & \begin{tabular}[c]{@{}l@{}}\begin{tabular}{@{\labelitemi\hspace{\dimexpr\labelsep+0.5\tabcolsep}}l}Replay\\ Spoofing \end{tabular}\end{tabular}   & No                                                                                               & Yes                                                                     & \begin{tabular}[c]{@{}l@{}}Load during\\ key exchange\textbf{ } \end{tabular} & \begin{tabular}[c]{@{}l@{}}Assume\\pre-installed\\ keys \end{tabular}                             & Analytical evaluation                                                                                       \\ 
		\hline
		\cite{Groza2017}
		& \begin{tabular}[c]{@{}l@{}}\begin{tabular}{@{\labelitemi\hspace{\dimexpr\labelsep+0.5\tabcolsep}}l}Symmetric key \\HMAC \end{tabular}\end{tabular}                    & \begin{tabular}[c]{@{}l@{}}\begin{tabular}{@{\labelitemi\hspace{\dimexpr\labelsep+0.5\tabcolsep}}l}Replay \\ Spoofing \end{tabular}\end{tabular}  & No                                                                                               & \begin{tabular}[c]{@{}l@{}}Delayed\\authentication \end{tabular}        & No                                                                            & \begin{tabular}[c]{@{}l@{}}Delayed \\authentication \end{tabular}                                 & \begin{tabular}[c]{@{}l@{}}S12 equipped with \\an XGATE\\ coprocessor\\ and Infineon TriCore \end{tabular}  \\ 
		\hline
		\cite{Bella2019}
		& \begin{tabular}[c]{@{}l@{}}\begin{tabular}{@{\labelitemi\hspace{\dimexpr\labelsep+0.5\tabcolsep}}l}AES-128\\Chasekey HMAC \end{tabular}\end{tabular}                  & \begin{tabular}[c]{@{}l@{}}\begin{tabular}{@{\labelitemi\hspace{\dimexpr\labelsep+0.5\tabcolsep}}l}Spoofing \\Replay \end{tabular}\end{tabular}   & No                                                                                               & Yes                                                                     & No                                                                            & \begin{tabular}[c]{@{}l@{}}24-bit MAC tag\\ and 40-bits \\ for the actual data \end{tabular}      & \begin{tabular}[c]{@{}l@{}}STM32F407\\CAN boards \end{tabular}                                              \\ 
		\hline
		\cite{Wu2016}
		& \begin{tabular}[c]{@{}l@{}}\begin{tabular}{@{\labelitemi\hspace{\dimexpr\labelsep+0.5\tabcolsep}}l}AES-128\\HMAC \\Compression algorithm \end{tabular}\end{tabular}   & \begin{tabular}[c]{@{}l@{}}\begin{tabular}{@{\labelitemi\hspace{\dimexpr\labelsep+0.5\tabcolsep}}l}Replay \\ Injection \end{tabular}\end{tabular} & No                                                                                               & Yes                                                                     & No                                                                            & \begin{tabular}[c]{@{}l@{}}No\\ \end{tabular} & \begin{tabular}[c]{@{}l@{}}CANoe \\simulator \end{tabular}                                                  \\ 
		\hline
		\cite{Lin2012}
		& \begin{tabular}[c]{@{}l@{}}\begin{tabular}{@{\labelitemi\hspace{\dimexpr\labelsep+0.5\tabcolsep}}l}MAC tables\\Pairwise key\\Symmetric key \end{tabular}\end{tabular} & \begin{tabular}[c]{@{}l@{}}\begin{tabular}{@{\labelitemi\hspace{\dimexpr\labelsep+0.5\tabcolsep}}l}Replay\\Spoofing \end{tabular}\end{tabular}    & ECU server                                                                                       & No                                                                      & Yes                                                                           & No                                                                                                & No                                                                                                          \\
		\hline
	\end{tabular}
\end{table}
\subsection{CAN Frame Encryption}


-In \cite{Dariz2017} the authors used a combination of encryption and authentication mechanisms to provide data confidentiality, integrity and authenticity. This approach provides prevention against sniffing and injection attacks. However, it sends more than one frame for a single CAN ID message which can lead to latency and increased bus load~\cite{Bella2019} .  
 
-In \cite{Siddiqui2017} the authors  used a hardware-based approach to provide authentication and encryption for a CAN bus. A dedicated hardware (ECU Server) was used to manage all the ECUs in the CAN bus to authenticate ECUs and distribute keys. A Xilinx Kintex KC705 FPGA Evaluation board and an embedded Physical Unlockable Function (PUF) was used in the testbed. This approach assumes that keys are registered for ECUs during manufacture, and assembling and CAN controller boards need to support the physical PUF function. This is likely to lead to less overhead, but it is infeasible to implement in current vehicle network due to the hardware modifications required. 

-CANTrack algorithm by Farag et al. \cite{Farag2017} uses a dynamic symmetric key to encrypt the 8byte data payload, but does not modify the Msg ID as it is used to access the bus during the arbitration mechanism. They have tested their approach with CANoe software, and it has shown to prevent sniffing, replay and spoofing attacks.
\begin{table}[ht!]
	\centering
	\tiny
	\caption{CAN Frame Encryption methods}
	\label{tbl:CAN Frame Encryption methods}
	\begin{tabular}{l|l|llllll} 
		\hline
		\textbf{ Authors }  & \textbf{Method }                                                                                                                                                   & \begin{tabular}[c]{@{}l@{}}\textbf{Attacks}\\\textbf{mitigated } \end{tabular}                                                                               & \begin{tabular}[c]{@{}l@{}}\textbf{Hardware}\\\textbf{Security }\\\textbf{Module } \end{tabular} & \begin{tabular}[c]{@{}l@{}}\textbf{Real }\\\textbf{time } \end{tabular} & \begin{tabular}[c]{@{}l@{}}\textbf{Bus}\\\textbf{load } \end{tabular}          & \begin{tabular}[c]{@{}l@{}}\textbf{Change }\\\textbf{CAN bus } \end{tabular}                                           & \begin{tabular}[c]{@{}l@{}}\textbf{Test }\\\textbf{bed }\\\textbf{environment } \end{tabular}               \\ 
		\hline
		\cite{Dariz2017}
		& \begin{tabular}[c]{@{}l@{}}\begin{tabular}{@{\labelitemi\hspace{\dimexpr\labelsep+0.5\tabcolsep}}l}HMAC\\SHA1\\AES\\ DES \end{tabular}\end{tabular}                & \begin{tabular}[c]{@{}l@{}}\begin{tabular}{@{\labelitemi\hspace{\dimexpr\labelsep+0.5\tabcolsep}}l}Sniffing\\Replay\\Spoofing \end{tabular}\end{tabular}     & No                                                                                               & Yes                                                                     & Yes                                                 & \begin{tabular}[c]{@{}l@{}}Split CAN\\Frames \end{tabular} & Simulator                                                                                                   \\ 
		\hline
		\cite{Siddiqui2017}
		& \begin{tabular}[c]{@{}l@{}}\begin{tabular}{@{\labelitemi\hspace{\dimexpr\labelsep+0.5\tabcolsep}}l}AES-128\\Asymmetric key \end{tabular}\end{tabular}              & \begin{tabular}[c]{@{}l@{}}\begin{tabular}{@{\labelitemi\hspace{\dimexpr\labelsep+0.5\tabcolsep}}l}Sniffing\\Replay \\spoofing \end{tabular}\end{tabular}    & \begin{tabular}[c]{@{}l@{}}Hardware PUF\\ ECU server \end{tabular}                               & Yes                                 & \begin{tabular}[c]{@{}l@{}}During \\initialisation\\ and session \end{tabular} & \begin{tabular}[c]{@{}l@{}}Change CAN \\transceiver \end{tabular}                           & \begin{tabular}[c]{@{}l@{}}Xilinx Kintex KC705 FPGA \\hardware embedded\textasciitilde{} PUF \end{tabular}  \\ 
		\hline
		\cite{Farag2017}
		& \begin{tabular}[c]{@{}l@{}}\begin{tabular}{@{\labelitemi\hspace{\dimexpr\labelsep+0.5\tabcolsep}}l}Dynamic symmetric key\\Key generator \end{tabular}\end{tabular} & \begin{tabular}[c]{@{}l@{}}\begin{tabular}{@{\labelitemi\hspace{\dimexpr\labelsep+0.5\tabcolsep}}l}Spoofing \\Replay \\Sniffing\end{tabular}\\ \end{tabular} & No                                                                                               & Yes                                                                     & No                                                                             & No                                                                                                                     & CANoe software                                                                                              \\
		\hline
	\end{tabular}
\end{table}

CAN FD was introduced to tackle the needs of higher speed and larger data payload size. The following approaches are focused on CAN FD.

- In \cite{Woo2016}, the authors introduced an architecture supporting key management, encryption and authentication for a CAN FD bus. They used symmetric key and Authenticated Key Exchange Protocol~2 (AKEP2) to ensure distribution of keys and key freshness. They provided 16 bytes of HMAC-SHA256 tags and AES-128 to encrypt the rest of the data (47 bytes). Also, they provided an access control gateway ECU to limit the number of nodes that can access the bus. They validated their approach using three types of CAN-FD boards and CANoe software.  

-Agrawal et al.~\cite{Agrawal2019} introduced a secure CAN FD bus which uses public, private keys and groups of ECUs connected through a Gateway ECU (GECU). The GECU is used to verify session keys and key freshness for each ECU, and to forward frames between different CAN sub-buses, e.g. the high and low speed CAN buses. This approach uses 36bytes for the data payload and 28bytes for the cryptographic tag. They used CANoe software and the LPC54618 microcontroller to validate their approach.
 
-Groza et al.~\cite{Groza2017} introduced an approach for supporting CAN FD authentication. They used CANoe software to validate their approach. Their approach makes use of a group-based key sharing and generation key algorithm, a MAC algorithm to produce tags and a verification algorithm to validate received messages.

-Carel et al.~\cite{Carel2018} used the lightweight Chaskey MAC algorithm over limited capacity computational resources such as a 32-bit microcontroller. They used this algorithm with a 128bit key and compared it with the HMAC-SHA1 algorithm. They found Chaskey has a lower latency comparing with HMAC-SHA1. They used 4bytes as message counters, 16 bytes for Chaskey MAC tag and 43 bytes of actual data payload. Their approach has focused on message authentication and ignores issues of data confidentiality. 
\begin{table}[ht!]
	\centering
	\tiny
	\caption{CAN FD encryption and authentication. Improved datapayload 64-bytes, allow more space for MAC signature along with Actual data}
	\label{tbl:CAN FD encryption and authentication}
	\begin{tabular}{l|l|llllll} 
		\hline
		\textbf{ Authors }  & \textbf{Method }                                                                                                                                                    & \begin{tabular}[c]{@{}l@{}}\textbf{Attacks}\\\textbf{mitigated } \end{tabular}                                                                               & \begin{tabular}[c]{@{}l@{}}\textbf{Hardware}\\\textbf{Security }\\\textbf{Module } \end{tabular} & \begin{tabular}[c]{@{}l@{}}\textbf{Real }\\\textbf{time } \end{tabular} & \begin{tabular}[c]{@{}l@{}}\textbf{Bus}\\\textbf{load} \end{tabular} & \begin{tabular}[c]{@{}l@{}}\textbf{Change }\\\textbf{CAN bus } \end{tabular}                                      & \begin{tabular}[c]{@{}l@{}}\textbf{Test }\\\textbf{bed }\\\textbf{environment } \end{tabular}                     \\ 
		\hline
		\cite{Woo2016}
		& \begin{tabular}[c]{@{}l@{}}\begin{tabular}{@{\labelitemi\hspace{\dimexpr\labelsep+0.5\tabcolsep}}l}AES-128\\HMAC\\SHA256 \end{tabular}\end{tabular}                 & \begin{tabular}[c]{@{}l@{}}\begin{tabular}{@{\labelitemi\hspace{\dimexpr\labelsep+0.5\tabcolsep}}l}Sniffing\\Replay \\Spoofing\end{tabular}\end{tabular}     & No                                                                                               & Yes                                                                     & No                     & No                                                                                                                & \begin{tabular}[c]{@{}l@{}}Threetypes of CAN-FD\\ boards and \\CANoe software \end{tabular}                       \\ 
		\hline
		\cite{Agrawal2019}
		& \begin{tabular}[c]{@{}l@{}}\begin{tabular}{@{\labelitemi\hspace{\dimexpr\labelsep+0.5\tabcolsep}}l}Gateway ECU\\Public and\\Private keys \end{tabular}\end{tabular} & \begin{tabular}[c]{@{}l@{}}\begin{tabular}{@{\labelitemi\hspace{\dimexpr\labelsep+0.5\tabcolsep}}l}Sniffing\\Replay \\spoofing \end{tabular}\end{tabular}    & No                                                                                               & Yes                                                                     & No                                                                   & \begin{tabular}[c]{@{}l@{}}Gateway ECU \\needed \end{tabular}                                    & \begin{tabular}[c]{@{}l@{}}CANoeVector \\and LPC54618 \\micro-controller \end{tabular}                            \\ 
		\hline
		\cite{Groza2017}
		& \begin{tabular}[c]{@{}l@{}}\begin{tabular}{@{\labelitemi\hspace{\dimexpr\labelsep+0.5\tabcolsep}}l}ECU Group\\Sharing keys \end{tabular}\end{tabular}               & \begin{tabular}[c]{@{}l@{}}\begin{tabular}{@{\labelitemi\hspace{\dimexpr\labelsep+0.5\tabcolsep}}l}Spoofing \\Replay \\Sniffing\end{tabular}\\ \end{tabular} & No                                                                                               & Yes                                                                     & NO                                                                   & \begin{tabular}[c]{@{}l@{}}Group based key \\sharing \end{tabular}                              & \begin{tabular}[c]{@{}l@{}}neonTriCore controllers\\contrasted with low-end\\ Freescale S12X cores \end{tabular}  \\ 
		\hline
		\cite{Carel2018}
		& \begin{tabular}[c]{@{}l@{}}\begin{tabular}{@{\labelitemi\hspace{\dimexpr\labelsep+0.5\tabcolsep}}l}ChaskeyMAC \\Pre-shared \\128 bit key \end{tabular}\end{tabular} & \begin{tabular}[c]{@{}l@{}}\begin{tabular}{@{\labelitemi\hspace{\dimexpr\labelsep+0.5\tabcolsep}}l}Replay \\Spoofing \end{tabular}\end{tabular}              & No                                                                                               & Yes                                                                     & No                                                                   & \begin{tabular}[c]{@{}l@{}}4 bytes counters\\ 16 bytes MAC tag 43 bytes actual data \end{tabular} & \begin{tabular}[c]{@{}l@{}}ArduinoUno Rev3 \\Arduino MPro \end{tabular}                                           \\
		\hline
	\end{tabular}
\end{table}
\subsection{In-Vehicle Intrusion Detection Systems}

Using IDS to detect malicious attacks is a key approach implemented inside vehicle networks. IDS can be signature based or anomaly-based systems~\cite{Hoppe2009}. The location of the IDS is also a key decision: Host-IDS (HIDS) based and Network-IDS (NIDS) based \cite{Wu2019a}. HIDS to detect attacks~\cite{Larson2008} may not be applicable for current vehicle networks and not cost effective, as this would require a change in ECUs. Therefore, installing a NIDS, as an additional node on the  CAN bus,  such as an OBD-2 dongle can be more feasible and practical and  does not need CAN bus modification \cite{Young2019}.

An IDS can be passive i.e. only reporting attacks, or active i.e. performing actions to prevent attacks. ECUs inside the vehicle have a fixed interval to generate CAN messages even if no change occurs~\cite{Miller2016}. Thus the implementation of an IDS relies on deviations from a constant CAN traffic behaviour. Another approach uses the characteristics of the physical layer of each ECU, such as its signal and voltage profile, and compares the changes in these characteristics to detect anomalies. Müter et al.~\cite{Muter2010} have categorised the features that an IDS can use to detect attacks on the bus using the following sensors:
\begin{itemize}
\item \textbf{Format sensor}: looks at different fields in the CAN frame such the correct size of the CAN message and the value of the check sum field. 
\item \textbf{Location sensor}: indicates whether the message comes from the right CAN subsystems.
\item \textbf{Payload range sensor}: It looks at the legitimate range of values (data payload) inside the payload field.
\item \textbf{Frequency sensor}: considers the timing of the CAN message, as ECUs have a fixed frequency of data exchange/ operation.
\item \textbf{Correlation sensor}: considers messages exchanged between multiple sub-domains within a vehicular networks. The gateway sensor is used to connect different sub-networks such as a high and low CAN domain. Thus, this sensor can use this feature to verify the legitimacy of the message that is transferred from one domain to another.
\item \textbf{Protocol sensor}: is used to monitor CAN traffic and detect changes in the protocol specification, such as the order of the messages and validity of the start and end time.
\item \textbf{Plausibility sensor}: checks if payload values are in the pre-defined range, and if there is no sudden, anomalous increase in the payload.
\item\textbf{Consistency sensor}: looks at the consistency of the values in the payload field. This sensor operates in contrast to the Plausibility, looking at additional sensors to verify the consistency of the messages transferred on the CAN bus. For instance, the rotation of a tyre would indicate that the vehicle is stopped, while the GPS sensor indicates that the vehicle is moving. This approach therefore checks for consistency across multiple sensor values. 
\end{itemize}

\begin{figure}[ht!]
  \includegraphics[width=.7\linewidth]{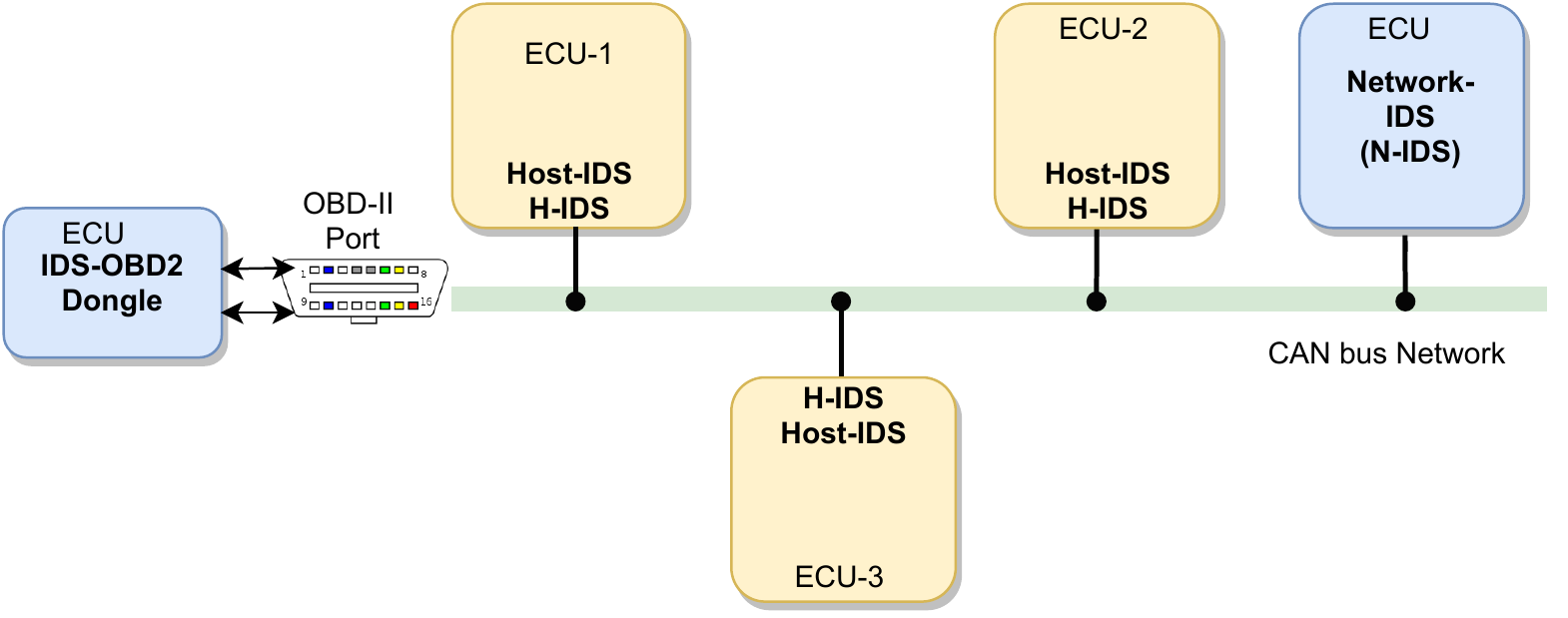}
  \caption{Positions of IDS inside automotive CAN network}
  	 \label{Fig:Position of IDS inside automotive CAN network}
\end{figure}

Below is a figure~\ref{Fig:Position of IDS inside automotive CAN network} illustrates the position of an IDS inside a CAN network. The IDS can use different features in data link layer CAN frames such as:
\begin{noindlist2}

\item \textbf{CAN identifier}:  11-bit or 29-bit value which determines the priority of the message on the bus and the content of the message. 
For example, CAN ID 0x000 is a malicious message since it can be used to occupy the bus and perform DoS attacks. Also, monitoring the broadcast intervals can be through the CAN ID frequency as it is unique across the network. 

\item \textbf{DLC}: Data Length Code is a 4-bit field and used to identify the length of the data payload. This also has a fixed value and range, as each ECU uses fixed byte size in the data payload.

\item \textbf{Data field}: It is 8 bytes maximum and it also has a fixed length and range which should not be exceeded. Anomalies can be detected if abnormal values and changes occur in the data field. The authorised identifier, the payload range and consistency, fixed DLC length and fixed rate are features that can be used to detect malicious and anomalous traffic.

\item \textbf{Timestamp}: Each CAN frame has a timestamp which can be either hardware or software based. Through this timestamp, an IDS can monitor the time intervals of CAN messages and observe any unusual behaviour. This approach is based on the observation that that ECUs have fixed broadcast intervals and thus an anomaly can be detected. 
\end{noindlist2}
\begin{figure}[ht!]
  \centering
  \includegraphics[width=.7\linewidth]{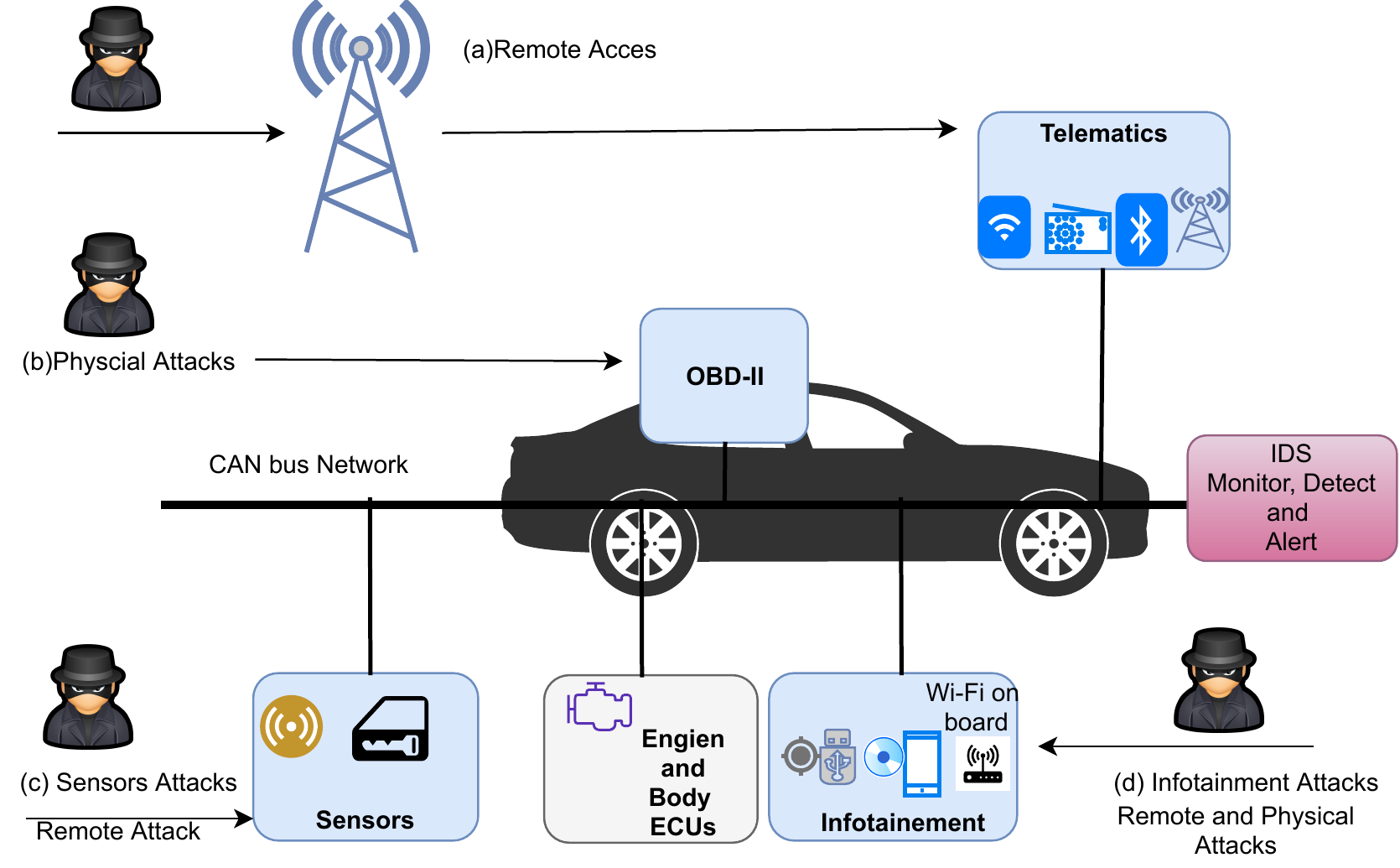}
  \caption{ IDS as ECU inside a vehicle -- based on~\cite{Song2020}}
  	 \label{Fig:IDS as ECU inside a vehicle}
\end{figure}

\subsection{\textbf{IDS based on signature}} This IDS is based on detecting a pre-defined list of attack signatures.  Although it has low false positive in the detection process, it needs to update its database signatures when new attacks emerge~\cite{S.SyedNavaz2013}. Also, the signature-based IDS needs to maintain a potentially large database of known attacks on in-vehicle networks (including potential variants of these)~\cite{Stabili2017a}. Extracting attack signatures in real time for a moving can also be a challenge and suffer from high latency. 

Studnia et al.~\cite{Studnia2018}  introduced a signature-based IDS which uses a list of signature derived from a CAN data set. However, this approach has limited benefit as the length of CAN bus words may not be known apriori.  Furthermore, this approach may fail to detect an attack if it does not sense the first part of the data exchanged as malicious packets~\cite{Avatefipour2019}. Larson et al.~\cite{Larson2008} introduced a Host IDS (HIDS) installed on each ECU and compares messages on the bus based on the CAN bus specification. This IDS monitors all incoming and outgoing traffic and compares them against the protocol specification. This approach requires changing the network topology and is not usable for real time applications. In \cite{Dagan2016}, the authors introduced an anti-spoofing system that detects malicious messages using each ECU -- by detecting CAN message ID that were not sent by the ECU itself. The ECU informs the IDS and an interrupt pulse is sent to the CAN bus to overwrite the spoofed message. 
\begin{table}[ht!]
	\centering
	\tiny
	\caption{IDS using signatures \& rules}
	\label{tbl:Signature and rules IDS}
	\begin{tabular}{l|l|lllll} 
		\hline
		\textbf{ Authors } & \textbf{Type }                                                               & \textbf{Layer }                                                              & \begin{tabular}[c]{@{}l@{}}\textbf{CAN ID / }\\\textbf{Data Payload }\end{tabular}                  & \begin{tabular}[c]{@{}l@{}}\textbf{Detection }\\\textbf{mechanism }\end{tabular}                                            & \begin{tabular}[c]{@{}l@{}}\textbf{Attacks }\\\textbf{Detected }\end{tabular} & \textbf{Prevention }  \\ 
		\hline
		\cite{Studnia2018}
		& \begin{tabular}[c]{@{}l@{}}\textbf{Signature }\\\textbf{based }\end{tabular} & \begin{tabular}[c]{@{}l@{}}DataLink Layer\\ (Controller layer) \end{tabular} & \begin{tabular}[c]{@{}l@{}}CAN frame ID\\ and dataflow \end{tabular}                                 & \begin{tabular}[c]{@{}l@{}}Derive \\signature and\\ rules match \end{tabular} & \begin{tabular}[c]{@{}l@{}}Malicious CAN ID\\ and false payload \end{tabular} & No                    \\ 
		\hline
		\cite{Larson2008}
		& \textbf{Specification }                                                      & DataLink                                                                     & \begin{tabular}[c]{@{}l@{}}Extract signature \\from CAN Open protocol\\specifications \end{tabular} & \begin{tabular}[c]{@{}l@{}}Detect attack based\\ on rules \end{tabular}                                       & \begin{tabular}[c]{@{}l@{}}Specification \\based attacks \end{tabular}        & No                    \\ 
		\hline
		\cite{Dagan2016}
		& \begin{tabular}[c]{@{}l@{}}\textbf{Access}\\\textbf{list }\end{tabular}      & Data link                                                                     & CAN ID                                                                                              & HIDS in each ECU                                                                                                           & \begin{tabular}[c]{@{}l@{}}Malicious \\CAN ID \end{tabular}                   & No                    \\
		\hline
	\end{tabular}
\end{table}

\subsection{\textbf{IDS based on Anomaly Detection}}
This method is implemented using statistical, machine learning, rule-based and physical fingerprint methods. It builds a learning model able to identify {\it abnormal} traffic, identify new patterns and predict attacks that have not been observed before.  

\subsection{\textbf{IDS using statistical approaches}} 
\begin{figure}[ht]
	\centering
	\includegraphics[width=\linewidth]{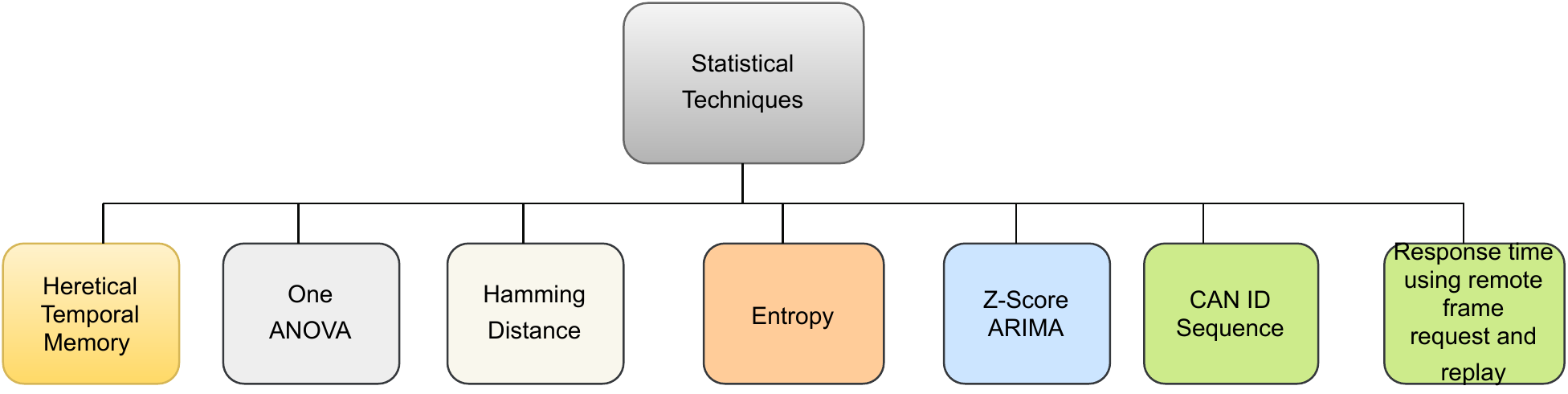}
	\caption{IDS based on Statistical Techniques}
	\label{tbl: IDS based on Statistical Techniques}
\end{figure}

This IDS learns {\it normal} behaviour of the system based on conditional statistical relationship analysis -- as outlined in figure~\ref{tbl: IDS based on Statistical Techniques}. A baseline pattern is then developed as a threshold, in case changes are detected from the norm. In CAN bus networks, statistical analysis uses CAN features such as CAN ID frequency and payload consistency. In general, ECUs have fixed intervals of time to send CAN frames. These CAN messages have a unique CAN identifier and used as a feature along with the time interval between frames, and the number of frames in each time unit~\cite{Tomlinson2018c}. Furthermore, the payloads inside CAN frames usually have consistent sequential values. A broader approach involves linking relationships between vehicular parameters such as the speed and RPM signals (under normal operation, there is a statistical correlation between RPM and speed). Finally, transmission frequency of messages, identification (ID) of messages, the number of packets received over a pre-determined time frame, message received sequence, and semantics of data fields can be used~\cite{Ji2018}.
Hence, this IDS can detect manipulated and incorrect payload values along with inconsistent use of CAN ID. The following statistical approaches have been reported in literature: 
\begin{itemize}
\item number of packets exchanged within a particular time period;
\item time interval between CAN frames; 
\item frequency of transmission using a particular CAN ID; 
\item throughput observed; 
\item response time using remote frame requests and message replay;
\item Hamming distance to compare messages;
\item analysis approach used, e.g. Entropy, Anova, Z-score, ARIMA (time wondow based moving average), Heretical Temporal Memory etc.
\end{itemize}

\noindent \textbf{CAN ID Frequency}:
-Ling and Feng~\cite{Ling2012} measure anomalies observed in traffic frequency --  to detect DoS and message injection attacks. However using their approach it is difficult to detect small volume attacks, payload manipulation attacks and impersonated ECU attacks where legitimate CAN ID messages are generated from the attacker ECU.

\noindent \textbf{Intervals between CAN frames}:
-Cho and Shin~\cite{Cho2016a}  introduced a clock-based IDS to fingerprint each ECU based on its message exchange interval. Their approach uses a least square cost function and sequential analysis technique called Cumulative Sum algorithm to detect anomalies. Similar to the previous approach, it is difficult to detect low volume injected messages and impersonated disabled ECUs. The testbed consists of an Arduino UNO board and a SeedStudio CAN shield. 

Other approaches identified in~\cite {Song2016} have used an algorithm to analyse and detect unusual time interval for specific CAN ID message transmissions. Their approach focused on CAN injection attacks.

\noindent \textbf{CAN ID frequency}:
-In \cite{Gmiden2017}, the authors have introduced an IDS that can monitor CAN message frequency for each CAN message ID used by ECUs. This approach can also detect CAN injection and DoS attacks, but small forged messages are difficult to detect since they might not alter the broadcast frequency of CAN ID. Furthermore, their approach does not consider data payload manipulation attack.

\noindent \textbf{CAN traffic behaviour over a time window}: In \cite{Taylor2015}, the author developed an IDS based on an analysis of anomalies in data flow. This work involves comparing statistical values of current CAN traffic, over a one second time window, with historical values. However, this anomaly detection over a time window cannot precisely detect small sized malicious messages.\cite{Stabili2017a}.

\noindent \textbf{Remote Frame Request and Reply intervals}: Lee et al ~\cite{Lee2018} used remote frames to detect anomalies based on the request and response intervals between frames. Since each ECU replies to a remote frame which has its CAN message id (and where the data payload field is empty), the authors calculated the average time between the request and reply to each ECU, and were able to detect anomalies based on time interval variation against a calculated average response time. They were able to detect CAN bus injection and ECU impersonation, as this would change the average response to a remote frame and in case of an impersonated ECU in the network, this would get response from both the legitimate and illegitimate ECU.

\noindent \textbf{Entropy of CAN ID and data payload behaviour}: In \cite{Muter2011}, the authors consider the CAN id and the payload as features for their approach. They measured the entropy associated with variation in CAN traffic compared to a baseline of normal CAN traffic. They have tested their approach against frame injection attacks, and they found that their approach cannot detect a small number of injected CAN messages.
 
\noindent \textbf{Entropy of CAN ID and data payload behaviour}: In \cite{Marchetti2016}, the authors evaluated an entropy-based anomaly detection IDS for in-vehicle networks, and they found that dividing CAN messages into classes and feeding them to an entropy-based anomaly detection algorithm would lead to a more accurate detection (compared to considering one class). Their anomaly detection approach calculates entropy of all CAN bus traffic (message ids) over a time window, compared to a baseline (normal) traffic over the same time window. Also, they calculated the entropy for each message id separately over a time window with a fixed number of messages. They found that measuring the entropy for each message id gives better performance in detecting smaller forged attacks, whereas considering all CAN traffic together would detect only larger sized attacks.

\noindent \textbf{Entropy of CAN ID and data payload behaviour}: Wu et al.~\cite{Wu2018} used an entropy-based IDS, with a fixed number of CAN frames over a sliding window as a baseline for their IDS. They improved the detection accuracy of the IDS based on the use of entropy calculation, by using the optimal sliding window size with a fixed number of messages. They were able to achieve a better accuracy compared to previous entropy based IDS.

\noindent \textbf{One-way ANOVA Function}: In \cite{Hsu2015}, the authors used a one-way ANOVA function to statistically determine the pattern within a data set and created a set of {\it normal} patterns to detect anomalies. They grouped CAN data set using vehicle parameters such as: fuel usage, gear ratio, engine parameters, etc to detect abnormal events for each group. 

\noindent \textbf{Hamming Distance}: Anomaly detection based on Hamming distance algorithms have also been considered by other authors, e.g. in \cite {Stabili2017a} the authors analysed CAN payload and recorded each bit in the data field. They calculate Hamming distance for each payload to each message id, and attacks were identified based on significant deviation from the calculated Hamming distance function.

\noindent \textbf{Quantized interval and the absolute Differences}: In \cite{Koyama2019}, the authors used an anomaly detection system based on quantized intervals for periodic CAN ID, and determined the absolute difference of the CAN payload values. Their approach was validated against message injection attacks and it showed positive results (using metrics such as True Positive/Negative Rates and False Positive/Negative Rates). However, they acknowledged that low volume injection attacks were difficult to detect using their approach.

\noindent \textbf{Cumulative Sum algorithm}: In \cite{Olufowobi2019}, the authors used an anomaly detection system based on the statistical cumulative sum algorithm. This is a sequential analysis technique used to support change detection. 

\noindent \textbf{ARIMA and  Z-SCORE in Defined Time Window}: In~\cite{Tomlinson2018a}, the authors used an average value for the number of times a CAN ID was broadcast mean over a time window. This was used to determine changes in the CAN ID broadcast intervals over different time windows. The authors used a Z-Score and ARIMA, along with a supervised method, to compare the mean broadcast intervals of CAN ID. They were able to detect CAN injections and dropped packet attacks.

\noindent \textbf{Heretical Temporal Memory (HTM)}: In~\cite{Wang2018}, the authors used a distributed IDS based on Heretical Temporal Memory (HTM), a technique (similar to recurrent neural networks) widely used in time series forecasting and analysis. 

\begin{table}[ht!]
	\centering
	\tiny
	\caption{IDS based on Statistical methods}
	\label{tbl:IDS based on statistical methods}
	\begin{tabular}{l|l|llll} 
		\hline
		\textbf{ Authors }                                                                                                                                                   & \textbf{Type } & \textbf{Layer }                                                        & \begin{tabular}[c]{@{}l@{}}\textbf{CANID / }\\\textbf{Data Payload }\end{tabular}                & \begin{tabular}[c]{@{}l@{}}\textbf{Detection }\\\textbf{mechanism }\end{tabular}                                                                              & \begin{tabular}[c]{@{}l@{}}\textbf{Attacks }\\\textbf{Detected }\end{tabular}                                                                                        \\ 
		\hline
		
		\begin{tabular}[c]{@{}l@{}}\textbf{CAN ID~}\\\textbf{Frequency}\\ \cite{Ling2012}\end{tabular}                                                                              & Statistics     & \begin{tabular}[c]{@{}l@{}}Data Link~\\(Controller layer)\end{tabular} & CAN ID behaviour                                                                                 & \begin{tabular}[c]{@{}l@{}}Detect malicious CAN ID\\Detect Unusual CAN frequency\end{tabular}                                                                 & \begin{tabular}[c]{@{}l@{}}\begin{tabular}{@{\labelitemi\hspace{\dimexpr\labelsep+0.5\tabcolsep}}l}Injection~\\DoS\end{tabular}\end{tabular}                         \\ 
		\hline
		\begin{tabular}[c]{@{}l@{}} \\\textbf{Entropy based}\\\textbf{anomaly}\\\textbf{ Detection }\\ \cite{Marchetti2016} \end{tabular}                                                            & Statistics     & Data Link ~                                                            & \begin{tabular}[c]{@{}l@{}}CAN ID frequency \\changed \end{tabular}                              & \begin{tabular}[c]{@{}l@{}}Provide independent \\variables\\ for entropy-based anomaly\\ detector\\ for each group or class of \\CAN messages\end{tabular}    & \begin{tabular}[c]{@{}l@{}}\begin{tabular}{@{\labelitemi\hspace{\dimexpr\labelsep+0.5\tabcolsep}}l}Malicious CAN ID\\Manipulated payload \end{tabular}\end{tabular}  \\ 
		\hline
		\begin{tabular}[c]{@{}l@{}} \\\textbf{Detecting attacks }\\\textbf{based on identifying }\\\textbf{Packet timing Anomalies}\\\textbf{ in Time Windows }\\ \cite{Tomlinson2018a} \end{tabular} & Statistics     & Data link                                                              & \begin{tabular}[c]{@{}l@{}}CAN ID broadcast mean \\in defined time window\textbf{ } \end{tabular} & \begin{tabular}[c]{@{}l@{}}Detect attack based\\ on specification rules \end{tabular}                                                                         & \begin{tabular}[c]{@{}l@{}}\begin{tabular}{@{\labelitemi\hspace{\dimexpr\labelsep+0.5\tabcolsep}}l}Injections\\DoS attacks\end{tabular}\end{tabular}                 \\ 
		\hline
		\begin{tabular}[c]{@{}l@{}} \textbf{}\\\textbf{Detecting attacks}\\\textbf{~through }\\\textbf{Hamming distance }\\\cite{Stabili2017a} \end{tabular}                                       & Statistics     & Data link                                                              & \begin{tabular}[c]{@{}l@{}}Consecutive data payloads\\ in certain CAN message ids \end{tabular}  & \begin{tabular}[c]{@{}l@{}}Compare the changes in \\Hamming distance\\ values in sequential data \\payloads of CAN message ID \end{tabular}                     & \begin{tabular}[c]{@{}l@{}}\begin{tabular}{@{\labelitemi\hspace{\dimexpr\labelsep+0.5\tabcolsep}}l}Injection\\Spoofing \end{tabular}\end{tabular}                    \\ 
		\hline
		\begin{tabular}[c]{@{}l@{}} \\\textbf{Anomaly detection}\\\textbf{~based}\\\textbf{ on ID sequence }\\ \cite{Marchetti2017a}\end{tabular}                                                    & Statistics     & Data link                                                              & \begin{tabular}[c]{@{}l@{}}Sequence between\\ CAN ID messages \end{tabular}                      & \begin{tabular}[c]{@{}l@{}}Compare the sequence \\of CAN ID\\ with the knowledge\\ acquired from real time model\textbf{ } \end{tabular}                        & \begin{tabular}[c]{@{}l@{}}\begin{tabular}{@{\labelitemi\hspace{\dimexpr\labelsep+0.5\tabcolsep}}l}Replay \\injection\end{tabular}\end{tabular}                      \\ 
		\hline
		\begin{tabular}[c]{@{}l@{}} \\\textbf{Entropy IDS based}\\\textbf{ on CAN ID}\\ \cite{Wang2019}\end{tabular}                                                                     & Statistics     & Data link                                                              & Entropy of each CAN ID                                                                           & \begin{tabular}[c]{@{}l@{}}Detect the changes\\ on each bit of the CAN ID \end{tabular}                                                                        & \begin{tabular}[c]{@{}l@{}}\begin{tabular}{@{\labelitemi\hspace{\dimexpr\labelsep+0.5\tabcolsep}}l}Flooding \\injection\end{tabular}\end{tabular}                    \\ 
		\hline
		\begin{tabular}[c]{@{}l@{}} \\\textbf{Time series algorithm. }\\\textbf{ARIMA and Z-Score }\\ \cite{Tomlinson2018a} \end{tabular}                                                             & Statistics     & Data link                                                              & \begin{tabular}[c]{@{}l@{}}Broadcast intervals\\ in time window \end{tabular}                    & \begin{tabular}[c]{@{}l@{}}Check the change \\in broadcast\\ intervals of CAN ID \end{tabular}                                                              & \begin{tabular}[c]{@{}l@{}}\begin{tabular}{@{\labelitemi\hspace{\dimexpr\labelsep+0.5\tabcolsep}}l}Drop \\ injection \end{tabular}\end{tabular}                      \\ 
		\hline
		\begin{tabular}[c]{@{}l@{}} \textbf{}\\\textbf{offset ratio and }\\\textbf{remote frame IDS }\\ \cite{Lee2018}	\end{tabular}                                                           & Statistics     & Data link                                                              & \begin{tabular}[c]{@{}l@{}}CAN request and response\\ intervals using remote frame \end{tabular} & \begin{tabular}[c]{@{}l@{}}~Time interval changes\\ and the derived change in\\ response to a remote frame \end{tabular}                                     & \begin{tabular}[c]{@{}l@{}}\begin{tabular}{@{\labelitemi\hspace{\dimexpr\labelsep+0.5\tabcolsep}}l}~Injection \\ ECU impersonation \end{tabular}\end{tabular}        \\ 
		\hline
		\begin{tabular}[c]{@{}l@{}} \\\textbf{One-Way ANOVA }\\ \cite{Hsu2015} \end{tabular}                                                                                                   & Statistics     & Data link                                                              & \begin{tabular}[c]{@{}l@{}}Data payload consistency\\ across multiple CAN signals \end{tabular}  & \begin{tabular}[c]{@{}l@{}}Compare the mean\\ of related CAN \\frames according to \\the normal\\statistical observation\\ eg. speed and engine \end{tabular} & \begin{tabular}[c]{@{}l@{}}\begin{tabular}{@{\labelitemi\hspace{\dimexpr\labelsep+0.5\tabcolsep}}l}Data payload \\manipulation \end{tabular}\end{tabular}            \\ 
		\hline
		\begin{tabular}[c]{@{}l@{}}\textbf{Cumulative Sum }\\\textbf{algorithm }\\\textbf{in defined time }\\\textbf{window}\\ \cite{Olufowobi2019}\end{tabular}                                  & Statistics~    & Data link                                                              & CAN ID sequence                                                                                  & CAN ID sequence behaviour                                                                                                                                     & \begin{tabular}[c]{@{}l@{}}\begin{tabular}{@{\labelitemi\hspace{\dimexpr\labelsep+0.5\tabcolsep}}l}Injection\\DoS attack\\Frame Fuzz. Attack \end{tabular}\end{tabular}                   \\
		\hline
	\end{tabular}
\end{table}

\vspace{0.1in}

\subsection{\bf Machine Learning-based Approaches} 

\noindent IDS based on Machine Learning (ML) can be a good choice in extracting and learning normal vs. anomalous behaviour and then providing a model to detect and predict  attacks. ML-IDS is widely used to handle large data volumes of CAN traffic with multiple features. It is useful to have a method to extract raw CAN data and pre-process it. This is particularly important as vehicle manufacturers tend not to publish detailed specification and provide guidance on how to decode raw data features. Supervised ML algorithms can be time consuming, as raw CAN data needs to be labelled, CAN attacks need to be identified and then the data needs to be labelled and classified as well. Whereas unsupervised ML approaches do not require labelled data sets, and the algorithms can find common patterns directly from data, and can use these patterns to classify traffic and identify anomalous behaviour. 
\begin{figure}[ht]
	\centering
	\includegraphics[width=.9\linewidth]{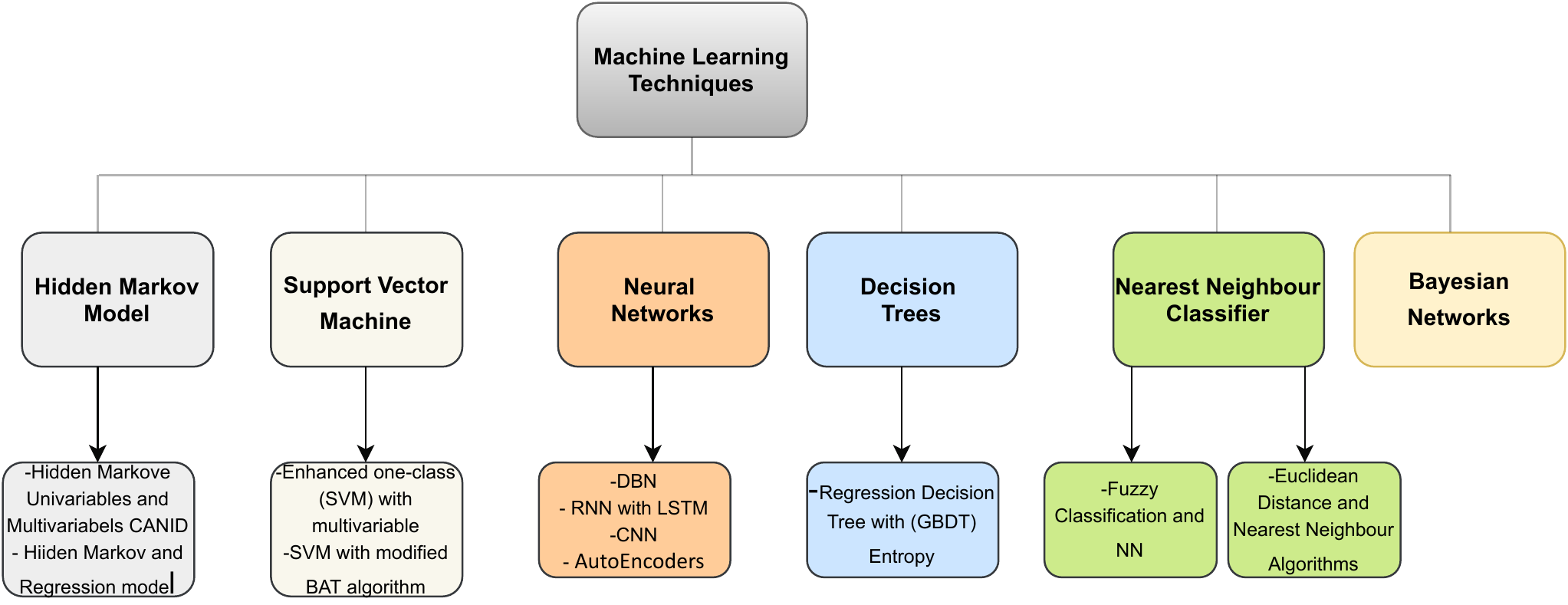}
	\caption{Machine learning based IDS}
	\label{Fig:Machine learning based IDS}
\end{figure} 

\textbf{Hidden Markov Models:} this approach works on time series data to detect anomalous behaviour. Narayanan et al.~\cite{Narayanan2016} used an IDS based on a Hidden Markov Model to build a model able to detect anomalies and raise alarms. They investigated the use of each ID separately and using multiple vehicle variables together, such as vehicle speed and RPM CAN ID messages. They evaluated their model using instant observations, against sudden changes in speed and RPM by injecting malicious message for the parameters separately. They evaluated multiple attacks by injecting malicious speed and RPM messages together. Similarly, in \cite{Levi2018} the authors used a Hidden Markov Model to learn normal vehicle behaviour and used a regression model to build a threshold for the probability of occurrence of events to identify anomalies. This is a hybrid IDS approach which the authors in~\cite{Levi2018} trained online during driving and stationary vehicle behaviour through captured data from the CAN bus. They tested the model with noise attacks to mimic a real environment.

\textbf{Support Vector Machines (SVM)}: In~\cite{Theissler2014}, the authors enhanced one-class Support Vector Machines (SVM) to work with multiple variables to classify CAN data using an unsupervised ML technique.  Their technique used unlabelled time series data from a vehicle to learn normal behaviour and detect anomalies based on deviations. Their approach used a training set from real vehicles with error free logs. They then used a model with noisy data to detect errors and anomalies in the recorded data. In \cite{Avatefipour2019}, the authors used one-class SVM, comparing their approach with a Random Forest and classical One-class SVM (leading to better detection accuracy using the True Positive Rate metric). 

\textbf{Neural Networks (NN)}: In \cite{Kang2016a} the authors used deep neural networks to learn normal patterns using unsupervised data sets, and to detect deviation from normal as anomalies. They have used an unsupervised Deep Believe Network (DBN) to pre-process the data and identify a normal pattern. To validate their approach, they inserted noise to their test data set to mimic real vehicle data. They simulated and generated CAN frames using a real world vehicle test bed and network experiments software \cite{Borazjani2014}.

In \cite{Taylor2016a}, the authors used a Recurrent Neural Network (RNN) with Long Short-Term Memory (LSTM) to detect attacks on the CAN bus. Their approach works with raw CAN bus data without the need to reduce and abstract data during the pre-processing phase of analysis.
 
In \cite{Seo2018}, the authors used Generative Advertorial Nets (GANs) to identify patterns of CAN data without classification. Their approach was able to detect anomalies based on the normal data provided. They tested their approach against Denial of Service (DoS), frame fuzzification and Spoofing attacks. Their work demonstrates that this technique is able to  detect attacks with high accuracy. 
  
In~\cite{Song2020}, the authors have used Convolutional Neural Networks (CNNs) to build an IDS able to detect sequential patterns of vehicle traffic to detect Spoofing and DoS attacks. Their approach is based on the idea that CAN traffic is fed directly to their model without the need for pre-processing. They tested their approach offline, and they acknowledged that it is difficult to use it online in current vehicles.
  
In~\cite{Pawelec2019}, the authors have used a Recurrent Neural Network (RNN) with three LSTM layers: a dropout layer and two dense layers. The former layer is used to prevent over fitting and the latter dense layer consists of 64 nodes to predict data payload for each CAN ID. Their approach uses an unsupervised technique where it does not need labelled free attack data, and trained on labelled attack data set. They argue that an ideal IDS should be able to plug inside existing vehicles to detect anomalies without the need for either reverse engineering CAN traffic or contacting the vehicle manufacturer to get CAN messages specification. 

In~\cite{Hanselmann2020}, the authors used a neural network model which consists of LSTM for CAN bus time series behaviour, auto-encoder to learn the normal behaviour of unlabelled data in unsupervised manner. Also, Exponential Linear Unit (ELU) is used for better classification as part of their  neural network model. This approach benefit from the LSTM ability to learn from previous events of CAN bus traffic as it is it is suitable for time series data and the auto-encoder ability to extract normal behaviour from unsupervised datasets. This is suitable for CAN bus data as the CAN bus data representation is not published and considered as confidential and private for car makers.

In~\cite{Lokman2019a}, the authors used unsupervised deep learning method known as Deep Contractive  Autoencoders (DCAE). Furthermore, they have evaluated their approach against DoS, frame fuzzification to impersonate attacks while they have used metrics such as Mean Square Error and Mean Absolute Error to compare between actual and predicted data. 	

In~\cite{Lokman2018}, the authors also used unsupervised deep learning method using multiple layers of Stacked Sparse Autoencoders (SSAEs). This SSAE finds meaningful data representation of CAN, which enables their model to classify attacks from normal CAN data points. Finally, their approach has shown better performance compared to basic Sparsed and Stacked Autoencoders.
   
In a different approach, authors in \cite{Loukas2017} have used a deep learning IDS models using Cloud computing to detect cyber-attacks on the CAN bus. This approach can benefit from the large number of computational resources in the cloud, while it can  be limited to provide offline detection. 
   
In \cite{Sharma2018}, Sharma and Moller have introduced an architecture for using IDS based on neural networks to detect attacks for in-vehicle networks, alert a manufacturer, surrounded connected cars and push updates to mitigate the attacks. However, this is a theoretical framework that the authors have not validated on real scenarios.
   
In \cite{Suda2018}, the authors have used CAN ID, data payload and intervals between CAN messages using an RNN algorithm. They tested their approach in a simulated environment using CAN data extracted from a real vehicle. They considered malfunction attack (false CAN ID and data payload) and flooding attack.
 
Another hybrid IDS is introduced in \cite{Weber2018},  based on a specification based IDS used to detect data payload consistency as a first stage. The authors then use an ML algorithms such as RNN, SVM and Lightweight online Detector to detect anomalies.

\textbf{Decision trees (DT):}
Decision tree-based approaches classify CAN data into two classes (normal, anomalous). DT needs a supervised labelled data set during the  training stage to be able to make decisions. In \cite{Tian2018}, authors have used a regression Decision Tree with Gradient Boosting (GBDT) technique to make a better classifier. They have used entropy to construct the decision algorithm in which they calculated the entropy of the CAN ID and the data payload time. Also, Gradient Boosting is a technique of using multiple trees and training them to get the optimal DT model. They validated their approach using real captured CAN data containing 750,000 messages. They changed the test data set by inserting random abnormal values as anomalies.

\textbf{Nearest Neighbour Classifier:}
In \cite{Martinelli2017}, the authors used fuzzy classification algorithms based on Nearest Neighbour classifiers (NNC) to discriminate attacks targeting the CAN bus. They used a data set available online which contains CAN attacks to validate their approach. They used the data payload, actual data (8bytes), as features to classify CAN traffic. They tested their approach on different attacks such as DoS, frame injection and frame fuzzification provided in the data set. They achieved a precision value between 0.85 to 1 using a neural network algorithm. However, this detector may fail to detect small forged injected messages and impersonated ECU attacks. 
In~\cite{Tomlinson2018b}, the authors used a combination of Euclidean distance and nearest neighbour algorithms. They improved the method of distance based nearest neighbour technique by categorising CAN data into four domains -- improving potential prediction accuracy by limiting to these four domains. They tested their approach against frame fuzzification attacks where they randomly change the data payload of the logged CAN messages. They considered CAN ID frequency, time between packets and data payload values as features. 

 \textbf{Bayesian Networks}: Bayesian networks is a graphical model that uses probabilistic relations between related variables. IDS based on this approach can utilise a variety of features such as the ability of Bayesian networks, e.g.: (1)  to predict the sequence (time evolution) of an event; (2) to integrate previous knowledge with probabilistic techniques, and (3) to handle missing data by encoding inter-dependencies between variables\cite{Patcha2007}
 
In~\cite{ Al-Khateeb2018}, authors have used a collection of sensor data e.g speed, geo-location and routes from a connected car. Their approach then makes use of a detection system using probabilistic Recursive Bayesian Estimation IDS. They have used three models in their approach: filtering (estimating the current event value), smoothing (estimating past event value) and prediction (estimation the likelihood of a future event).
  
In~\cite{Bezemskij2018}, the authors looked at various attack vectors on autonomous vehicles using a Bayesian network to detect and classify the type and source of the attack e.g cyber-physical attacks using sensor data from autonomous vehicle. They have used Hill-Climbing algorithm to construct a Direct Acyclic Graph (DAG) to learn the behaviour of all data sources, classify these sources and detect cyber-attacks (remote) and physical attacks based on the source of the data.
  
In~\cite{Casillo2019}, the authors used a series of probabilistic approaches based on a Bayesian network to detect attacks. They implemented a test bed based on the CARLA simulator along with support for accelerator, steer, brake sensors and IDS connected to the simulator as ECUs. Furthermore, they  evaluated their IDS based on various metrics such as true positive and true negative rates, precision, recall and F1 score.
\begin{table}[ht!]
	\centering
	\tiny
	\caption{Machine Learning based IDS}
	\label{tbl:ML based IDS}
	\begin{tabular}{l|l|l|l|l} 
		\hline
		\textbf{ Authors }                          & \textbf{Type}                                                                                                                                                                  & \begin{tabular}[c]{@{}l@{}}\textbf{Detecting}\\\textbf{threshold }\end{tabular}                                                                       & \begin{tabular}[c]{@{}l@{}}\textbf{Detection }\\\textbf{mechanism }\end{tabular}                                   & \begin{tabular}[c]{@{}l@{}}\textbf{Attacks }\\\textbf{Detected }\end{tabular}                                                                                       \\ 
		\hline
		\cite{Narayanan2016}
		& \begin{tabular}[c]{@{}l@{}}\textbf{Hidden Markov Model}\\  \end{tabular}                                                           & \begin{tabular}[c]{@{}l@{}}Univariante CAN signal\\Multivariant CAN signals~\\e.g RPM and speed\end{tabular}                                          & \begin{tabular}[c]{@{}l@{}}Deviation from the~\\sequence behaviour\end{tabular}                                    & \begin{tabular}[c]{@{}l@{}}\begin{tabular}{@{\labelitemi\hspace{\dimexpr\labelsep+0.5\tabcolsep}}l}Single injection~\\Multiple injection\end{tabular}\end{tabular}  \\ 
		\hline
		\cite{Levi2018}
		& \begin{tabular}[c]{@{}l@{}}\textbf{Hidden Markov Model~}\\\textbf{Regression Model}\\  \end{tabular}                                    & \begin{tabular}[c]{@{}l@{}}~HMM and regression \\model to build \\a threshold for \\the log probabilities\end{tabular}                                & \begin{tabular}[c]{@{}l@{}}Offline learning from~\\dataset and online~\\learning\end{tabular}                      & \begin{tabular}{@{\labelitemi\hspace{\dimexpr\labelsep+0.5\tabcolsep}}l}Noise Attack\end{tabular}                                                                   \\ 
		\hline
		\cite{Theissler2014}
		& \begin{tabular}[c]{@{}l@{}}\textbf{Enhaced SVM}\\  \end{tabular}                                                                   & \begin{tabular}[c]{@{}l@{}}CAN ID and data payload~\\Multivarinete CAN signals\end{tabular}                                                           & \begin{tabular}[c]{@{}l@{}}Deviation from \\ESVM Enhanced one-class\\Support Vector Machine~\end{tabular}          & \begin{tabular}[c]{@{}l@{}}\begin{tabular}{@{\labelitemi\hspace{\dimexpr\labelsep+0.5\tabcolsep}}l}Error and~\\Signal faults\end{tabular}\end{tabular}              \\ 
		\hline
		\cite{Chockalingam2017}
		& \begin{tabular}[c]{@{}l@{}}\textbf{O-SVM}\\  \end{tabular}                                                                      & \begin{tabular}[c]{@{}l@{}}CAN message intervals~\\and frequencies\\One Class support Vector~\\based Anomaly IDS\end{tabular}                         & \begin{tabular}[c]{@{}l@{}}Anomaly based on~\\One SVM~\\class detection\end{tabular}                               & \begin{tabular}{@{\labelitemi\hspace{\dimexpr\labelsep+0.5\tabcolsep}}l}Fuzzing\end{tabular}                                                                        \\ 
		\hline
		\cite{Avatefipour2019}
		& \begin{tabular}[c]{@{}l@{}}\textbf{~O-SVM with modified}\\\textbf{BAT alogrithm}\\  \end{tabular}                       &                                                                                                                                                       & \begin{tabular}[c]{@{}l@{}}One-class SVM \\ algorithm\end{tabular}                                & \begin{tabular}[c]{@{}l@{}}\begin{tabular}{@{\labelitemi\hspace{\dimexpr\labelsep+0.5\tabcolsep}}l}Injection\\DoS\end{tabular}\end{tabular}                         \\ 
		\hline
		\cite{Taylor2016a}
		& \begin{tabular}[c]{@{}l@{}}\textbf{Recurrent Neural Network}\\\textbf{(RNN)with}\\\textbf{long short-term memory}\\  \end{tabular}  & \begin{tabular}[c]{@{}l@{}}CAN ID and data payload\\behaviour\end{tabular}                                                                            & \begin{tabular}[c]{@{}l@{}}Devaition from the RRN\\model and observations~\\learned in LSTM mechanism\end{tabular} & \begin{tabular}[c]{@{}l@{}}\begin{tabular}{@{\labelitemi\hspace{\dimexpr\labelsep+0.5\tabcolsep}}l}Injection\\DoS\end{tabular}\end{tabular}                         \\ 
		\hline
		\cite{Kang2016a}
		& \begin{tabular}[c]{@{}l@{}}\textbf{Deep Believe~}\\\textbf{Neural Network}\\\textbf{with Probability~feature}\\  \end{tabular}        & \begin{tabular}[c]{@{}l@{}}CAN ID and data payload\\behaviour\end{tabular}                                                                            & \begin{tabular}[c]{@{}l@{}}Change from the~\\NN model patteren\end{tabular}                                        & \begin{tabular}{@{\labelitemi\hspace{\dimexpr\labelsep+0.5\tabcolsep}}l}Injection\end{tabular}                                                                      \\ 
		\hline
		\cite{Martinelli2017}
		& \begin{tabular}[c]{@{}l@{}}\textbf{Fuzzy classification}\\\textbf{Nearest Neighbor~}\\\textbf{Classification}\\  \end{tabular}    & CAN ID and data payload                                                                                                                                & \begin{tabular}[c]{@{}l@{}}Checking each byte of the~\\datapayload as features\\to detect anomalies\end{tabular}   & \begin{tabular}[c]{@{}l@{}}\begin{tabular}{@{\labelitemi\hspace{\dimexpr\labelsep+0.5\tabcolsep}}l}DoS~\\Injection\\Fuzzy\end{tabular}\end{tabular}                 \\ 
		\hline
		\cite{Tomlinson2018b}
		& \begin{tabular}[c]{@{}l@{}}\textbf{Euclidean distance~}\\\textbf{and nearest~}\\\textbf{neighbor algorithms}\\  \end{tabular}     & CAN ID frequency in time window~ ~~                                                                                                                   & \begin{tabular}[c]{@{}l@{}}Change in CAN ID~\\broadcast\\data payload\end{tabular}                                 & \begin{tabular}{@{\labelitemi\hspace{\dimexpr\labelsep+0.5\tabcolsep}}l}Fuzzy\end{tabular}                                                                          \\ 
		\hline
		\cite{Tian2018}
		& \begin{tabular}[c]{@{}l@{}}\textbf{Regression Decision Tree}\\\textbf{with Gradient Boosting}\\\textbf{(GBDT) Entropy}\\  \end{tabular} & \begin{tabular}[c]{@{}l@{}}CAN ID and data payload~\\entropy change\end{tabular}                                                                      & \begin{tabular}[c]{@{}l@{}}Entropy change\\of CAN traffic\end{tabular}                                             & \begin{tabular}[c]{@{}l@{}}\begin{tabular}{@{\labelitemi\hspace{\dimexpr\labelsep+0.5\tabcolsep}}l}Injection\\DoS\end{tabular}\end{tabular}                         \\ 
		\hline
		\cite{Weber2018}
		\begin{tabular}[c]{@{}l@{}}\\ \end{tabular} & \begin{tabular}[c]{@{}l@{}}\textbf{MLHybrid-IDS }\\  \end{tabular}                                                                  & \begin{tabular}[c]{@{}l@{}}CAN messages\\ payload sequence in \\static check module.\\RNN based IDS\\ OCSVM and Online \\Algorithm LODA \end{tabular} & \begin{tabular}[c]{@{}l@{}}Detection in time window \\and payload consistency \end{tabular}                        & \begin{tabular}[c]{@{}l@{}}\begin{tabular}{@{\labelitemi\hspace{\dimexpr\labelsep+0.5\tabcolsep}}l}Injection\\DoS \end{tabular}\end{tabular}                        \\ 
		\hline
		\cite{Wang2018}
		& \begin{tabular}[c]{@{}l@{}}\textbf{Multiple }\\\textbf{Anomaly IDS }\\\textbf{based on HMS }\\  \end{tabular}                           & \begin{tabular}[c]{@{}l@{}}Data sequence \\anomaly based \\on HMS \end{tabular}                                                                      & \begin{tabular}[c]{@{}l@{}}Multiple HMS-IDS \\for each CAN signal~\\~learn from online stream \end{tabular}        & \begin{tabular}[c]{@{}l@{}}\begin{tabular}{@{\labelitemi\hspace{\dimexpr\labelsep+0.5\tabcolsep}}l}Injection \\DoS \end{tabular}\end{tabular}                       \\
		\hline
	\end{tabular}
\end{table}

\subsection{\textbf{IDS based on Physical Characteristics}} 
This approach works at the physical layer of the CAN bus, as it builds a profile of signals and voltage signature for each ECU. It then compares the traffic with the profile for abnormal traffic. In~\cite{Sagong2019}, the authors introduced a hardware-based Intrusion Response System (IRS). This is a signal and voltage based physical layer (transceiver layer) IDS which detects attacks based on changes in characteristics for each ECU. This approach can be used to detect unusual signals at the physical layer to overcome attacks such as over current, DoS and error frame re-transmission. In \cite{Cho2017}, the authors proposed a clock-based IDS to detect anomalies. They build a fingerprint for each ECU based on measuring and extracting the periodic frequency of messages sent by ECUs. They have used the fingerprint of each ECU to build a baseline behaviour of the ECU clock using Recursive Least Square (RLS) algorithm.  Their approach uses a Cumulative Sum (CUSUM) to detect any significant deviation from the normal fingerprint baseline. In \cite{Choi2018}, the authors introduced a Voltage-IDS which is based on the use of electrical CAN signals as a fingerprint for ECUs. Their approach was also used to detect an off-bus attack where an ECU is blocked and the attacker mimics a disabled ECU.
\begin{table}
	\centering
	\tiny
	\caption{IDS based on Physical Characteristics}
	\label{tbl:IDS based on Physical Characteristics}
	\begin{tabular}{l|l|lllll} 
		\hline
		\textbf{ Authors } & \textbf{Type }                                                                                                                      & \textbf{Layer }                                                                  & \begin{tabular}[c]{@{}l@{}}\textbf{Detecting}\\\textbf{threshold }\end{tabular}                                         & \begin{tabular}[c]{@{}l@{}}\textbf{Detection }\\\textbf{mechanism }\end{tabular}                                                      & \begin{tabular}[c]{@{}l@{}}\textbf{Attacks }\\\textbf{Detected }\end{tabular}                                                                                                      & \textbf{Prevention }  \\ 
		\hline
		\cite{Sagong2019}
		& \begin{tabular}[c]{@{}l@{}}\textbf{Signal and}\\\textbf{ Voltage based } \end{tabular}                                              & \begin{tabular}[c]{@{}l@{}}Physical\\layer \\(Transceiver \\layer) \end{tabular} & \begin{tabular}[c]{@{}l@{}}Detect attacks based \\on changes on \\(signal and voltage) \\characteristics \end{tabular} & \begin{tabular}[c]{@{}l@{}}measure the unique \\signal for \\each ECU and detect \\unusual behaviour \end{tabular}            & \begin{tabular}[c]{@{}l@{}}physical layer attacks such as\\over-current\\DoS bus idle and error frame \\re-transmission. Will not work in\\ ECU impersonation attack \end{tabular} & Yes                   \\ 
		\hline
		\cite{Cho2017}
		& \begin{tabular}[c]{@{}l@{}}\textbf{Voltage Profile}\\\textbf{ for each ECU }\end{tabular}                                           & \begin{tabular}[c]{@{}l@{}}Physical \\layer \end{tabular}                        & \begin{tabular}[c]{@{}l@{}}The changes\\ of voltage \\on the line \end{tabular}                                          & \begin{tabular}[c]{@{}l@{}}each ECU has its own\\ unique voltage profile \end{tabular}                                                & \begin{tabular}[c]{@{}l@{}}Any data generated \\from unfamiliar\\ ECU voltage will \\be detected\end{tabular}                                                                      & Yes                   \\ 
		\hline
		
		\cite{Choi2018} 
		& \begin{tabular}[c]{@{}l@{}}\textbf{Electrical CAN}\\\textbf{ signals }\\\textbf{as a fingerprint }\\\textbf{for ECUs }\end{tabular} & \begin{tabular}[c]{@{}l@{}}Physical\\layer \end{tabular}                         & \begin{tabular}[c]{@{}l@{}}Change in the electric\\signal of each ECU \end{tabular}                                      & \begin{tabular}[c]{@{}l@{}}Change in the electrical signals\\ on the bus and comparing  \\the fingerprint of the ECU \end{tabular} & Off bus attack                                                                                                                                                                     & Yes                   \\
		\hline
	\end{tabular}
\end{table}

\textbf{Comparing IDS:} as mentioned previously, an IDS is used to detect malicious CAN attacks. Signature based IDS has been shown to detect attacks with low false positives however an attack signature needs to be extracted from the CAN bus. Therefore new CAN attacks can be difficult to identify. It is also difficult to detect attacks on a moving car to extract attack signature and CAN messages.  Anomaly or behavior-based IDS has the advantage that it can detect and predict attacks based on the training and learning process and can in some cases be used without the need for more training. IDS based on machine learning can benefit from raw data directly extracted from vehicles. Machine learning (ML) approaches can also handle multiple variable instances as vehicles generate large amounts of data which needs to be pre-processed in order to be meaningful. This problem can be overcome using unsupervised ML which can classify patterns and detect anomalies in unlabeled raw data. 

\section{Limitations with Current Approaches}
\label{Sec7:MitigationStrategy-Evaluation}

Based on the survey in previous sections, we now describe limitations with current approaches for securing in-vehicle systems (particularly focusing on the CAN bus). The implementation of a CAN cryptographic algorithm should consider the unique nature of the protocol, the limited network infrastructure (with support for limited data payload) and computationally constrained ECU specification. The algorithms should also consider the broadcast nature of the CAN bus, key distribution and freshness to avoid replay attacks.  Real time sensitivity is a concern inside vehicles since critical services and functions are sensitive to latency. Therefore, any countermeasure should consider these criteria.
\begin{figure}[ht!]
  \includegraphics[width=\linewidth]{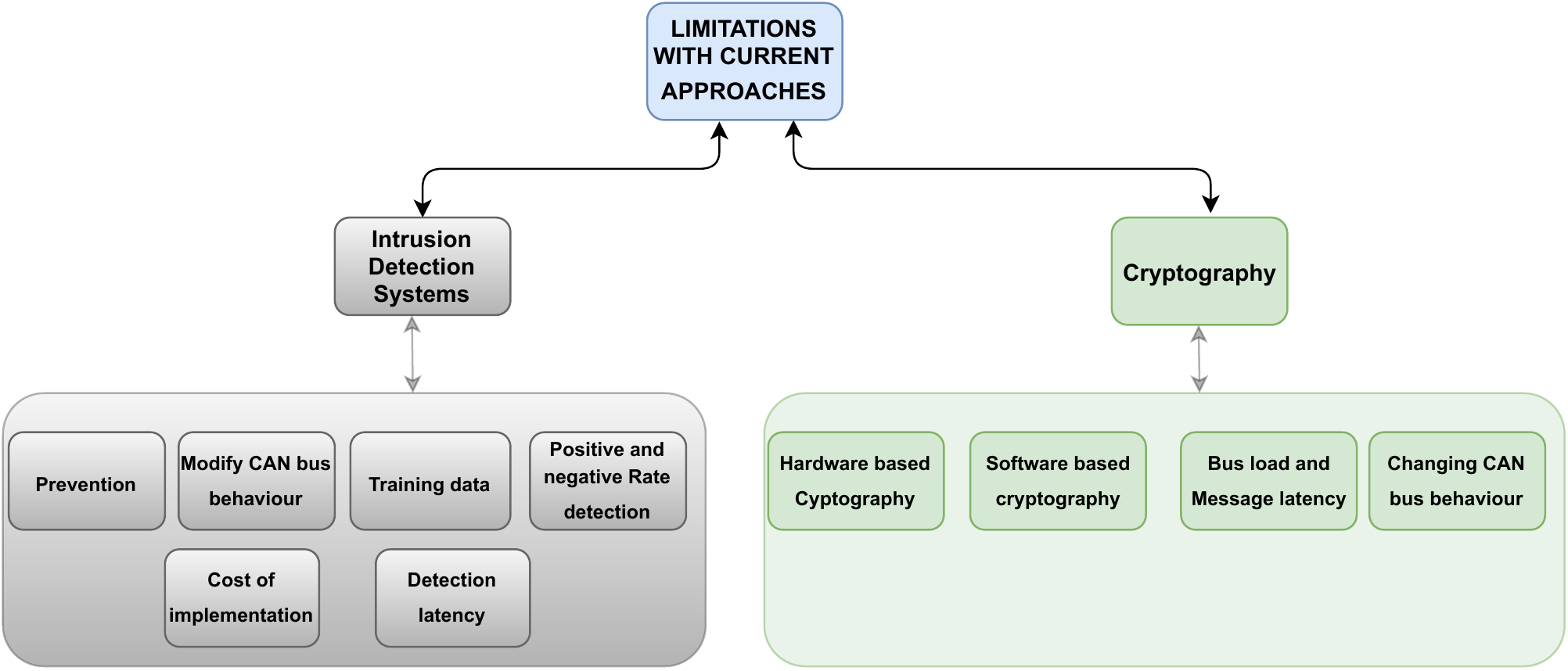}
  \caption{Research Challenges: Cryptography and IDS based approaches}
  	 \label{Fig:Challenges-Future-Directions}
\end{figure}

\subsection{Cryptography}
\textbf{Hardware based cryptography:} hardware mechanisms can be used to speed up the process of generating cryptographic functions for in-vehicle networks. Depending on the number of ECUs (typically 70), each ECU would need to be updated to include hardware-based cryptography. While this approach can increase and speed up the process of cryptographic mechanisms to meet real time needs, it is not compatible with current vehicles and the cost of implementation can be significant.  Future CAN FD boards are expected to be embedded with hardware supported security mechanisms such as AES and embedded authentication~\cite{Pfeiffer2018}. Also, better computational resources are expected in the next generation of CAN FD ECUs to handle the higher bit rate and data payload size needed. 

\textbf{Software based cryptography:} the main approaches to provide security for a CAN bus is using encryption and authentication mechanisms without the need for additional hardware or modification to existing ECUs. Authentication approaches are less computationally expensive, as they add additional information within an existing data payload. However, with the large number of ECUs inside vehicles, current approaches have not validated their approach using this large number. In \cite{Nowdehi2018}, the authors evaluated 10 CAN MAC approaches against industrial criteria and they indicated that some of these approaches are applicable to a subset of the network with a very limited number of ECUs.    
 
\textbf{Message latency:} vehicles utilise several real time functions for which latency can threaten safety on the road.  Therefore, encryption mechanisms should provide security and reduce message latency to a minimum.  This process overcomes the limited computational resources inside current vehicle ECUs. Further, the payload size of a classical CAN bus makes it difficult to add secure tags and signatures along with actual data. Therefore authentication and encryption focus on lightweight mechanisms using MAC tags without encrypting the whole payload. The CAN bus should not be loaded with extra security related messages, e.g. by splitting CAN bus messages, one message for the actual data and the other for authentication of the message. This can increase bus load two folds, and therefore affect the quality of the network. 
	
\textbf{Changing CAN bus behaviour:} some approaches change the CAN protocol by either changing the payload size or splitting a message into data and authentication messages. Furthermore, other approaches have changed CAN frame structure by replacing and inserting MAC tags and signatures inside CRC fields and CAN identifier fields. This can lead to incompatibility issues with ECUs and add additional complexity to the current CAN bus. Other approaches extended CAN payload size to 16 byte CAN+ which also causes incompatibility, as ECU controller and transceiver needs to be changed in order to support this extended frame size.	


\subsection{Intrusion Detection System}  
\textbf{Complexity:} since there is no global attack signature database, an IDS needs to collect and analyse CAN network data in order to build an attacks signature database. Furthermore, it is dangerous to perform attacks on moving vehicle to extract attack signatures and maintain them.  

\textbf{Computational resources:} very few approaches have validated their approach in a testbed that contains resources with representative computational capabilities to a vehicular network. ML based IDS may have high computational resource requirements, however ECU resources which exist inside vehicles may not be able to handle this workload. In~\cite{Wu2018} and \cite{Koyama2019} the authors suggest that dedicated hardware is needed to deploy such approaches. Furthermore, their idea is that a statistical based IDS can be a lighter approach that can be applicable in current vehicle networks. Similarly, in \cite{Jichici2019} authors have examined the use of IDS based on neural networks and they recommend that it is difficult to use them in current vehicle networks due to the large memory and computational time needed and they have suggested a dedicated hardware. In other approach \cite{Narayanan2016}, authors have inserted an IDS in OBD-2 port using raspberry pi board and they said that this approach can be used in current vehicles and embed in future vehicles. This can be a good approach to avoid the constrained power in current vehicles.
  
\textbf{Modify CAN bus behaviour}:  IDS approaches work in a passive manner where they don’t require to change the CAN bus protocol behaviour. They only monitor, detect malicious attacks and report them, for example, to the driver and fleet management centre. 

\textbf{Detection latency:} In current vehicle networks, ECUs have low computational power, thereby limiting the potential to implement IDS based on deep learning. In \cite{Loukas2017}, authors indicate that an IDS based on deep learning incurs a high latency due to increased processing time. They therefore located their deep learning model in the cloud and provide offline detection for vehicles from a central point. However, they acknowledged that offloading data from a vehicle to a cloud platform requires a stable network connection. Also, real time detection is needed for passenger safety, for which an IDS based on the cloud may not be able to offer. Alternatively, authors in \cite{Guo2019} have suggested to put Edge E-IDS in OBD-2 port as a plug in dongle to detect attacks and process data at the network edge before utilising a cloud platform. 

\textbf{Cost of implementation:} It is worth noting that an IDS can be installed in each ECU, as a Host-IDS, to detect attacks~\cite{Larson2008} -- however, this can lead to incompatibility and high deployment cost. Therefore, installing an IDS as a network node such as an IDS-OBD-2 dongle can be more feasible and practical and  does not need CAN bus modification \cite{Young2019}.  
     
\textbf{Positive and Negative Rate detection:}  IDS should work and detect attacks at runtime with the intention to minimise false positives. According to \cite{Tomlinson2018c}, a false positive rate of 0.0001 percent can cause 5 false positives every 1 hour in a CAN network broadcasts 1500 frames per second. Therefore, an IDS should carefully verify their decisions. In \cite{Ji2018}, authors evaluated four types of IDS, information entropy, CAN ID sequence, message frequency and   throughput based IDSs. They evaluated these IDSs based on the positive and negative detection rate. They tested them offline against four known attacks, packet dropping, spoofing, replay and flooding attacks. They found different negative and positive rates for each attack in each type of IDS they evaluated.
 
\textbf{Training data:}  IDS need either a database of CAN attacks (signatures) to be able to detect malicious attacks or by analysing a CAN data set offline to extract normal behaviour. While there is no global signature database of attacks, a signature database should be built by analysing normal CAN traffic along with generating various CAN attacks. However, attacks on vehicles continue to emerge and signatures of all known attacks are difficult to be maintained and updated to detect new attacks. 



\textbf{Prevention:} Some approaches focus on preventing attacks rather than passively detecting them. A combination of hardware and software-based IDS techniques are needed to be more effective to prevent attacks. 

\section{Conclusion} 
\label{Sec8:Conclusion}

Cryptographic mechanisms have been used to secure the CAN bus from attacks that originate from inside the vehicle, or when an external attacker can get access to the CAN bus. However, it may be difficult to use encryption because of the lack of computational resources in current ECUs, and the small data payload size and low data bit rate of the CAN bus network. Additionally, decision making within vehicles requires real time data analysis, and any delay due to data encryption can lead to safety issues on the road. In contrast, IDS operates as a countermeasure inside a vehicles and works in a passive manner. An IDS does not require a change in the network and protocol specifications compared to some cryptographic methods. However, some IDS based on deep learning, for instance, requires significant computational resources not available within a vehicle. 

For the current classical CAN bus vehicle networks, edge ECU devices can be used to manage countermeasures e.g (1)  monitoring and management of message authentication and encryption mechanisms, while in (2) IDS approaches, edge ECU devices can be used as a plug in edge device e.g inside OBD-2, telematics and infotainment interfaces to support edge IDS mechanisms, detect attacks, process vehicular data and push it for further analysis (e.g OEM and fleet management clouds) such as diagnostics and attack analysis. These edge ECUs devices can provide suitable resources to support countermeasures and can be used in current vehicles with limited CAN bus modifications. If ECU modification are needed, for example to support authentication and contact with edge ECU monitoring devices, OEMs and car makers should consider update ECU capabilities, e.g. over-the-air updates.
 
As CAN FD is the improved version of the classical CAN bus, it is already implemented inside many new vehicles. Many OEMs are expected to use CAN FD by 2022 in the US and Europe~\cite{ReinerZitzmann2020}. The next generation CAN FD is expected to provide better resources e.g. higher data bit rate, payload size and support for embedded encryption methods. As a result, vehicles based on CAN FD can overcome current ECU shortcomings. Other additional capabilities incude: (1) Better ECUs to handle higher data payload, (2) embedded cryptographic algorithms such as Advanced Encryption Standard (AES) and better data bit rate. Therefore, suppliers, car makers and OEMs can embed countermeasures e.g IDS, encryption and message authentication along with firewall and access control lists for external CAN bus connections.
 
Communication outside CAN bus such as telematics, infotainment  and wireless sensor interfaces along with Dedicated Short-Range Communications (DSRC) for Vehicle-2-Vehicle (V2V) and Vehicle-2-Infrastructure (V2I) should be protected, as they are entry points for data injection to the internal vehicle CAN bus network.
 
Although a number of bus architectures have been introduced -- e.g. FlexRay and LIN, it is important to highlight that the CAN bus remains the most widely used standard in the automotive industry. As outlined in this paper, a number of improvements have been made by vehicle manufacturers to the CAN bus over the years --e.g. to support higher data rates for connected and autonomous cars -- such as in CAN FD and CAN XL. Also, since CAN bus protocol is used inside both electric and autonomous cars \cite{Horton2019}, and due to the significant interest in these types of vehicles, interest in cybersecurity of the CAN bus protocol will continue to grow.

\bibliographystyle{unsrt}  
\bibliography{Survey}
\end{document}